\documentclass[aps,twocolumn,showpacs,preprintnumbers,amsmath,amssymb,nofootinbib,superscriptaddress,showkeys,noeprint]{revtex4-1}

\usepackage{hyperref}
\usepackage{float}
\usepackage{epsfig}
\usepackage{graphicx}
\usepackage{combinedgraphics}
\usepackage{epsf}
\usepackage{epstopdf}
\usepackage{mathptmx}      
\usepackage{latexsym}
\usepackage{natbib}
\usepackage{longtable}
\usepackage{dcolumn}
\usepackage[utf8]{inputenc}
\usepackage{amsmath}
\usepackage{physics}
\usepackage{url}
\usepackage{xcolor}
\usepackage{fdsymbol}
\usepackage{oplotsymbl}

\def\XXint#1#2#3{{\setbox0=\hbox{$#1{#2#3}{\int}$ }
\vcenter{\hbox{$#2#3$ }}\kern-.6\wd0}}

\newcommand{\nk}{{\bf k}}

\newcommand{\np}{{\bf p}}
\newcommand{\nK}{{\bf K}}
\newcommand{\nr}{{\bf r}}
\newcommand{\nR}{{\bf R}}

\DeclareMathVersion{nxbold}
\SetSymbolFont{operators}{nxbold}{OT1}{cmr} {b}{n}
\SetSymbolFont{letters}  {nxbold}{OML}{cmm} {b}{it}
\SetSymbolFont{symbols}  {nxbold}{OMS}{cmsy}{b}{n}

\DeclareMathAlphabet{\mathcal}{OMS}{cmsy}{m}{n}
\begin{document}

\title{Center-of-mass momentum dependence of short-range correlations
  with the coarse-grained Granada potential}

\author{P.R. Casale}
\affiliation{Departamento de F\'{\i}sica At\'omica, Molecular y Nuclear, 
Universidad de Granada, E-18071 Granada, Spain.}

\author{J.E. Amaro}
\email{amaro@ugr.es}
  \affiliation{Departamento de
  F\'{\i}sica At\'omica, Molecular y Nuclear \\ and Instituto 
  Interuniversitario Carlos I
  de F{\'\i}sica Te\'orica y Computacional \\ Universidad de Granada,
  E-18071 Granada, Spain.} 

\author{E. Ruiz Arriola}
  \affiliation{Departamento de
  F\'{\i}sica At\'omica, Molecular y Nuclear \\ and Instituto 
  Interuniversitario Carlos I
  de F{\'\i}sica Te\'orica y Computacional \\ Universidad de Granada,
  E-18071 Granada, Spain.} 

\author{I. Ruiz Simo}
  \email[Corresponding author: ]{ruizsig@ugr.es} 
  \affiliation{Departamento de
  F\'{\i}sica At\'omica, Molecular y Nuclear \\ and Instituto 
  Interuniversitario Carlos I
  de F{\'\i}sica Te\'orica y Computacional \\ Universidad de Granada,
  E-18071 Granada, Spain.}

\date{\today}

\begin{abstract} 
The effect of the center of mass motion on the high-momentum
distributions of correlated nucleon pairs is studied by solving the
Bethe-Goldstone equation in nuclear matter with the Granada
nucleon-nucleon potential.  We show that this coarse-grained potential
reduces the problem to an algebraic linear system of five (ten)
equations for uncoupled (coupled) partial waves that can be easily
solved. The corresponding relative wave functions of correlated pn, pp
and nn pairs are computed for different values of their CM momentum.
We find that the pn pairs dominate the high-momentum tail of the
relative momentum distribution, and that this only depends marginally
on center of mass momentum. Our results provide further justification
and agreement for the factorization approximation commonly used in the
literature.  This approximation assumes that the momentum distribution
of nucleon pairs can be factorized as the product of the center of
mass momentum distribution and the relative momentum distribution.
\end{abstract}


\maketitle

\section{Introduction}
The nucleon-nucleon (N-N) correlations and the attempts to reduce it
to a problem of self-consistent fields, similar to the Hartree method
\cite{Hartree:1928a, Hartree:1928b} of Atomic Physics constitute an
old topic in Nuclear Physics \cite{Jastrow:1950zz,Jastrow:1951vyc,
  Jastrow:1955zz, Brueckner:1955zzd, Bethe:1956zz, bethe1957effect,
  Brueckner:1958zz, Bethe:1965zz}.  It has experienced a revival in
the last two decades due to the advent of high energy electron beams
accelerators facilities such as the Continuous Electron Beam
Accelerator Facility (CEBAF) at Jefferson Lab (JLab)
\cite{Leemann:2001dg, Mecking:2003zu}, both from the experimental
\cite{Kester:1995zz, vanLeeuwe:1995nv, Onderwater:1998zz,
  Blomqvist:1998gq, Starink:2000qhh, Tang:2002ww, Egiyan:2005hs,
  Shneor:2007tu, Subedi:2008zz, Baghdasaryan:2010nv, Fomin:2011ng,
  Boeglin:2011mt, Hen:2012yva, Korover:2014dma, Hen:2014nza,
  Ye:2017mvo, Cohen:2018gzh} and theoretical
\cite{Schiavilla:1985gb,Ramos:1989hqs, Vonderfecht:1993qrl,
  Frankfurt:1993sp, Stoitsov:1993zz, Muther:1995zz, Muther:1995bk,
  Ryckebusch:1995usx, Giusti:1999sv, Dewulf:2003nj, Schiavilla:2006xx,
  Kortelainen:2007rn, Alvioli:2007zz, Wiringa:2008dn,
  Frankfurt:2008zv, Weinstein:2010rt, Feldmeier:2011qy,
  Sargsian:2012sm, Alvioli:2012qa, Vanhalst:2012ur, Wiringa:2013ala,
  White:2013xxa, Sammarruca:2014roa, Cai:2015xga, Sammarruca:2015hba,
  Neff:2015xda, Colle:2015ena, Alvioli:2016wwp, Weiss:2016obx,
  Chen:2016bde, Mosel:2016uge, Artiles:2016akj, Ding:2016oxp,
  RuizSimo:2016vsh, Cruz-Torres:2017sjy, Stevens:2017orj,
  Rios:2017muz, CiofidegliAtti:2017tnm, RuizSimo:2017tcb} points of
view (for recent reviews the reader is referred to
\cite{Arrington:2011xs, Atti:2015eda, Hen:2016kwk, Fomin:2017ydn}).
 
 From the theoretical side, the short-range N-N correlations (SRCs)
 are very important in different contexts of Nuclear Physics, covering
 aspects from fundamental to applied Nuclear Physics: properties of
 nuclear matter \cite{Bethe:1971xm, Jeukenne:1976uy, Ramos:1989hqs,
   Dewulf:2003nj, Vonderfecht:1991zz}; high momentum components in the
 nuclear wave function \cite{Fantoni:1984zz, Muther:1995zz,
   Wiringa:2013ala, Benhar:1986jha, VanOrden:1979mt, Sargsian:2012sm};
 implications in nuclear astrophysics and evolution of neutron stars
 through the equation of state of nuclear and neutron matter
 \cite{Riffert:1996jf, Frankfurt:2008zv, Mukherjee:2008un,
   Shen:2011kr, Ropke:2014fia, Hen:2016ysx}; calculations of symmetry
 energy and pairing gaps in nuclear and neutron matter
 \cite{Hen:2014yfa, Cai:2015xga, Ding:2016oxp, Rios:2017muz}; models
 of relativistic heavy-ion collisions \cite{Broniowski:2010jd};
 calculations of nuclear matrix elements for neutrino-less double beta
 decay \cite{Simkovic:2009pp, Kortelainen:2007mn}; description of
 $(e,e^\prime)$, $(e,e^\prime N)$ and $(e,e^\prime NN)$ reactions
 \cite{Frankfurt:1993sp, Weinstein:2010rt, Giusti:1999sv,
   Ryckebusch:1995usx, Colle:2015ena}; and, recently, the universality
 of the N-N SRCs and its connection with factorization properties of
 the nuclear wave functions and momentum distributions, and with the
 nuclear contacts \cite{Tan:2008a, Tan:2008b,Tan:2008c,
   Alvioli:2011aa, Alvioli:2013qyz, Weiss:2015mba, Alvioli:2016wwp,
   Weiss:2016obx, Weiss:2017huz}, just to mention a few of them.
 
 The main methods to tackle this complex problem have been
 traditionally two: the use of Jastrow correlation functions with
 adequate behaviors at short and long inter-nucleon distances applied
 to Slater determinants of single-particle wave-functions within
 variational approaches \cite{Jastrow:1955zz, Fantoni:1974jv,
   Fantoni:1975a, Guardiola:1980ma, Guardiola:1981ujn, Benhar:1991iw,
   Stoitsov:1993zz, Benhar:1994hw, Ryckebusch:1995usx,
   Guardiola:1996dq, Bishop:1998za, Vanhalst:2012ur}; and the
 Brueckner theory of nuclear matter \cite{Brueckner:1958zz,
   Brueckner:1954zz, Brueckner:1955zze} by solving the Bethe-Goldstone
 (B-G) equation \cite{Bethe:1956zz, Goldstone:1957zz, Dahll:1969hmo}
 or the effective interaction encoded in the G-matrix formalism
 \cite{Kohler:1961a, Haftel:1970zz, Jeukenne:1974zz, Nakayama:1984xgi,
   Hosaka:1985xwy, Nakayama:1987czg, Boersma:1993yy}.

There are also other useful methods to deal with this problem: the
similarity renormalization group (SRG) methods, which can provide
phase equivalent potentials that soften the short-range interaction,
thus avoiding the problems related with the hard core
\cite{Bogner:2006pc, Timoteo:2011tt, Neff:2015xda}; and the \textit{ab
  initio} variational Monte Carlo methods, which solve exactly the
non-relativistic many-body problem for light nuclei when a particular
N-N interaction is given \cite{Carlson:1993zz, Forest:1995zz,
  Quaglioni:2009mn, Hagen:2010gd, Leidemann:2012hr, Barrett:2013nh}.

Our aim in this work is to extend our two previous papers
\cite{RuizSimo:2016vsh, RuizSimo:2017tcb} on the short-range
correlations in the independent pair approximation picture
\cite{Viollier:1976ab} for the case when the total center-of-mass (CM)
momentum of the nucleon pair is different from zero, $\nK_{\rm CM}\ne
\mathbf{0}$, and to study its effect on the high-momentum components
of the relative wave function in momentum space. To this end we make
use of the coarse-grained Granada potential of
Ref. \cite{Perez:2013mwa}, and we use the angular average of the
Pauli-blocking operator appearing in the B-G equation. This
approximation has been widely used in the past by many other authors
\cite{Brueckner:1958zz, Bhargava:1967a, Kallio:1969mis, Haftel:1970zz,
  Jeukenne:1974zz, Muther:1995bk, Alonso:2003aq}.  Other successful
attempts to solve this problem without resorting to the approximation
of the angular average of the Pauli-blocking operator have been
explored in Refs. \cite{werner1959solution, Cheon:1988hn,
  Schiller:1998ff, Suzuki:1999jb, Sammarruca:2000dd,
  Stephenson:2004xs, White:2014oca}.
  
Given the fact that the coordinate space method is not widely used and
it is essential for our coarse-grained treatment of the N-N
interaction, we provide in two appendices all the necessary material
to make the paper as self-contained as possible, in order to target it
at a wider audience.

Therefore, the structure of this paper is as follows: in
Section~\ref{sect:framework} we describe the formalism to solve the
B-G equation with the angular average of the Pauli-blocking operator
by performing a partial wave expansion for the radial part of the
correlated relative wave function; in Section~\ref{Sec:mom_space} we
derive the correlated wave function in momentum space by applying the
Fourier transform to the wave function in coordinate representation,
and obtaining the high-momentum components in the relative wave
function induced by the SRCs; in Section~\ref{Sec:results} we present
our results and discuss them in depth; in
Section~\ref{Sec:conclusions} we draw our conclusions; finally, we
provide two final appendices~\ref{formal_derivation} and
\ref{derivation_radial} at the end of the paper.

\section{Theoretical framework}\label{sect:framework}
\subsection{General Formalism}\label{sect:general}
The Brueckner reaction matrix G plays a crucial role in describing
nucleon-nucleon scattering within the nuclear medium. It is a
fundamental concept in nuclear many-body theory, and its properties
are closely related to the Bethe-Goldstone equation. The G matrix is a
solution to the Bethe-Goldstone equation and is essentially a modified
nucleon-nucleon scattering matrix that takes into account the
influence of the nuclear medium on nucleon interactions. It can be
thought of as a generalization of the Lippmann-Schwinger equation,
which is commonly used to describe scattering in vacuum.

The Brueckner G-matrix is usually represented in operator form as the
well-known B-G equation:
\begin{equation}\label{G-operator}
G = V + V\, \frac{Q}{E-H_0}\, G
\end{equation}
where $G$ is the G-matrix or effective interaction; $V$ represents the
nucleon two-body potential; $Q$ is the Pauli- blocking operator that
prevents scattering over two-particle occupied states; $E$ represents
the energy eigenvalue of the two-nucleon system; and, finally, $H_0$
is the unperturbed or free Hamiltonian containing the sum of the
kinetic energies of the two independent particles.  The action of the
Pauli-blocking operator over uncorrelated two-particle states
$\left|\nk_1,\nk_2\right\rangle$ is given by
\begin{equation}\label{action_Q_operator}
Q\left|\nk_1,\nk_2\right\rangle=
\left\{
\begin{array}{c}
\left|\nk_1,\nk_2\right\rangle \qquad \textrm{if both} \quad \left|\nk_i\right|>k_F\\
0 \qquad\qquad\quad\;\; \textrm{otherwise}
\end{array}
\right.
\end{equation}

It is well-known that due to translational invariance
symmetry~\cite{Walecka1995}, if the N-N potential only depends on the
relative coordinate $\nr$ of the two-nucleon system and not on the CM
coordinate $\nR_{\rm CM}$, then the CM momentum of the two-nucleon
system is conserved, i.e, it is a constant of motion. This means in
practice that the CM motion can be described by a plane wave in
nuclear matter, and that the correlated total wave function is
separable into a product of a plane wave for the CM motion and a
correlated relative wave function, $\psi_{\,\nK_{\rm CM},\nk}(\nr)$,
depending explicitly on the relative coordinate $\nr$, the initial
relative momentum $\nk$, but also on the total momentum $\nK_{\rm CM}$
of the nucleon pair (see, for instance, Refs. \cite{RuizSimo:2016vsh,
  Walecka1995}).  The dependence on the total momentum $\nK_{\rm CM}$
of the nucleon pair in the relative wave function can be understood if
one observes that the Pauli-blocking operator $Q$ explicitly depends
on the CM momentum in the eigenket representation of CM and relative
momenta for the two-nucleon system $\left|\nK_{\rm
  CM},\nk\right\rangle$. The relationships between these two different
representations for the two-nucleon system are given by
\begin{eqnarray}
\nR_{\rm CM}&=&\frac12(\nr_1+\nr_2), \qquad \nK_{\rm CM}=\nk_1+\nk_2,
\nonumber\\
\nr&=&\nr_1-\nr_2, \qquad \nk=\frac12(\nk_1-\nk_2),\nonumber\\
\left\langle\nR_{\rm CM},\nr\right.\left|\nr_1,\nr_2\right\rangle&=&
\delta^3\left(\nR_{\rm CM}-\frac12(\nr_1+\nr_2) \right) 
\delta^3\left(\nr-\left(\nr_1-\nr_2 \right) \right),\nonumber\\
\left\langle\nK_{\rm CM},\nk\right.\left|\nk_1,\nk_2\right\rangle&=&
\delta^3\left(\nk_1-\nk-\frac{\nK_{\rm CM}}{2} \right) 
\delta^3\left(\nk_2+\nk-\frac{\nK_{\rm CM}}{2} \right)\nonumber\\
&=&\delta^3\left(\nK_{\rm CM}-\left(\nk_1+\nk_2 \right) \right)
\delta^3\left(\nk-\frac12\left(\nk_1-\nk_2\right)\right),\nonumber\\
\left\langle\nR_{\rm CM},\nr\right.\left|\nK_{\rm CM},\nk\right\rangle&=&
\left\langle\nr_1,\nr_2\right.\left|\nk_1,\nk_2\right\rangle=
\frac{e^{i\, \nk_1\cdot\nr_1}}{(2\pi)^{\frac32}}
\frac{e^{i\, \nk_2\cdot\nr_2}}{(2\pi)^{\frac32}}=\nonumber\\
&=&\frac{e^{i\, \nK_{\rm CM}\cdot\nR_{\rm CM}}}{(2\pi)^{\frac32}}
\frac{e^{i\, \nk\cdot\nr}}{(2\pi)^{\frac32}}.\label{CM_single_part_represent}
\end{eqnarray}

The advantage of using the CM and relative momenta representation for
initial and final two-nucleon states $\left|\nK_{\rm
  CM},\nk\right\rangle$ is based on the fact that then the Brueckner
G-matrix can be solved solely for the relative wave function, at the
price of introducing a dependence on the total momentum $\nK_{\rm CM}$
through the Pauli-blocking operator $Q$. But the SRCs are completely
incorporated in the relative wave function $\psi_{\,\nK_{\rm
    CM},\nk}(\nr)$.

On the other hand, if one insists on working with the two-nucleon
momenta eigenket representation $\left|\nk_1,\nk_2\right\rangle$, the
action of the Pauli-blocking operator on these states is much simpler
(see eq.(\ref{action_Q_operator})), but then one spoils the simplicity
of the N-N potential matrix elements in the CM and relative
coordinates representation
\begin{equation}\label{pot_mat_elem}
\left\langle\nR^\prime_{\rm CM},\nr^\prime\right|\left.V\right|
\left.\nR_{\rm CM},\nr\right\rangle=\delta^3(\nR^\prime_{\rm CM}-
\nR_{\rm CM})\, V(\nr)\, \delta^3(\nr^\prime-\nr)
\end{equation}
if the potential is, additionally, local in the relative coordinate,
as the one we use in this work and in our previous ones
\cite{RuizSimo:2016vsh, Perez:2013mwa, Perez:2013jpa,
  RuizSimo:2017tcb}. Furthermore, with the latter approach one has to
self-consistently solve the B-G equation for a correlated two-body
wave function, $\Psi(\nr_1,\nr_2)$, depending on the coordinates and
quantum numbers of the single nucleons (see for example eq. (3) of
Ref. \cite{RuizSimo:2016vsh}), instead of solving a one-body relative
wave function with external inputs $(\nK_{\rm CM},\nk)$ in a single
relative coordinate $\psi_{\,\nK_{\rm CM},\nk}(\nr)$.
 
 The B-G equation in operator form, given in eq. (\ref{G-operator}),
 is equivalent to the following equation for the perturbed or
 correlated two-nucleon state
 \begin{eqnarray}
 \left| \Psi_{\nK_{\rm CM},\nk} \right\rangle &=&
 \left| \nK_{\rm CM},\nk \right\rangle + 
 \int d^3K^{\prime}_{\rm CM}\, d^3k^\prime\; 
 \frac{Q(\nK^{\prime}_{\rm CM},\nk^\prime)}{\frac{\left(
 \nK^2_{\rm CM}-\nK^{\prime\,2}_{\rm CM}\right)}{2M_T}
 +\frac{\left(\nk^2-\nk^{\prime\,2}\right)}{2\mu}}\nonumber\\
 && \left| \frac{\nK^\prime_{\rm CM}}{2}+\nk^\prime,
 \frac{\nK^\prime_{\rm CM}}{2}-\nk^\prime \right\rangle
\left\langle \nK^\prime_{\rm CM},\nk^\prime\right| V 
 \left| \Psi_{\nK_{\rm CM},\nk} \right\rangle \label{corr_2nucleon_state}
 \end{eqnarray}
 where $ \left| \nK_{\rm CM},\nk \right\rangle$ is the unperturbed or
 uncorrelated state, $Q(\nK^\prime_{\rm CM},\nk^\prime)$ is the
 Pauli-blocking operator depending on the CM and relative momenta and
 is given by
 \begin{equation}\label{Q_CM_rel_momenta}
 Q(\nK^\prime_{\rm CM},\nk^\prime)=
 \theta\left(\left|\frac{\nK^\prime_{\rm CM}}{2}+\nk^\prime \right| - k_F \right)
 \theta\left(\left|\frac{\nK^\prime_{\rm CM}}{2}-\nk^\prime \right| - k_F \right),
 \end{equation}
 with $\theta(x)$ the Heaviside or step function. Additionally, in
 eq. (\ref{corr_2nucleon_state}), $\mu=\frac{M_N}{2}$ is the reduced
 mass of the two-nucleon system and $M_T=2M_N$ is the total mass of
 it. Finally, the integration over the off-shell states runs over the
 total CM and relative momenta of the two-nucleon pair. It is also
 important to notice that, despite its dependence, the ket $\left|
 \frac{\nK^\prime_{\rm CM}}{2}+\nk^\prime, \frac{\nK^\prime_{\rm
     CM}}{2}-\nk^\prime \right\rangle$ is not a ket belonging to the
 CM and relative momenta representation as they are, for instance,
 $\left| \nK_{\rm CM},\nk \right\rangle$ or $\left\langle
 \nK^\prime_{\rm CM},\nk^\prime\right|$, but it is a ket belonging to
 the two-nucleon single momenta representation $\left| \nk_1,\nk_2
 \right\rangle$ with $\nk_1=\frac{\nK^\prime_{\rm CM}}{2}+\nk^\prime$
 and $\nk_2= \frac{\nK^\prime_{\rm CM}}{2}-\nk^\prime$.

 The formal derivation of eqs.~(\ref{corr_2nucleon_state}) and
 (\ref{corr_relative_ket_equation}) (see below) starting from
 eq.~(\ref{G-operator}) is deferred to
 appendix~\ref{formal_derivation}.
 
 Now, to get rid of the CM momentum in
 eq. (\ref{corr_2nucleon_state}), it is completely necessary to assume
 that the potential is of the form given by eq. (\ref{pot_mat_elem}),
 i.e, a local potential not depending on the CM
 coordinate~\footnote{The assumption of locality is also exploited
 specifically in other computational frameworks such as the Monte
 Carlo approach~ \cite{Carlson:1993zz, Forest:1995zz,
   Quaglioni:2009mn, Hagen:2010gd, Leidemann:2012hr, Barrett:2013nh}.
 As already mentioned in the introduction, the SRG method reduces the
 core at the expense of introducing strong non-localities. On the
 contrary, the Monte Carlo method needs a strong repulsive core below
 $0.5$ fm within a purely local interaction scheme. The main advantage
 of the coarse graining approach is that the quality of the N-N
 interaction fits is compatible with the assumption that possible
 non-localities take place at distances below the coarse graining
 scale of $\Delta r=0.6$ fm, and {\it simultaneously} reduces the
 short distance core. As it will be shown, this has the further
 practical advantage of reducing tremendously the computational
 effort.}.  With this assumption, which is right for the kind of
 coarse-grained potential used in this work, one can obtain a similar
 equation to that given in (\ref{corr_2nucleon_state}) but for the
 relative ket, removing as much as possible the dependence on the CM
 momentum. The final result is
 \begin{equation}\label{corr_relative_ket_equation}
 \left| \psi_{\nK_{\rm CM},\nk} \right\rangle=
 \left| \nk \right\rangle + \int d^3k^\prime \; 
 \frac{Q(\nK_{\rm CM},\nk^\prime)}{k^2-k^{\prime\,2}}\,
 \left| \nk^\prime \right\rangle \left\langle \nk^\prime \right|
 2\mu V \left| \psi_{\nK_{\rm CM},\nk} \right\rangle
 \end{equation}
 where $\left| \psi_{\nK_{\rm CM},\nk} \right\rangle$ is the relative
 part of the perturbed $\left| \Psi_{\nK_{\rm CM},\nk} \right\rangle$
 state of eq.  (\ref{corr_2nucleon_state}); $\left| \nk \right\rangle$
 is the plane wave state with definite relative momentum $\nk$; and
 the integral over the off-shell states $\left| \nk^\prime
 \right\rangle$ now only runs over the relative two-nucleon
 momentum. It is not possible to remove completely all the dependence
 on the total momentum $\nK_{\rm CM}$ because of the presence of the
 Pauli-blocking operator, as it is obvious from
 eq. (\ref{corr_relative_ket_equation}).

\subsection{Angular average of the Pauli-blocking operator}
\label{ang_average}

The main problem when solving eq. (\ref{corr_relative_ket_equation}),
besides the form of the N-N potential, is the additional angular
dependence introduced by the Pauli-blocking function given in
eq. (\ref{Q_CM_rel_momenta}). Indeed, the function $Q(\nK_{\rm
  CM},\nk^\prime)$ depends explicitly on the polar angle between the
vectors $\nK_{\rm CM}$ and $\nk^\prime$ and, therefore, breaks
rotational invariance in eq.  (\ref{corr_relative_ket_equation}) even
for a central N-N potential \cite{werner1959solution}, causing a
mixing among different partial waves if one tries to perform a partial
wave expansion to solve eq.
(\ref{corr_relative_ket_equation}). Although some few authors have
solved the problem in general for different (non-central) N-N
potentials \cite{Cheon:1988hn, Schiller:1998ff, Suzuki:1999jb,
  Sammarruca:2000dd, Stephenson:2004xs, White:2014oca}, we are going
to use in this work the approximation, first proposed by Brueckner
\cite{Brueckner:1958zz} and also taken by many other authors
\cite{Bhargava:1967a, Kallio:1969mis, Haftel:1970zz, Jeukenne:1974zz,
  Muther:1995bk, Alonso:2003aq}, of substituting the angle-dependent
Pauli-blocking function $Q(\nK_{\rm CM},\nk^\prime)$ by its angular
average around the direction defined by the CM momentum.  This
approximation amounts to perform the replacement
\begin{equation}\label{def_angular_average}
Q(\nK_{\rm CM},\nk^\prime)\longrightarrow\,
\overline{Q}(K_{\rm CM},k^\prime)\equiv
\frac{1}{4\pi}\int d\Omega_{\hat{k}^\prime}\;
Q(\nK_{\rm CM},\nk^\prime)
\end{equation}
in eq. (\ref{corr_relative_ket_equation}). With this replacement, now
the angle-averaged Pauli-blocking function $\overline{Q}(K_{\rm
  CM},k^\prime)$ only depends on the magnitude of both CM and relative
momenta, but not on the angle between both vectors $(\widehat{\nK_{\rm
    CM},\nk^\prime})$.
 
\begin{figure}[ht]
\includegraphics[width=9.15cm]{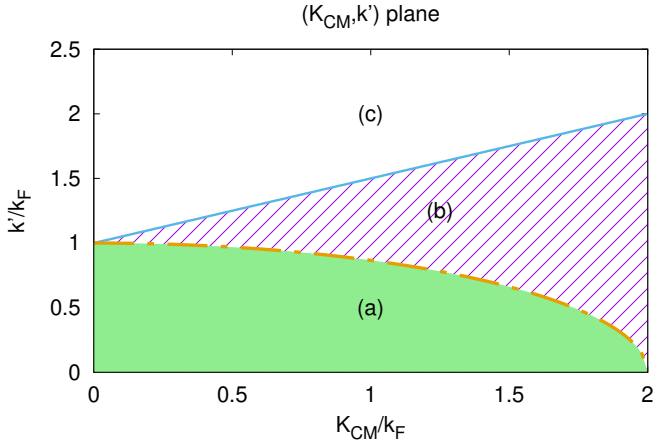}
\caption{Representation of the different zones of the phase space in
  the variables $(K_{\rm CM},k^\prime)$ where the angular average of
  the Pauli-blocking operator $\overline{Q}(K_{\rm CM},k^\prime)$
  takes different values. To avoid specifying a definite value for the
  Fermi momentum $k_F$, the axes of the plot are represented in units
  of the Fermi momentum.}
\label{Fig:zones}
\end{figure}

Obviously, the angular average of a function (see
eq. (\ref{Q_CM_rel_momenta})) that can only take the values $1$ or $0$
is another function that can \emph{continuously} reach values between
$0$ and $1$ depending on the different zones of the $(K_{\rm
  CM},k^\prime)$-plane, as depicted in Fig.  \ref{Fig:zones}. The
functional form of $\overline{Q}(K_{\rm CM},k^\prime)$ for $K_{\rm
  CM}<2\, k_F$ is well-known since the times of Brueckner
\cite{Brueckner:1958zz},

\begin{equation}\label{Qbar_function}
\overline{Q}(K_{\rm CM},k^\prime)=
\left\{
\begin{array}{c}
0 \qquad \quad \quad {\rm if} \quad 0 \leqslant k^\prime \leqslant 
\sqrt{k^2_F - \frac{K^2_{\rm CM}}{4}} \\
\frac{\frac{K^2_{\rm CM}}{4}+k^{\prime\,2}-k^2_F}{K_{\rm CM}\, k^\prime}
\;\;\; {\rm if} \;\; \sqrt{k^2_F - \frac{K^2_{\rm CM}}{4}}
< k^\prime \leqslant k_F+\frac{K_{\rm CM}}{2}  \\
1 \qquad \qquad \qquad {\rm if} \qquad k^\prime > 
k_F+\frac{K_{\rm CM}}{2}.
\end{array}
\right.
\end{equation}
We restrict our study in this work to the zone $K_{\rm CM}<2\, k_F$,
corresponding to the abscissas axis range of Fig. \ref{Fig:zones},
because this is the maximum total (CM) momentum of an uncorrelated
nucleon pair in the ground state of nuclear matter, i.e, when both
nucleons have their largest single momenta, $k_F$, in parallel
direction.

The region labeled by (a) in Fig. \ref{Fig:zones}, where the
angle-averaged Pauli-blocking function $\overline{Q}(K_{\rm
  CM},k^\prime)=0$, corresponds to the forbidden region for two
nucleons to scatter below the Fermi momentum $k_F$, i.e, this region
limited by the quarter of an ellipse with semi-major and semi-minor
axes $2\,k_F$ and $k_F$, respectively, corresponds to the region where
the single off-shell nucleon momenta satisfy that both
$\left|\nk^\prime_i\right|<k_F$, and thus this region is excluded by
the Pauli-blocking operator (cf. eq. (\ref{action_Q_operator})).

The region labeled by (c) in Fig. \ref{Fig:zones}, where the function
$\overline{Q}(K_{\rm CM},k^\prime)$ takes the value $1$, corresponds
to the totally allowed region for two off-shell nucleons to scatter
above the Fermi momentum $k_F$, i.e, this region bounded from below by
the straight line $k^\prime=k_F+\frac{K_{\rm CM}}{2}$ is the region
where the single off-shell nucleon momenta always satisfy that both
$\left|\nk^\prime_i\right|>k_F$, and thus this region is fully
included by the Pauli-blocking operator
(cf. eq. (\ref{action_Q_operator})).

\begin{figure*}[!ht]
\centering
\begin{tabular}{cc}
\includegraphics[width=7.5cm]{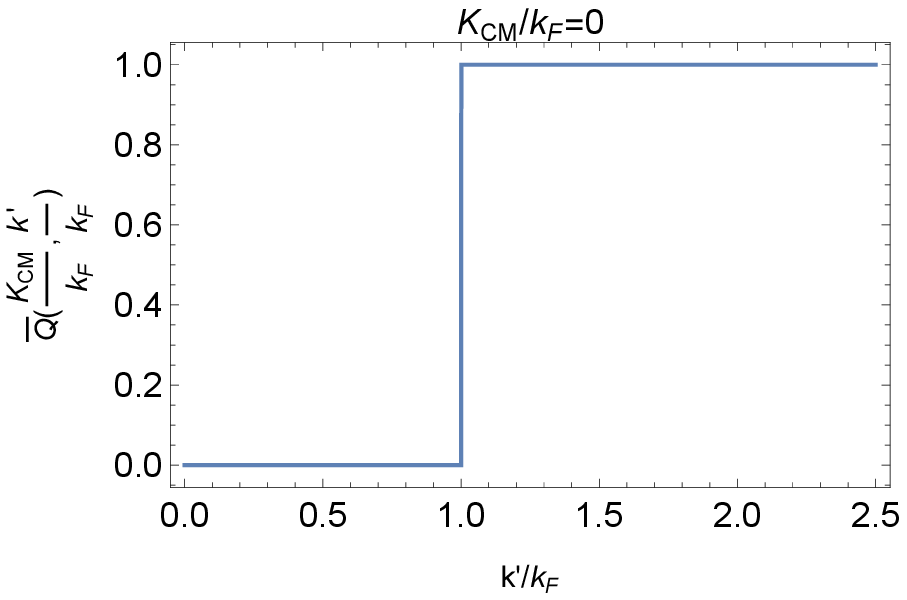}&
\includegraphics[width=7.5cm]{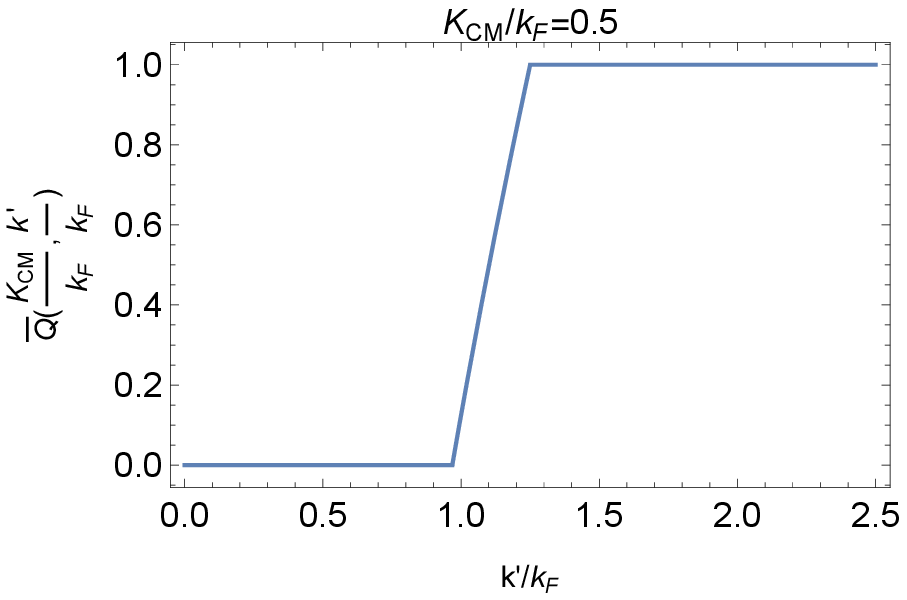}\\
\includegraphics[width=7.5cm]{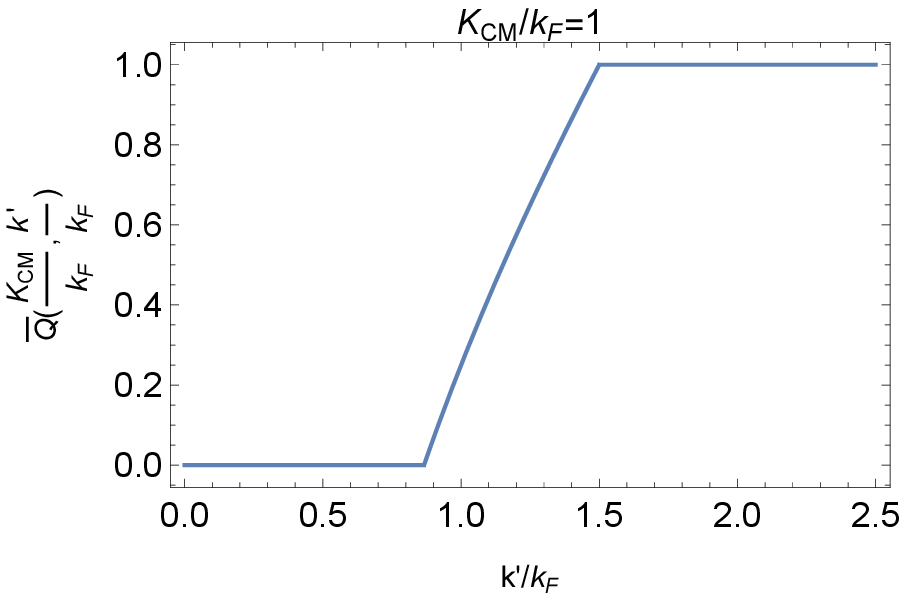}&
\includegraphics[width=7.5cm]{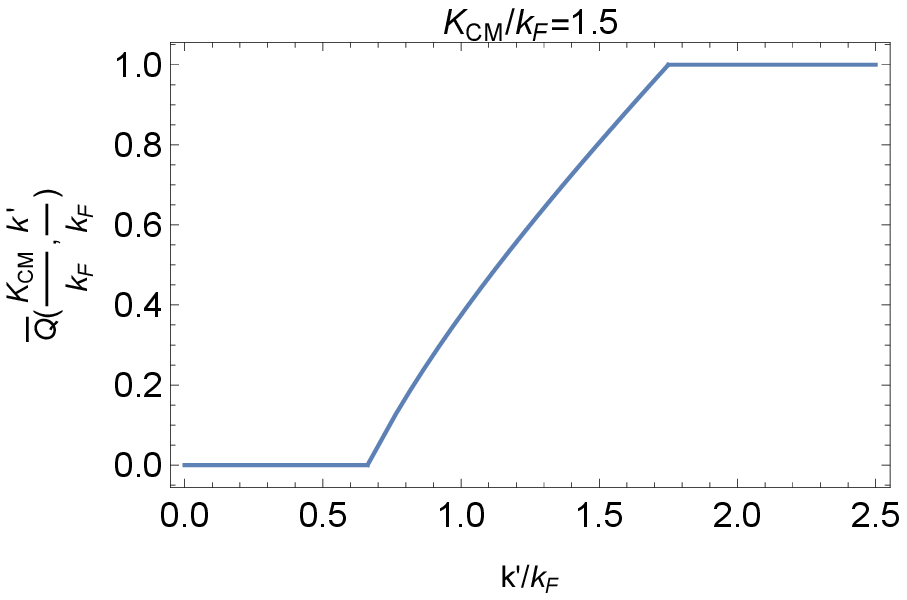}\\
\includegraphics[width=7.5cm]{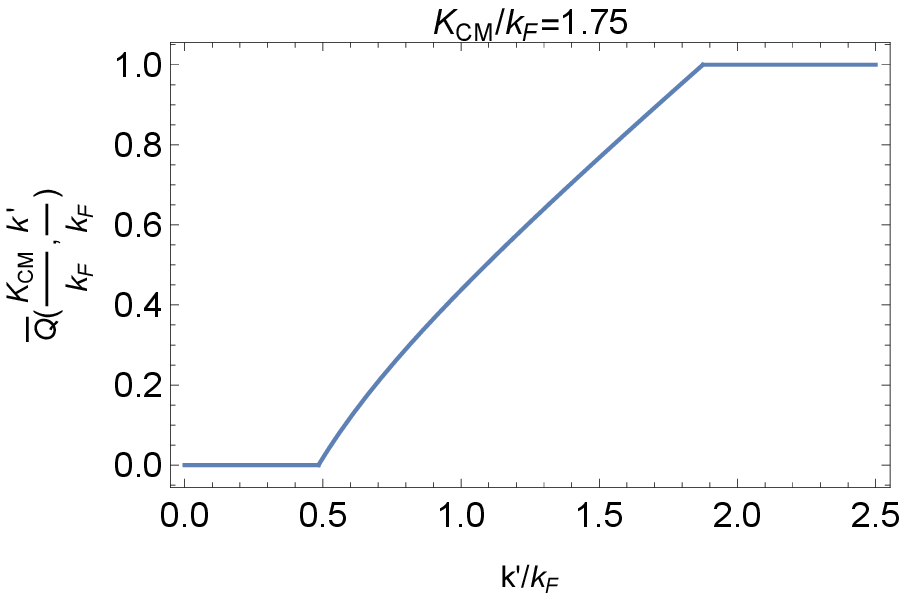}&
\includegraphics[width=7.5cm]{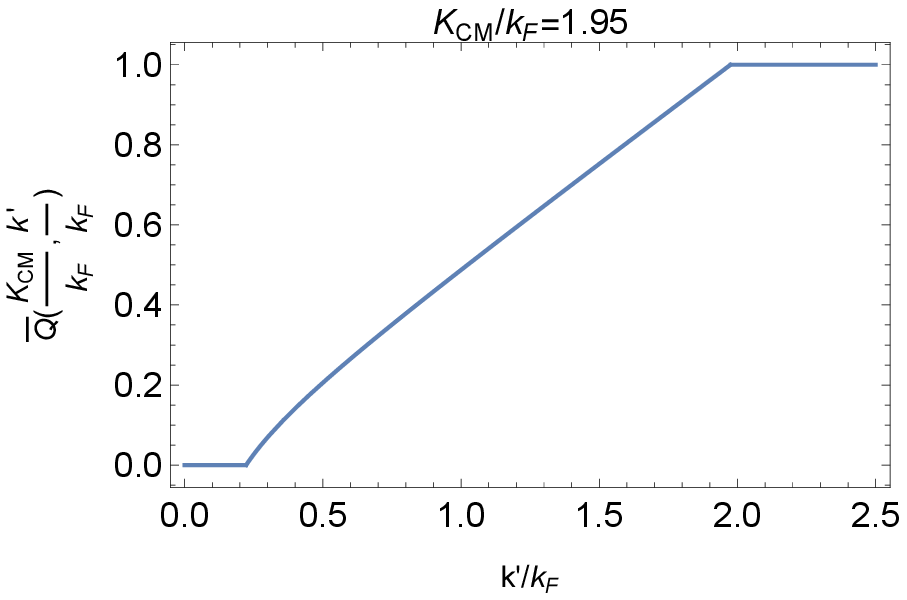}\\
\end{tabular}
\includegraphics[width=7.25cm]{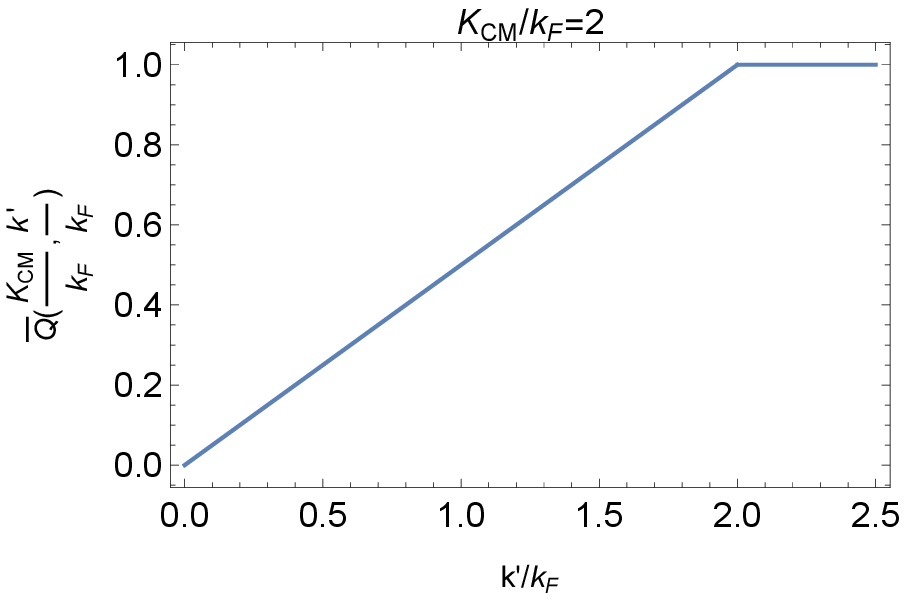}
\caption{Plots of the angle-averaged Pauli-blocking function
  $\overline{Q}(K_{\rm CM},k^\prime)$ for different values of the
  total CM momentum of the nucleon pair, ranging from $0$ up to
  $2\,k_F$. The different panels are labeled by the CM momentum in
  units of the Fermi momentum $k_F$. The abscissa axis corresponds to
  the relative $k^\prime$ momentum of the pair, in units of the Fermi
  momentum as well. The range spanned in the variables $(K_{\rm
    CM},k^\prime)$ is the same as that also displayed in
  Fig. \ref{Fig:zones}.}
\label{Fig:Qbarra}
\end{figure*}

Finally, the region labeled by (b) in Fig. \ref{Fig:zones}, bounded
by the ellipse from below and by the straight line from above,
corresponds to the transition region between both extreme situations
of zones (a) and (c). In this region, (b), the angle-averaged
Pauli-blocking function $\overline{Q}(K_{\rm CM},k^\prime)$ takes
intermediate values between $0$ and $1$, depending of course on the
values of $K_{\rm CM}$ and $k^\prime$ within this region.  Physically,
the picture of this region represents situations where, when
performing the angular average of eq.  (\ref{def_angular_average}),
for some values of the angle between $\nK_{\rm CM}$ and $\nk^\prime$
both single nucleon momenta are above the Fermi momentum
($\left|\nk^\prime_i\right|>k_F$), thus contributing the maximum to
the integral of eq. (\ref{def_angular_average}); while for other
values of the angle $(\widehat{\nK_{\rm CM},\nk^\prime})$, one or both
single momenta of the nucleon pair are below the Fermi momentum
($\left|\nk^\prime_i\right|<k_F$), thus contributing $0$ to the
integral.  Therefore, the final result is an intermediate-valued
function between $0$ and $1$, as shown in Fig. \ref{Fig:Qbarra}.

What is being shown in Fig. \ref{Fig:Qbarra} is exactly the piece-wise
function $\overline{Q}(K_{\rm CM},k^\prime)$ of
eq. (\ref{Qbar_function}) when the momenta are expressed in units of
the Fermi momentum $k_F$, for the same range in the variables
$(\frac{K_{\rm CM}}{k_F},\frac{k^\prime}{k_F})$ as that plotted in
Fig. \ref{Fig:zones}. Each panel corresponds to a selected value for
the CM momentum of the nucleon pair, and the $\overline{Q}$ function
is represented in terms of the relative momentum of the pair. Several
properties of this function can be commented:
\begin{enumerate}
\item For $K_{\rm CM}=0$ the $\overline{Q}$ function is exactly a step
  function, namely $\theta(k^\prime-k_F)$ and therefore it presents a
  discontinuity at the point $k^\prime=k_F$.  This situation
  corresponds to moving along the y-axis in Fig.  \ref{Fig:zones}. In
  this case the function is zero on the green region until one reaches
  the point $k^\prime=k_F$ (where the ellipse and the straight line
  cut each other), and beyond that point, for $k^\prime>k_F$, the
  $\overline{Q}$ function is always equal to $1$.  Therefore, this
  corresponds to the situation when the region (b) of Fig.
  \ref{Fig:zones} reduces to a single point when moving along the
  y-axis.
\item For relatively low values of the CM momentum, $K_{\rm
  CM}\lesssim k_F$, the joining $\overline{Q}$ function between
  regions (a) and (c) of Fig. \ref{Fig:zones}, i.e, along the region
  (b) of the same figure for a definite CM momentum (moving along a
  vertical straight line in Fig. \ref{Fig:zones}), is almost a
  straight line with large slope. This is because the $\overline{Q}$
  function has to increase from $0$ to $1$ in a relatively short range
  of $k^\prime$ values, thus making a larger slope.
\item For intermediate values of the CM momentum, $k_F \lesssim K_{\rm
  CM} \lesssim 2\,k_F$, a clear curvature in the joining
  $\overline{Q}$ function is evident, especially at the lowest
  $k^\prime$ values, i.e, in region (b) of Fig. \ref{Fig:zones} but
  close to the ellipse. In addition, as the $\overline{Q}$ function
  has a longer range in $k^\prime$ values to rise from $0$ to $1$, the
  average slope is much less steep than in the case discussed in the
  previous point.
\item For the maximum allowed CM momentum of two nucleons below the
  Fermi momentum, $K_{\rm CM}=2\,k_F$, the region (a) of
  Fig. \ref{Fig:zones} reduces to a single point ($k^\prime=0$), and
  the $\overline{Q}$ function in region (b) of the same figure is
  exactly a straight line in the $\frac{k^\prime}{k_F}$ variable with
  slope $\frac12$. This can be analytically proven from the second
  line of eq. (\ref{Qbar_function}) by simply substituting $K_{\rm
    CM}=2\,k_F$.
\item Finally, it is worth noticing that, except for $K_{\rm CM}=0$,
  the piece-wise function of eq. (\ref{Qbar_function}) is a continuous
  function even at the curves separating the three different regions
  of Fig. \ref{Fig:zones}, i.e, at the ellipse and the straight line
  of the same figure. However, this $\overline{Q}(K_{\rm
    CM},k^\prime)$ function has no continuous derivatives with respect
  to the $k^\prime$ variable precisely along the ellipse and the
  straight line of Fig. \ref{Fig:zones}. This behavior is clearly
  observable from Fig. \ref{Fig:Qbarra}, where at the joining points
  where $\overline{Q}$ is $0$ or $1$, the slopes are different if one
  approaches that point from below or above it. This last behavior for
  the derivative at the curves delimiting the different regions in
  Fig. \ref{Fig:zones} was already pointed out in Ref.
  \cite{Haftel:1970zz}, and it will be very relevant to interpret the
  high-momentum components of the relative wave functions that will be
  shown in Sect. \ref{subsec:radial_wf_momentum}.
\end{enumerate}

\subsection{B-G integral equation for the radial wave function}
\label{subsec:radial}
In principle, performing a partial wave expansion of eq.
(\ref{corr_relative_ket_equation}) along the lines of that carried out
in Ref. \cite{Brueckner:1958zz}, a set of coupled integral equations
are obtained for the radial components of the relative wave function.
The formal derivation of this last equation is also deferred to
Appendix \ref{derivation_radial}. In this work we extend the system of
equations presented in ~\cite{RuizSimo:2017tcb}. This previous study
considered these equations in the specific case where the relative
momentum is oriented along the Z-axis and the CM momentum was zero.
However, in this paper we investigate the general case where the
relative momentum can point in any direction, and the CM momentum is
non-zero.  This modification only impacts the radial functions in the
coupled channels, which in Ref.~\cite{RuizSimo:2017tcb} depended on a
single angular momentum label, $\widetilde{u}_l$. In the general case
considered here, these functions now depend on two angular momentum
indices, $\widetilde{u}_{l\,l^\prime}$.  Another difference with
respect to what was done in Ref.~\cite{RuizSimo:2017tcb} is the form
of the in-medium Green's function for the problem when $K_{\rm CM}\neq
0$.

Following the same normalization for the perturbed radial wave
function as in Ref.  \cite{RuizSimo:2017tcb}, the result is
\begin{equation}\label{radial_wf}
\widetilde{u}^{SJ}_{k,l\,l^\prime}(r)=
\hat{j}_l(kr)\delta_{ll^\prime}+ \int^{\infty}_0 
dr^\prime \; 
\widetilde{G}^{K_{\rm CM}}_{k,l^\prime}(r,r^\prime)
\sum_{l^{\prime\prime}} U^{SJ}_{l^\prime,l^{\prime\prime}}(r^\prime)\;
\widetilde{u}^{SJ}_{k,l\,l^{\prime\prime}}(r^\prime),
\end{equation}
where $\widetilde{u}^{SJ}_{k,l\,l^\prime}(r)$ is the perturbed radial
wave function; $\hat{j}_l(kr)=(kr)\,j_l(kr)$ is the reduced spherical
Bessel function of the first kind; $U^{SJ}_{l,l^\prime}(r)=2\mu
V^{SJ}_{l,l^\prime}(r)$ is the reduced potential matrix element for
the channel with total spin $S=0,1$ and total angular momentum $J$
between partial waves with different (or equal) orbital angular
momenta $(l,l^\prime)$; and, finally, $\widetilde{G}^{K_{\rm
    CM}}_{k,l}(r,r^\prime)$ is the Green's function for the radial B-G
equation and it is given by
\begin{equation}\label{Green_function_Kcm}
\widetilde{G}^{K_{\rm CM}}_{k,l}(r,r^\prime)=
\frac{2}{\pi} \int^{\infty}_0 dk^\prime\; \hat{j}_l(k^\prime r)\,
\frac{\overline{Q}(K_{\rm CM},k^\prime)}{k^2-k^{\prime\,2}}\,
\hat{j}_l(k^\prime r^\prime).
\end{equation}

In general, the radial wave functions of eq.~(\ref{radial_wf}) depend
on two angular momentum indices, $\widetilde{u}_{l\,l^\prime}$.  Due
to the tensor force, the channels with $l,l^\prime=J\pm 1$ are
coupled, while in the uncoupled channels one has $l=l^\prime$ (see
appendix~\ref{derivation_radial}).

Note that the radial wave functions
$\widetilde{u}^{SJ}_{k,l\,l^\prime}(r)$ also depend implicitly on the
value of the CM momentum $K_{\rm CM}$, although this dependence has
not been written explicitly in order to shorten the notation.

The Green's function of eq.~(\ref{Green_function_Kcm}) is a symmetric
function and reduces to the Green's function given in eq. (13) of
Ref.~\cite{RuizSimo:2017tcb} for the particular case when $K_{\rm
  CM}=0$. The integral of the oscillatory integrand of
eq.~(\ref{Green_function_Kcm}) over the infinite interval for
$k^\prime > k_F + \frac{K_{\rm CM}}{2}$ is carried out with Levin-type
integration methods ~\cite{Levin1994,Levin1996}.

At first sight one could foresee a divergence in the integrand of
eq.~(\ref{Green_function_Kcm}) when $k'=k$.  However, for the
calculations carried out in this work, one should have in mind that
the initial relative momentum $k$ of the nucleon pair is restricted to
lie in the region (a) of Fig.~\ref{Fig:zones}, because only in this
region both initial nucleons have individual momenta below the Fermi
momentum $k_F$.  Despite the general limits of integration in
eq.~(\ref{Green_function_Kcm}), the averaged Pauli-blocking operator
is zero unless $k^\prime>\sqrt{k^2_F-\frac{K^2_{\rm CM}}{4}}$.
Therefore, the true lower limit in the integral of
eq.~(\ref{Green_function_Kcm}) is $k^\prime=\sqrt{k^2_F-\frac{K^2_{\rm
      CM}}{4}}$ instead of zero, for a general total momentum of the
nucleon pair satisfying $K_{\rm CM}\le 2k_F$. The only point where
there could be some divergence in the integrand is when the initial
relative momentum of the pair, $k$, lies exactly in the ellipse of
Fig.~\ref{Fig:zones}. This would mean that there could be a
singularity exactly at the truly initial point of the integration
interval in eq.~(\ref{Green_function_Kcm}). However, at this point,
$k=k^\prime= \sqrt{k^2_F-\frac{K^2_{\rm CM}}{4}}$, the integrand is,
in general, finite, as it can be proven below by taking the limit
$k^\prime \rightarrow k^{+}$.  The only possible source of divergence
is the quotient $\frac{\overline{Q}(K_{\rm
    CM},k^\prime)}{k^2-k^{\prime\,2}}$.

If we calculate the limit of this quotient when $k^\prime \rightarrow
k^{+}\equiv \sqrt{k^2_F-\frac{K^2_{\rm CM}}{4}}$, we obtain the
result:
\begin{equation*}
 \lim_{k^\prime \rightarrow k^{+}}
 \frac{\overline{Q}(K_{\rm CM},k^\prime)}{k^2-k^{\prime\,2}}=
 \lim_{k^\prime \rightarrow k^{+}}
 \frac{\frac{\frac{K^2_{\rm CM}}{4}+k^{\prime\,2}-k^2_F}{K_{\rm CM}\,k^\prime}}{k^2_F-\frac{K^2_{\rm CM}}{4}-k^{\prime\,2}}=
 \lim_{k^\prime \rightarrow k^{+}}
 \frac{-1}{K_{\rm CM}\,k^\prime}.
\end{equation*}

Therefore, at the end, the only point of possible divergence
corresponds to the case $K_{\rm CM}=0$. In this particular case for
the value of the total momentum of the nucleon pair, we are
integrating over $k^\prime$ in eq.~(\ref{Green_function_Kcm}) along
the Y-axis of Fig.~\ref{Fig:zones}, and the only possible point of
divergence corresponds to the case when $k=k^\prime=k_F$. Note that in
this case (see Fig.~\ref{Fig:zones}), the region (b) gets reduced to a
single point where the averaged Pauli-blocking operator has a sudden
discontinuity at $k^\prime=k_F$, passing from zero to one, as the top
left panel of Fig.~\ref{Fig:Qbarra} shows. In this case, effectively
the integrand of eq.~(\ref{Green_function_Kcm}) has a discontinuity,
but only when the initial relative momentum $k$ of the pair reaches
its maximum allowed value $k_F$.

This situation physically corresponds to two back-to-back nucleons
carrying each one of them the maximum single momentum $k_F$. In the
calculations carried out here for $K_{\rm CM}=0$, or in those
performed in Ref.~\cite{RuizSimo:2017tcb}, we always have taken
$k<k_F$, thus avoiding any problem related with this singularity.

Now, we can introduce in eq. (\ref{radial_wf}) the form of the
coarse-grained Granada potential given by the sum of delta-shells for
each channel, defined by the values of the total spin $S$ and total
angular momentum $J$:
\begin{equation}\label{delta-shells}
U^{SJ}_{l,l^\prime}(r)= \sum^{N_\delta}_{i=1}
\left(\lambda_i \right)^{SJ}_{l,l^\prime}\; \delta(r-r_i)\;,
\end{equation}
where the five ($N_\delta=5$) delta-shells strengths $\left(\lambda_i
\right)^{SJ}_{l,l^\prime}$ are given in Table I of
Ref. \cite{Perez:2013mwa}, and they were fitted to reproduce the
phase-shifts of N-N scattering below the pion production threshold.
In this calculation we neglect the one-pion exchange (OPE)
contribution, which starts at distances larger than $3$ fm. While this
contribution is essential to describe the physical scattering data
with a high quality fit (particularly for the peripheral waves), its
influence becomes marginal for the study of short distance
correlations and makes the calculation unnecessarily more cumbersome.

The whole point of our framework has been to realize in previous works
that, even though in the current and traditional jargon of nuclear
physics, short distance effects are thought to imply extremely small
wavelengths, this is actually not so.  Including more delta-shells
does not improve the description of the scattering data in the elastic
regime. In fact, from a statistical point of view, the fits to the N-N
data do not improve but the statistical correlation among fitting
parameters increases and, hence, these additional deltas are largely
redundant.

At the present stage it is difficult to ponder on the impact on
3,4,5-body excitations within our approach.  There have been attempts
where mostly the 3-body interaction is included \cite{Moeini:2022fsf,
  HeYeTongLang:2013ihv, Kohno:2012vj, Lovato:2010ef,
  Holt:2009ty,Barnea:2004ac} as an effective (averaged) 2-body
one. Our expectation would be that these terms may modify the total
strength of the wave function but not the asymptotic behavior.

With this kind of potential given in eq. (\ref{delta-shells}), one can
easily perform the integral over the radial coordinate in
eq. (\ref{radial_wf}), thus obtaining the following algebraic equation
\begin{equation}\label{algebraic_radial_wf}
\widetilde{u}^{SJ}_{k,l\,l^\prime}(r)=
\hat{j}_l(kr)\delta_{ll^\prime} + \sum^{N_\delta}_{i=1}\;
\widetilde{G}^{K_{\rm CM}}_{k,l^\prime}(r,r_i)
\sum_{l^{\prime\prime}} \left(\lambda_i \right)^{SJ}_{l^\prime,l^{\prime\prime}} \;
\widetilde{u}^{SJ}_{k,l\,l^{\prime\prime}}(r_i).
\end{equation}

The form of the potential, eq. (\ref{delta-shells}), has allowed us to
transform a, in general, coupled integral equation for the radial wave
functions, eq. (\ref{radial_wf}), into a linear system of coupled
algebraic equations for the radial wave functions \emph{at the grid
points} $r_i$. Indeed, if we now take $r=r_j$ with $j=1,2\dots
N_\delta$, then eq.  (\ref{algebraic_radial_wf}) transforms into the
coupled linear system given by
\begin{equation}\label{algebraic_coupled_system}
\widetilde{u}^{SJ}_{k,l\,l^\prime}(r_j)=
\hat{j}_l(kr_j)\delta_{ll^\prime} + \sum^{N_\delta}_{i=1}\;
\widetilde{G}^{K_{\rm CM}}_{k,l^\prime}(r_j,r_i)
\sum_{l^{\prime\prime}} \left(\lambda_i \right)^{SJ}_{l^\prime,l^{\prime\prime}} \;
\widetilde{u}^{SJ}_{k,l\,l^{\prime\prime}}(r_i).
\end{equation}

Once the values of the radial wave functions at the grid points are
obtained, the B-G equation itself, namely
eq.~(\ref{algebraic_radial_wf}), directly allows for a sensible
interpolation of the wave function to any point between the grid
points.

The total spin $S$ of the two-nucleon system is known to be conserved
by the N-N interaction. When the two nucleons are in a singlet spin
state, $S=0$, then the tensor force does not couple states with
different orbital angular momentum, and therefore
$l=l^\prime=l^{\prime\prime}=J$ in
eq.~(\ref{algebraic_coupled_system}). In this case we have an
inhomogeneous linear system of $N_\delta=5$ equations (one for each
one of the possible values of $r_j$) with 5 unknowns, which are the 5
values of the radial wave functions $\widetilde{u}^{0J}_{k,J}(r_i)$ at
the five grid points $r_i$.

When the two nucleons are coupled to total spin $S=1$, for a given
total angular momentum $J$ of the partial wave, there are three
possibilities for the orbital angular momentum, $l=J-1,J,J+1$, except
for $J=0$, where only $l=1$ (P-state) is allowed. Due to the
conservation of parity in the N-N interaction, in the triplet channels
($S=1$), partial waves with angular momenta $l=l^\prime=J$ and parity
$P=(-1)^J$ are decoupled from those with $l,l^\prime=J\pm1$ and parity
$P=(-1)^{J+1}$.  In the former case, $l=l^\prime=J$,
eq.~(\ref{algebraic_coupled_system}) reduces again to 5 equations for
the radial wave function values at the grid points.

However, the partial waves for the case $S=1$ and
$(l,l^\prime)=J-1,J+1$ are known to be coupled due to the tensor part
of the N-N interaction, which has off-diagonal components in the
orbital angular momentum basis. In this case, we have to
simultaneously solve a coupled system for four different radial wave
functions at the grid points of the form given in eq.
(\ref{algebraic_coupled_system}). Now the sum over $l^{\prime\prime}$
in eq. (\ref{algebraic_coupled_system}) runs over two values
$l^{\prime\prime}=J-1,J+1$ for each pair of $(l,l^\prime)$
values. Therefore, the linear system we have to solve in this case is,
\emph{a priori}, a coupled inhomogeneous one of 20 equations with 20
unknowns. These unknowns are precisely the four coupled radial wave
functions at the 5 grid points.

Finally, once the values of the perturbed radial wave function for
each partial wave ${}^{2S+1}l_J$ are known at the grid points,
$\widetilde{u}^{SJ}_{k,l\,l^\prime}(r_i)$, then the wave function can
be known at any other point $r$ by means of eq.
(\ref{algebraic_radial_wf}).

Just to see the difference with respect to eqs.~(16--17) of
Ref.~\cite{RuizSimo:2017tcb}, which are valid only when the relative
momentum $\nk$ defines the Z axis, we write below the general coupled
equations for the ${}^3$S${}_1-{}^3$D${}_1$ coupled channels:
\begin{eqnarray}
&&\widetilde{u}^{11}_{k,0\,0}(r)=
\hat{j}_0(kr)+ \int^{\infty}_0 dr^\prime \; 
\widetilde{G}^{K_{\rm CM}}_{k,0}(r,r^\prime)
\left[U^{11}_{0,0}(r^\prime)\;
\widetilde{u}^{11}_{k,0\,0}(r^\prime)\right.\nonumber\\
&&\left. + U^{11}_{0,2}(r^\prime)\;
\widetilde{u}^{11}_{k,0\,2}(r^\prime)\right],\label{eqs:coupled_3S1-3D1-radial_1}\\
&&\widetilde{u}^{11}_{k,0\,2}(r)= \int^{\infty}_0 dr^\prime \; 
\widetilde{G}^{K_{\rm CM}}_{k,2}(r,r^\prime)
\left[U^{11}_{2,0}(r^\prime)\;
\widetilde{u}^{11}_{k,0\,0}(r^\prime)\right.\nonumber\\
&&\left. + U^{11}_{2,2}(r^\prime)\;
\widetilde{u}^{11}_{k,0\,2}(r^\prime)\right],\label{eqs:coupled_3S1-3D1-radial_2}\\
&&\widetilde{u}^{11}_{k,2\,0}(r)= \int^{\infty}_0 dr^\prime \; 
\widetilde{G}^{K_{\rm CM}}_{k,0}(r,r^\prime)
\left[U^{11}_{0,0}(r^\prime)\;
\widetilde{u}^{11}_{k,2\,0}(r^\prime)\right.\nonumber\\
&&\left. + U^{11}_{0,2}(r^\prime)\;
\widetilde{u}^{11}_{k,2\,2}(r^\prime)\right],\label{eqs:coupled_3S1-3D1-radial_3}\\
&&\widetilde{u}^{11}_{k,2\,2}(r)= 
\hat{j}_2(kr) + \int^{\infty}_0 dr^\prime \; 
\widetilde{G}^{K_{\rm CM}}_{k,2}(r,r^\prime)
\left[U^{11}_{2,0}(r^\prime)\;
\widetilde{u}^{11}_{k,2\,0}(r^\prime)\right.\nonumber\\
&&\left. + U^{11}_{2,2}(r^\prime)\;
\widetilde{u}^{11}_{k,2\,2}(r^\prime)\right].\label{eqs:coupled_3S1-3D1-radial_4}
\end{eqnarray}

Note that eqs.~(\ref{eqs:coupled_3S1-3D1-radial_1},
\ref{eqs:coupled_3S1-3D1-radial_2}) involve only the components
$(l,l^\prime)=(00),(02)$ of the radial wave functions; while
eqs.~(\ref{eqs:coupled_3S1-3D1-radial_3},
\ref{eqs:coupled_3S1-3D1-radial_4}) involve $(l,l^\prime)=(20),(22)$.
Therefore, these two pairs of equations can be solved separately as
two linear systems of 10 equations with 10 unknowns when using the
coarse-grained potential with five delta-shells.

\section{Relative wave function in momentum space}
\label{Sec:mom_space}
The derivation of the results of this section follows almost the same
lines as those of Sect. III C of Ref. \cite{RuizSimo:2017tcb}, with
caution because in general the perturbed radial wave functions depend
now on two angular momentum labels, $l\,l^\prime$, (as sketched in
Sect.  \ref{subsec:radial} and shown in Appendix
\ref{derivation_radial}), but for the general case when $K_{\rm
  CM}\neq 0$. At the end of Sect. \ref{sect:general} we wrote the B-G
equation that satisfies the relative perturbed wave function $\left|
\psi_{\nK_{\rm CM},\nk} \right\rangle$ without mentioning the spin of
the two-nucleon pair, cf. eq. (\ref{corr_relative_ket_equation}).
When the approximation of performing the angular average of the
Pauli-blocking operator is taken into account, the relative ket
$\left| \psi_{\nK_{\rm CM},\nk} \right\rangle$ no longer depends on
the direction of the CM momentum $\hat{K}_{\rm CM}$ of the two-nucleon
system, i.e, all directions of the CM momentum are equivalent in
infinite nuclear matter. Or, to say it in other words, there is an
isotropy property for the direction of the CM momentum $\hat{K}_{\rm
  CM}$.

If, finally, we also add the spin state of the two-nucleon pair to the
relative ket state we have a new ket state, labeled as $\left|
\psi_{\nk}, S M_S \right\rangle_{K_{\rm CM}}$, whose meaning is the
following one: it is the perturbed ket state with initial unperturbed
relative momentum $\nk$ of the two-nucleon system, and with total spin
$S$ and third component of spin $M_S$. This relative ket state can be
projected over the bra $\left\langle \np \right|$ to obtain the
probability amplitude of finding the state $\left| \psi_{\nk}, S M_S
\right\rangle_{K_{\rm CM}}$ in other one with relative momentum $\np$
due to the N-N interaction and the medium (angular average of the
Pauli-blocking operator) effects.

If we had an unperturbed state with relative momentum $\nk$ and spin
state $(S,M_S)$, its wave function in coordinate representation would
be
\begin{eqnarray}
&&\left\langle \nr \right| \left. \nk; S M_S\right\rangle =
\frac{e^{i\, \nk \cdot \nr}}{(2\pi)^{\frac32}}\; \chi_{{}_{S M_S}}
\nonumber\\
&=&\frac{4\pi}{(2\pi)^{\frac32}}\sum_{J,M}\; 
\sum_{l,m} 
i^{l}\, 
j_{l}(kr)\,
Y^*_{l m}(\hat{k})
\left\langle lm; S M_S\right| \left. J M \right\rangle\,
\mathcal{Y}_{l S J M} (\hat{r}) \nonumber\\
&=&\frac{4\pi}{(2\pi)^{\frac32}}\sum_{J,M}\,
\sum_{l,l^\prime,m} 
i^{l}\, j_{l}(kr)\, Y^*_{l^\prime m}(\hat{k})\,\delta_{l^\prime l}\,
\left\langle l^\prime m; S M_S\right| \left. J M \right\rangle\,
\mathcal{Y}_{l S J M} (\hat{r}), \nonumber\\
 \label{unperturbed_wf_S_Ms}
\end{eqnarray}
where in the last step we have added an additional sum over $l^\prime$
with the Kronecker delta $\delta_{l^\prime l}$ in order to match the
expansion of the perturbed or correlated state (see below).

In eq.~(\ref{unperturbed_wf_S_Ms}) we have coupled the orbital angular
momentum and spin angular momentum states, i.e., $l\otimes S$ to
obtain the spin-angular eigenfunctions, $\mathcal{Y}_{lSJM}(\hat{r})$,
with well-defined total angular momentum $J$ and third component $M$,
defined by:
\begin{equation}\label{angular-spin_eigenfunctions}
\mathcal{Y}_{l S J M} (\hat{r})=\sum_{m, M_S}
\left\langle l m; S M_S\right| \left. J M \right\rangle
Y_{lm}(\hat{r})\; \chi_{{}_{S M_S}}.
\end{equation}

A similar expansion in partial waves to that of eq.
(\ref{unperturbed_wf_S_Ms}) also holds for the perturbed state $\left|
\psi_{\nk}, S M_S \right\rangle_{K_{\rm CM}}$ in coordinates
representation,
\begin{eqnarray}
&&\left\langle \nr \right. \left| \psi_{\nk}, S M_S \right\rangle_{K_{\rm CM}}
= \frac{4\pi}{(2\pi)^{\frac32}}\sum_{J,M}\; 
\sum_{l,l^\prime,m} i^{l^\prime}\, Y^*_{l^\prime m}(\hat{k})\;
 u^{SJ}_{k,l^\prime\,l}(r) \nonumber\\
&\times&
\left\langle l^\prime m; S M_S\right| \left. J M \right\rangle\,
\mathcal{Y}_{l S J M} (\hat{r}), \label{perturbed_wf_S_Ms}
\end{eqnarray}
where the perturbed radial wave functions $u_{k,l^\prime \,
  l}^{SJ}(r)$ are normalized with respect to those appearing in sect.
\ref{subsec:radial} as:
\begin{equation}\label{normalization_radial_wf}
u_{k,l^\prime \, l}^{SJ}(r)=\frac{\widetilde{u}^{SJ}_{k, l^\prime\,l}(r)}{k\,r},
\end{equation}
in order to approach the free solution, the spherical Bessel functions
of the first kind, at long distances for the diagonal case,
$l=l^\prime$.

With these two partial wave expansions for the unperturbed and
perturbed states, eqs. (\ref{unperturbed_wf_S_Ms}) and
(\ref{perturbed_wf_S_Ms}), we can calculate the bra-ket product
$\left\langle \np \right| \left. \psi_{\nk}, S M_S
\right\rangle_{K_{\rm CM}}$ in momentum space by performing the
Fourier transform:
\begin{eqnarray}
&&\left\langle \np \right| \left. \psi_{\nk},SM_S \right\rangle_{K_{\rm CM}} \,
= \int \frac{d^3r}{(2\pi)^{\frac32}}e^{-i\,\np \cdot \nr}
\left\langle \nr\right| \left. \psi_{\nk},SM_S\right\rangle_{K_{\rm CM}}
\nonumber\\
&=&\frac{2}{\pi} \sum_{J,M}\,\sum_{l,l^\prime,m} i^{l^\prime-l}\,
Y^*_{l^\prime m}(\hat{k}) 
\left\langle l^\prime m; SM_S\right|\left.JM\right\rangle\,
\mathcal{Y}_{lSJM}(\hat{p})\nonumber\\
&\times&\frac{1}{p\,k}\int^\infty_0 dr\,  \hat{j}_l(pr)\; 
\widetilde{u}^{SJ}_{k,l^\prime\,l}(r).\label{perturbed_wf_mom_space}
\end{eqnarray}
To obtain the final expression of eq.(\ref{perturbed_wf_mom_space}) we
have used: the expansion of the plane wave in spherical harmonics
(Rayleigh's formula); the equation (\ref{perturbed_wf_S_Ms}) for the
perturbed wave function in coordinate representation; we have also
carried out the angular integration over $\hat{r}$ between a spherical
harmonic and a spin-angular eigenfunction, with the aid of
\begin{equation}\label{int_spherical_harm_spin_ang_eigenfunc}
\int d\Omega_{\hat{r}}\; Y^*_{l^\prime m^\prime}(\hat{r})\;
\mathcal{Y}_{lSJM}(\hat{r})=\delta_{l,l^\prime}\sum_{m_s}
\left\langle l m^\prime; S m_s\right|\left. JM\right\rangle\,
\chi_{Sm_s}.
\end{equation} 
Finally, we have performed the sum over the orbital angular momentum
label with the Kronecker delta; and we have coupled again one
spherical harmonic with the spinor wave function appearing in
eq.~(\ref{int_spherical_harm_spin_ang_eigenfunc}) to obtain the
spin-angular eigenfunction $\mathcal{Y}_{lSJM}(\hat{p})$.

The form of eq.~(\ref{perturbed_wf_mom_space}), apart from the
normalization factors, is completely equivalent to that of the
perturbed ket in position representation, given by
eq.~(\ref{perturbed_wf_S_Ms}). In this case, we identify the ``radial"
partial wave function in momentum representation as
\begin{equation}\label{radial_wf_momentum_rep}
\phi_{k,l^\prime\,l}^{SJ}(p)= \frac{2}{\pi} \, \frac{1}{p\,k}
\int^{\infty}_0 dr\; \hat{j}_{l}(pr)\; \widetilde{u}^{SJ}_{k,l^\prime\,l}(r).
\end{equation}

It is also worth noticing that the ``radial" wave function
$\phi_{k,l^\prime\,l}^{SJ}(p)$ for each partial wave also depends on
the magnitude of the CM momentum of the nucleon pair, $K_{\rm CM}$,
via the dependence on it of the radial wave function
$\widetilde{u}^{SJ}_{k,l^\prime\,l}(r)$
(cf. eq.~(\ref{algebraic_radial_wf})), as it has already been
mentioned in the discussion given in Sect.
\ref{subsec:radial}. However, this dependence has not been explicitly
written here to avoid a very cumbersome notation.

In the next step, to obtain an analytical expression for the ``radial"
wave function $\phi_{k,l^\prime\,l}^{SJ}(p)$, one needs to substitute
the radial wave function $\widetilde{u}^{SJ}_{k,l^\prime\,l}(r)$ from
eq.~(\ref{algebraic_radial_wf}) into
eq.~(\ref{radial_wf_momentum_rep}), and to use the explicit expression
of the Green's function, $\widetilde{G}^{K_{\rm CM}}_{k,l}(r,r')$,
given in eq. (\ref{Green_function_Kcm}), to carry out the integration
over the radial variable in eq. (\ref{radial_wf_momentum_rep}).  It is
also necessary to use the orthogonality property of the reduced
spherical Bessel functions
\begin{equation}\label{orthogonality_bessel}
\int^{\infty}_0 dr \; \hat{j}_l(pr) \; \hat{j}_l(kr)= \frac{\pi}{2}\; \delta(p-k),
\end{equation}
to obtain the final result:
\begin{equation}\label{analytical_radia_wf_mom}
 \phi_{k,l^\prime\,l}^{SJ}(p) = \delta_{l^\prime l}\,\frac{1}{p\,k}\; \delta(p-k)
  + \Delta\phi^{SJ}_{k,l^\prime\,l}(p),
\end{equation}
where
\begin{equation}\label{high_mom_components}
\Delta\phi^{SJ}_{k,l^\prime\,l}(p)=\frac{2}{\pi}\,\frac{1}{p\,k}\,
\frac{\overline{Q}(K_{\rm CM},p)}{k^2-p^2}\sum^{N_\delta}_{i=1}
\hat{j}_l(pr_i)\sum_{l^{\prime\prime}} (\lambda_i)^{SJ}_{l,l^{\prime\prime}}\;
\widetilde{u}^{SJ}_{k,l^\prime\,l^{\prime\prime}}(r_i).
\end{equation}

The first term of eq.~(\ref{analytical_radia_wf_mom}) corresponds to
the unperturbed ``radial" component of the state $\left|
\psi_{\nK_{\rm CM},\nk} \right\rangle$ of
eq.~(\ref{corr_relative_ket_equation}), coming from the bra-ket
product $\left\langle \np \right| \left. \nk \right\rangle$; while the
second term, given explicitly in eq.~(\ref{high_mom_components}),
corresponds genuinely to the high momentum components induced in the
perturbed relative wave function by the N-N interaction and the
medium.

For the ground state of an uncorrelated two-nucleon system in nuclear
matter with single momenta $\left| \nk_i \right| \leqslant k_F$, their
relative momentum $k$ is constrained to lie in region (a) of
Fig.~\ref{Fig:zones}. Therefore, there is not any divergence problem
in the second term of eq.~(\ref{analytical_radia_wf_mom}) when $p$
approaches $k$ from above, because in that case the angle-averaged
Pauli-blocking function $\overline{Q}(K_{\rm CM},p)$ is exactly $0$ on
the ellipse delimiting region (a) from (b) in Fig.~\ref{Fig:zones},
and below the ellipse as well (cf. eq.~(\ref{Qbar_function})).

Another interesting check corresponds to the case when the CM momentum
of the two-nucleon system is zero, $K_{\rm CM}=0$. In this case,
eq.~(\ref{high_mom_components}) should reduce to eq. (30) of
Ref.~\cite{RuizSimo:2017tcb}.  And indeed this is the case, because
for $K_{\rm CM}=0$ the angle-averaged Pauli-blocking function
$\overline{Q}(0,p)$ reduces to the step function $\theta(p-k_F)$, as
it can be deduced from the discussion given at point 1 of Sect.
\ref{ang_average}.

Some words of caution must be given again: in general, the radial wave
functions, either in coordinate or momentum representation, depend on
two labels for the orbital angular momenta, except for the uncoupled
nucleon-nucleon partial waves, where $l=l^\prime$ and there are no
off-diagonal wave functions.

\subsection{High-momentum density distribution}\label{subsec:high_mom_dens_dist}
We are going to obtain the high-momentum density distributions for a
given total spin $S=0,1$ of the nucleon pair.  To this end, we have to
integrate the modulus squared of the \emph{probability amplitude},
given in eq.~(\ref{perturbed_wf_mom_space}), of finding the perturbed
wave function with momentum $\np$, over the solid angle of $\hat{p}$,
assuming that we do not measure the direction of this momentum with
respect to the fixed CM momentum, $\nK_{\rm CM}$.  This quantity is
given by:
\begin{eqnarray}
&&\rho^{SM_S}_{\nk,K_{\rm CM}}(p)=\int d\Omega_{\hat{p}}
\left| \left\langle \np\right|\left. \psi_{\nk},SM_S\right\rangle_{K_{\rm CM}}
\right|^2=\nonumber\\
&&=\sum_{l,l^\prime,m}\;\sum_{l^{\prime\prime},m^{\prime\prime}}\,\sum_{JM}
i^{l^\prime-l^{\prime\prime}}\;
Y^*_{l^\prime m}(\hat{k})\, 
Y_{l^{\prime\prime} m^{\prime\prime}}(\hat{k})\nonumber\\
&\times&\left\langle l^\prime m;SM_S\right|\left.JM\right\rangle
\left\langle l^{\prime\prime} m^{\prime\prime};SM_S\right|\left.J M\right\rangle
\phi^{SJ *}_{k, l^{\prime\prime}\,l}(p)\;\phi^{SJ}_{k,l^\prime\,l}(p),
\label{high_dens_mom_dist_SM_S}\nonumber\\
\end{eqnarray}
where we have used the orthogonality property of the spin-angular
eigenfunctions $\mathcal{Y}_{l S J M}(\hat{p})$ to integrate over the
directions of $\hat{p}$
\begin{equation}\label{orthogonality_spin_angular}
\int d\Omega_{\hat{p}}\; 
\mathcal{Y}^*_{l^\prime S J^\prime M^\prime}(\hat{p})\;
\mathcal{Y}_{l S J M}(\hat{p})=\delta_{l, l^\prime}\;
\delta_{J, J^\prime}\; \delta_{M, M^\prime}\,,
\end{equation}
to carry out some discrete sums appearing when taking the modulus
squared of eq.~(\ref{perturbed_wf_mom_space}).

If, in addition, we do not measure the third component of the spin of
the pair of nucleons along the quantization axis defined by $\nK_{\rm
  CM}$, we have to perform again a sum over $M_S$ in
eq.~(\ref{high_dens_mom_dist_SM_S}), and an average over the number of
different $M_S$ values for each total spin $S$. We thus obtain:
\begin{eqnarray}
&&\rho^{S}_{\nk,K_{\rm CM}}(p) = \frac{1}{2S+1} \sum^{S}_{M_S=-S}
\rho^{SM_S}_{\nk,K_{\rm CM}}(p)=\nonumber\\
&=& \frac{1}{2S+1} \sum_{l,l^\prime,m}\;
\sum_{l^{\prime\prime},m^{\prime\prime}}\,\sum_J
i^{l^\prime-l^{\prime\prime}}\;Y^*_{l^\prime m}(\hat{k})\, 
Y_{l^{\prime\prime} m^{\prime\prime}}(\hat{k})\nonumber\\
&\times&\phi^{SJ *}_{k, l^{\prime\prime}\,l}(p)\;\phi^{SJ}_{k,l^\prime\,l}(p)
\sum_{M,M_S}
\left\langle l^\prime m;SM_S\right|\left.JM\right\rangle
\left\langle l^{\prime\prime} m^{\prime\prime};SM_S\right|\left.J M\right\rangle.
\nonumber\\
\label{high_dens_mom_dist_S}
\end{eqnarray}

Note that the final sum over the third components of angular momenta
of the product of two Clebsch-Gordan coefficients can be carried out
with the aid of the symmetry properties of these coefficients when
changing the order of coupling, and using their orthonormality
properties.  The symmetry property that we need here is to change the
order of coupling from $\left[l\otimes S \right]_J$ to $\left[J\otimes
  S \right]_{l}$, where $l$ stands for anyone of the two orbital
angular momenta appearing in eq.~(\ref{high_dens_mom_dist_S}):
\begin{equation}\label{C-G_symmetry_property}
\left\langle lm; SM_S\right|\left.JM\right\rangle=(-1)^{S+M_S}
\sqrt{\frac{2J+1}{2l+1}}\left\langle J,-M;SM_S\right|\left.l,-m\right\rangle.
\end{equation}
Using the above symmetry property of the C-G coefficients in the last
sum of eq.~(\ref{high_dens_mom_dist_S}), we obtain:
\begin{eqnarray}
&&\rho^{S}_{\nk,K_{\rm CM}}(p)= \frac{1}{2S+1}\sum_{l,l^\prime,m}
\sum_J \, Y^*_{l^\prime m}(\hat{k})\; Y_{l^\prime m}(\hat{k})\;
\Delta(JSl^\prime)\nonumber\\
&\times&\frac{(2J+1)}{(2l^\prime+1)}\;\left| \phi^{SJ}_{k, l^\prime\,l}(p)
\right|^2.\label{high_dens_mom_dist_S_2}
\end{eqnarray}
To obtain the above equation we have used the following facts in the
final sum of eq.~(\ref{high_dens_mom_dist_S}) over the third
components $M,M_S$: that the factor $(-1)^{2(S+M_S)}$ is always
positive regardless of the spin of the nucleon pair being integer or
half-integer (of course it is always integer, but the factor would
also be positive in the case of half-integer spin); that the sum over
$M\equiv-M^\prime$ can be carried out in reverse order without
changing anything; using the orthonormality property of the C-G
coefficients, which when summed over $M,M_S$ give
$\Delta(JSl^\prime)\,\delta_{l^\prime,l^{\prime\prime}}\,
\delta_{m,m^{\prime\prime}}$; and finally performing the sums over
$l^{\prime\prime}$ and $m^{\prime\prime}$ with the aid of the
Kronecker deltas.

Finally, notice that in eq.~(\ref{high_dens_mom_dist_S_2}) the sum
over $m$ only affects the spherical harmonics, and this can be
simplified a lot by using
\begin{equation}\label{completeness_spher_harm}
\sum_m\; Y^*_{l^\prime m}(\hat{k})\; Y_{l^\prime m}(\hat{k})=
\frac{(2l^\prime+1)}{4\pi}, 
\end{equation}
thus obtaining the final result
\begin{equation}\label{high_dens_mom_dist_spinS_final}
\rho^{S}_{k,K_{\rm CM}}(p)=\frac{1}{2S+1}\sum_{l,l^\prime,J}
\Delta(JSl^\prime)\;\frac{(2J+1)}{4\pi}\;
\left| \phi^{SJ}_{k, l^\prime\,l}(p)\right|^2
\end{equation}

It is also worth noting that $\Delta(JSl^\prime)$ is the triangular
inequality for the coupling of two angular momenta to a third one,
meaning that the sum over $J$ and $l^\prime$ in
eq.~(\ref{high_dens_mom_dist_spinS_final}) is restricted to run over
those values of $J$ and $l^\prime$ which are compatible to couple to a
total spin of the two-nucleon system of $S=0$ or $S=1$. To be more
specific, for a given total spin $S$ and total angular momentum $J$
for the partial wave, the sum over $l^\prime$ runs from $\left| J-S
\right|$ to $J+S$, with another restriction coming from the
antisymmetry of the relative wave function for a system of two
identical fermions such as the proton-proton (pp) or neutron-neutron
(nn) pair. For these cases, if $S=0$ (antisymmetric spin state in
terms of the single nucleon spin states) then only even orbital
angular momenta contribute; while if $S=1$ (symmetric spin state in
terms of the single nucleon spin states) only odd values of $l$ and
$l^\prime$ contribute in the sum of
eq.~(\ref{high_dens_mom_dist_spinS_final}).  However, this is not the
case for a neutron-proton (np) pair, where all the $(l,l^\prime)$
values compatible with the rules of angular momentum coupling (from
$\left| J-S \right|$ to $J+S$), and coupling of partial waves due to
the tensor force of the N-N potential are allowed in the sum of
eq.~(\ref{high_dens_mom_dist_spinS_final}).

Finally, it is also worth warning the reader that
eq.~(\ref{high_dens_mom_dist_spinS_final}) is the general equation
instead of equation (34) of Ref.~\cite{RuizSimo:2017tcb}, where the
relative momentum $\nk$ was chosen to lie along the Z-axis. Although
not explicitly written, the ``radial" momentum wave functions
$\phi_{k,l^\prime\,l}^{SJ}(p)$ depend on the magnitude of the CM
momentum of the two nucleon system, while in
Ref.~\cite{RuizSimo:2017tcb} the results were obtained for $K_{\rm
  CM}=0$ only.  However, in order to facilitate the comparisons with
the results of Ref.~\cite{RuizSimo:2017tcb} for the high-momentum
density distributions for a given total spin $S$ of the nucleon pair,
and to see the differences, in the results section
(Sect.~\ref{Sec:results}) we are going to use the same normalization
for the high-momentum density distributions as defined in
Ref.~\cite{RuizSimo:2017tcb}, i.e, we are going to adopt the
convention of eq. (43) of that reference to plot the high-momentum
density distributions for a given spin $S$.  This convention amounts
to plot $4\pi(2S+1)\rho^{S}_{k,K_{\rm CM}}(p)$, with
$\rho^{S}_{k,K_{\rm CM}}(p)$ defined by
eq.~(\ref{high_dens_mom_dist_spinS_final}), in order for a
straightforward comparison of figures 8, 9(a) and 10 of
Ref.~\cite{RuizSimo:2017tcb} with the ones obtained from
eq.~(\ref{high_dens_mom_dist_spinS_final}) (see
Sect.~\ref{Sec:results} for the discussion).

\section{Results and discussion}\label{Sec:results}
In this section we provide results for the perturbed radial wave
functions in coordinate and momentum representations. We show the
results for a Fermi momentum of $k_F=250$ MeV/c and an initial
relative momentum of the pair of $k=140$ MeV/c, in order to compare
with what was done in Ref.~\cite{RuizSimo:2017tcb}. The results of
this section have been calculated for different CM momenta. All the
pairs $(K_{\rm CM},k)$ belong to the region (a) of
Fig.~\ref{Fig:zones}, thus we are always in the region where the
single nucleon momenta are below the Fermi momentum ($\left| \nk_i
\right| \leqslant k_F$), i.e, in the ground state of nuclear matter.

While we show for definiteness results for $k=140$ MeV/c of relative
momentum, halfway the Fermi momentum, we have verified that our
conclusions regarding the CM do not depend strongly on the particular
$k$-value. Actually, for zero CM momentum, the universality of the
particle pair distribution was explicitly verified in our previous
work for $k=40,140,200$ MeV/c (see figure 10 (a) in
Ref.~\cite{RuizSimo:2017tcb}).  However, the wave function with
initial momentum $k$ does depend directly on the CM momentum as a
direct consequence of the B-G equation, as can be seen explicitly in
Eq.~(\ref{high_mom_components}).  Regarding the Fermi momentum
dependence, as a direct consequence of the Pauli blocking kernel, the
momentum distribution is shifted above the Fermi momentum.

\subsection{Perturbed radial wave functions in coordinate 
representation}\label{subsec:perturbed_radial_wf_coordinates}

In this subsection, for the uncoupled partial waves, $l=l^\prime$, we
use the notation $\widetilde{u}^{SJ}_{k,l}\equiv
\widetilde{u}^{SJ}_{k,l\,l}$ for the radial wave functions in the
figures.

\begin{figure*}[!ht]
\begin{tabular}{cc}
\includegraphics[width=7.5cm]{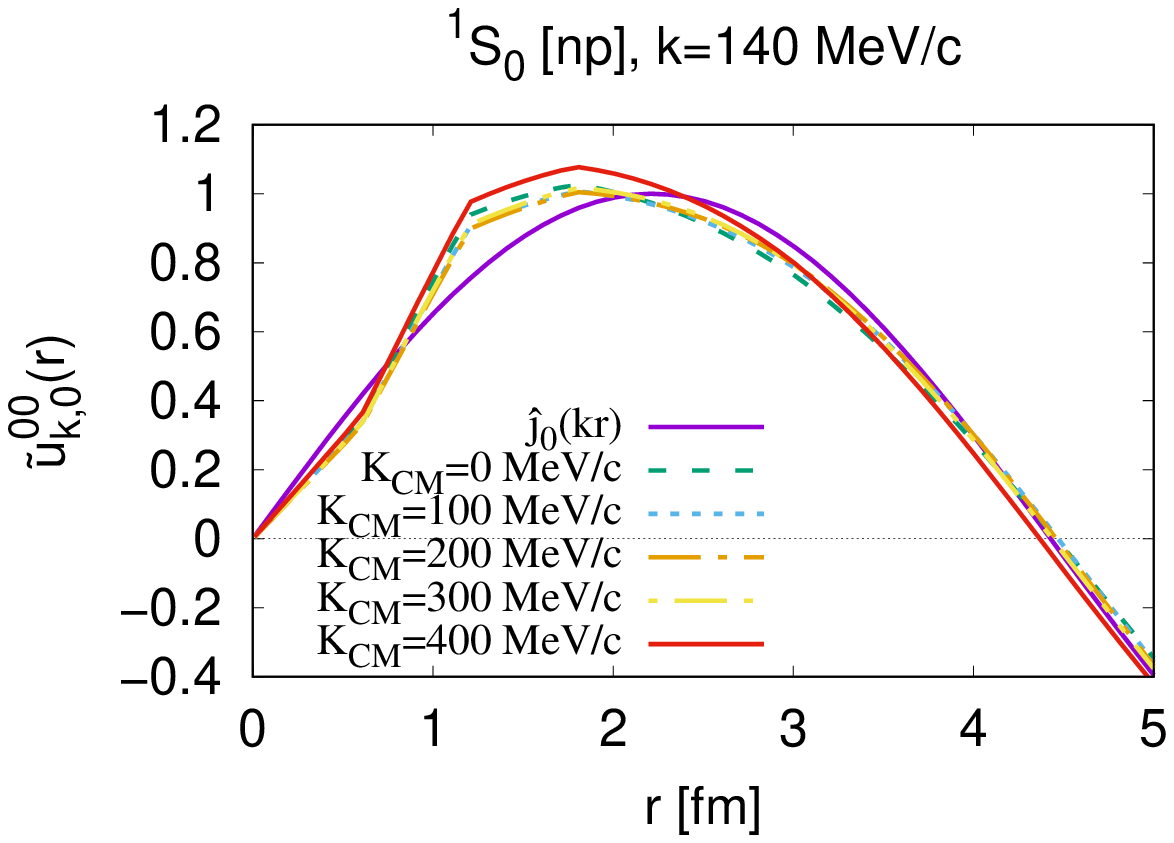}&
\includegraphics[width=7.5cm]{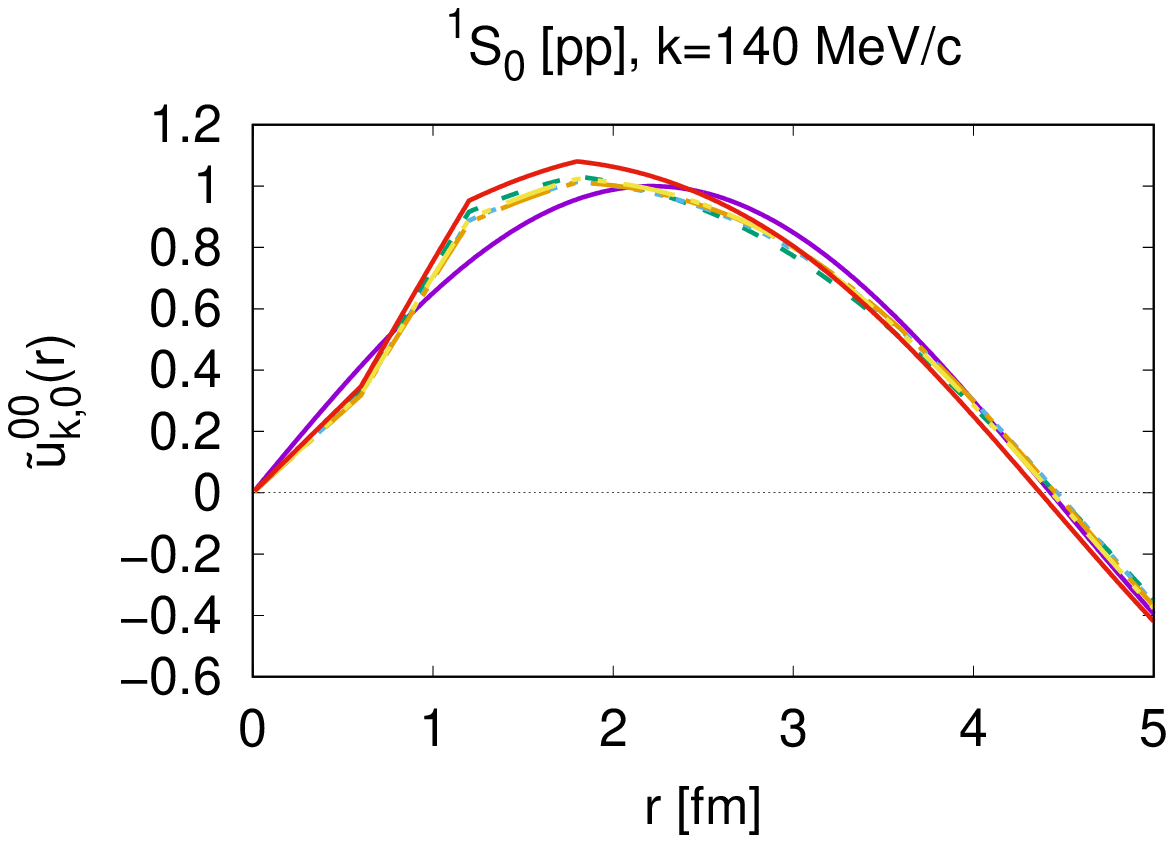}\\
\includegraphics[width=7.5cm]{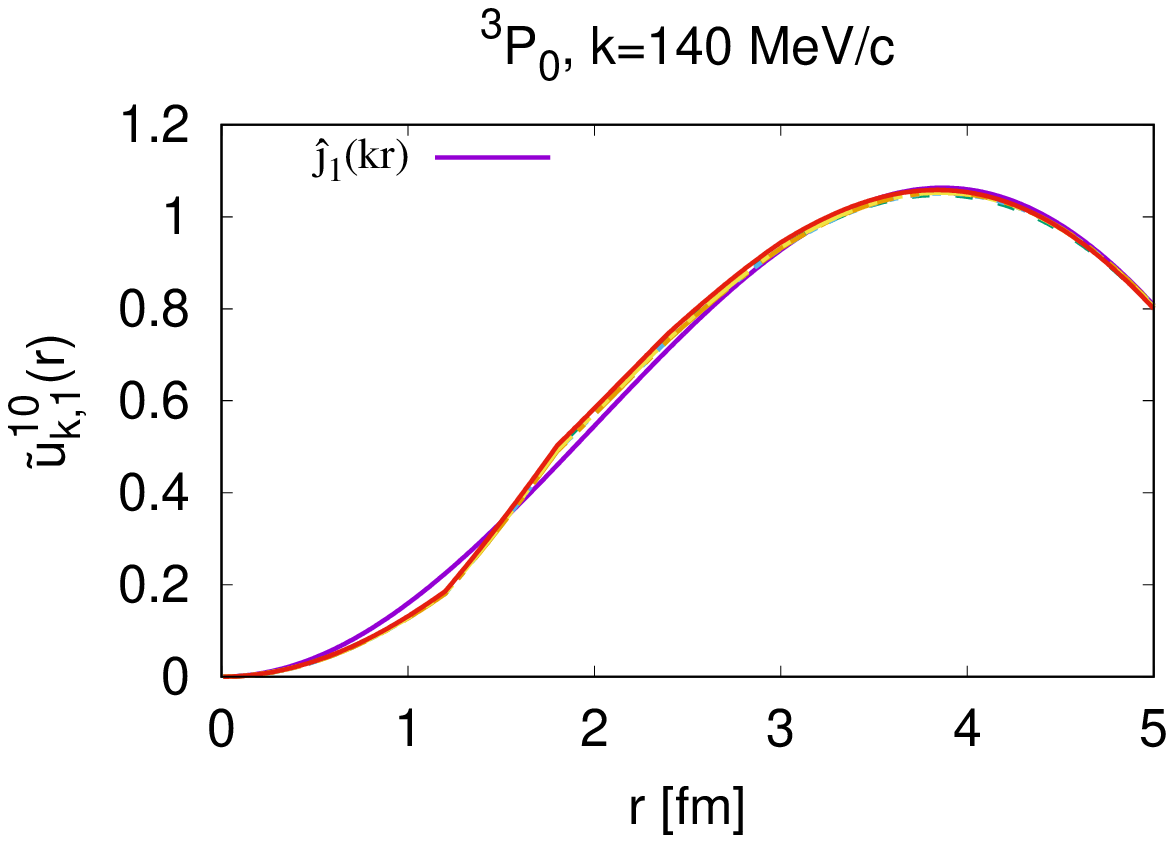}&
\includegraphics[width=7.5cm]{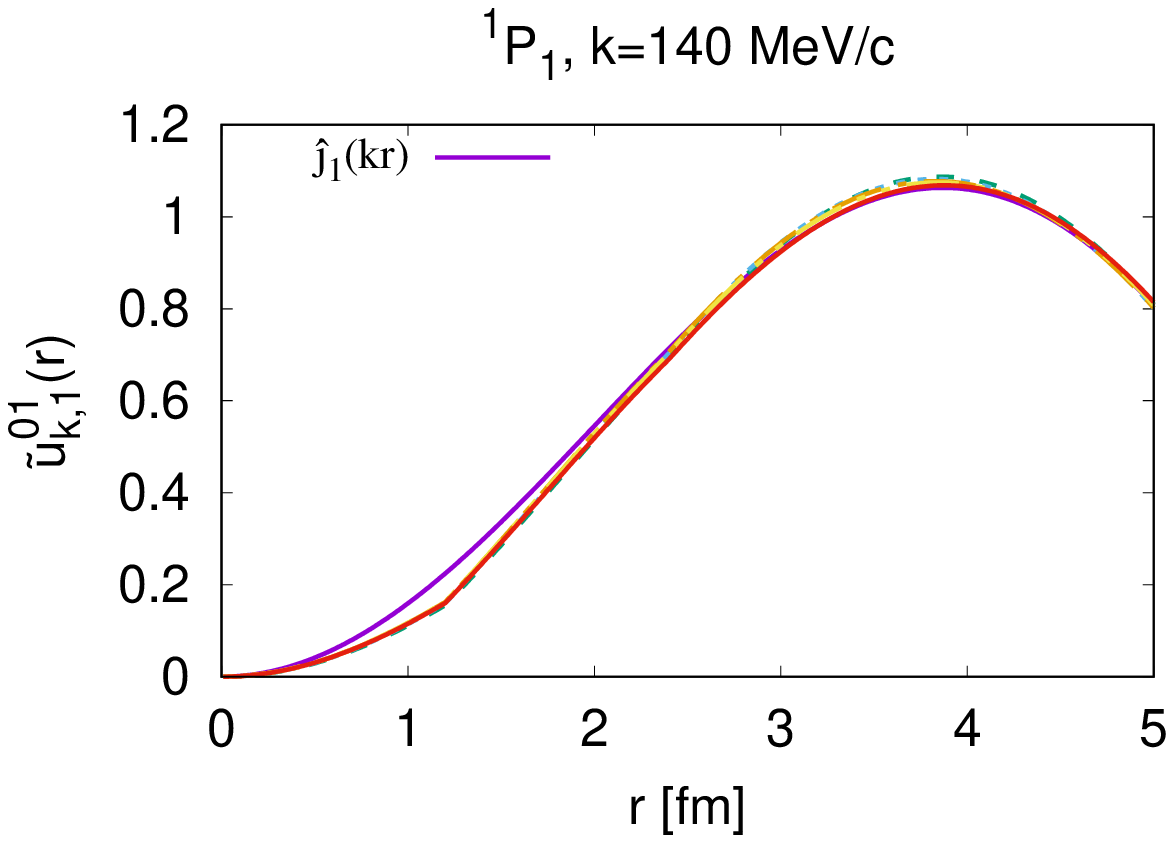}\\
\includegraphics[width=7.5cm]{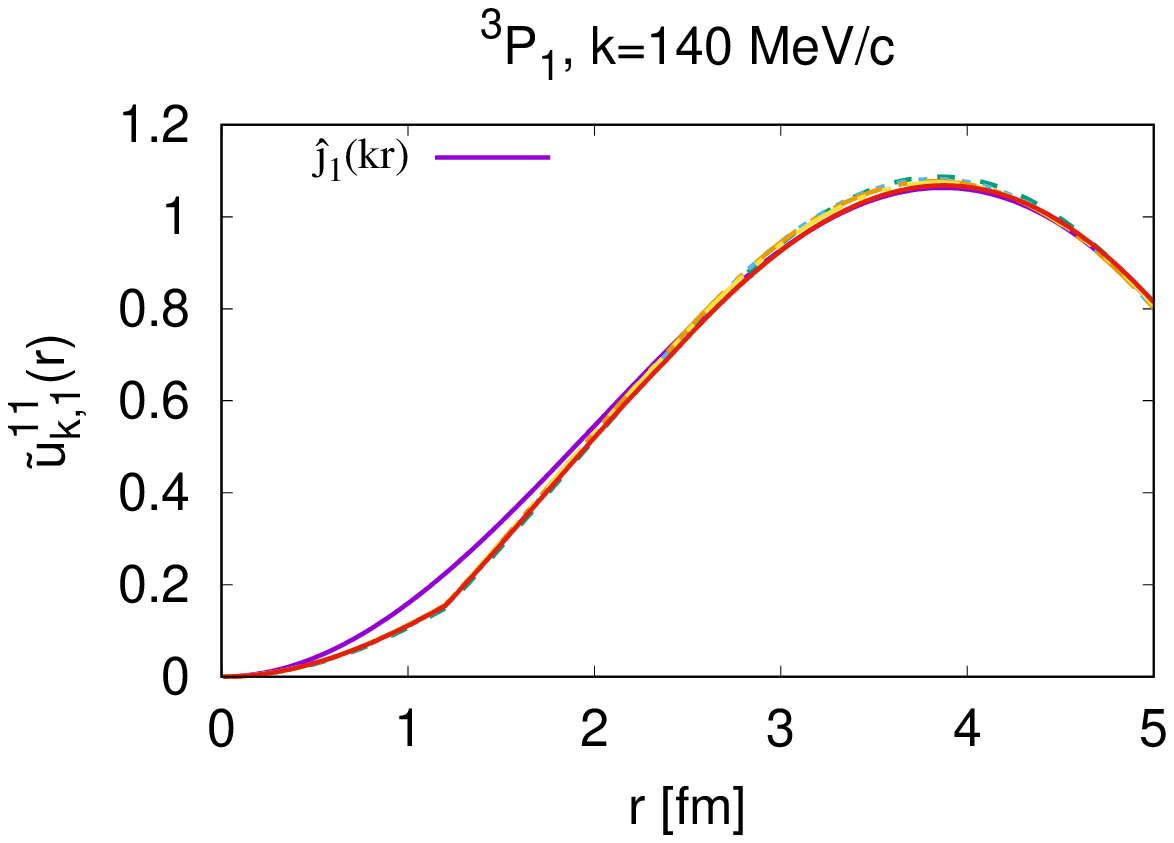}&
\includegraphics[width=7.5cm]{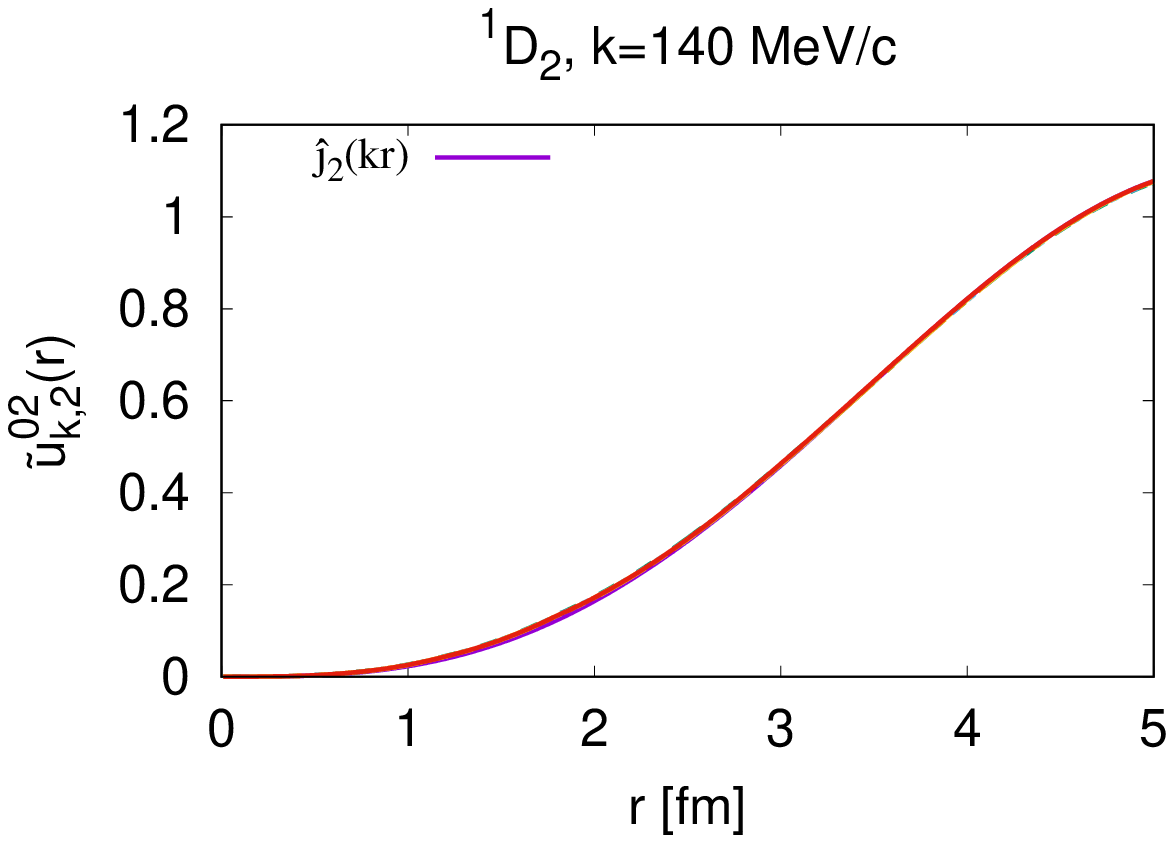}\\
\includegraphics[width=7.5cm]{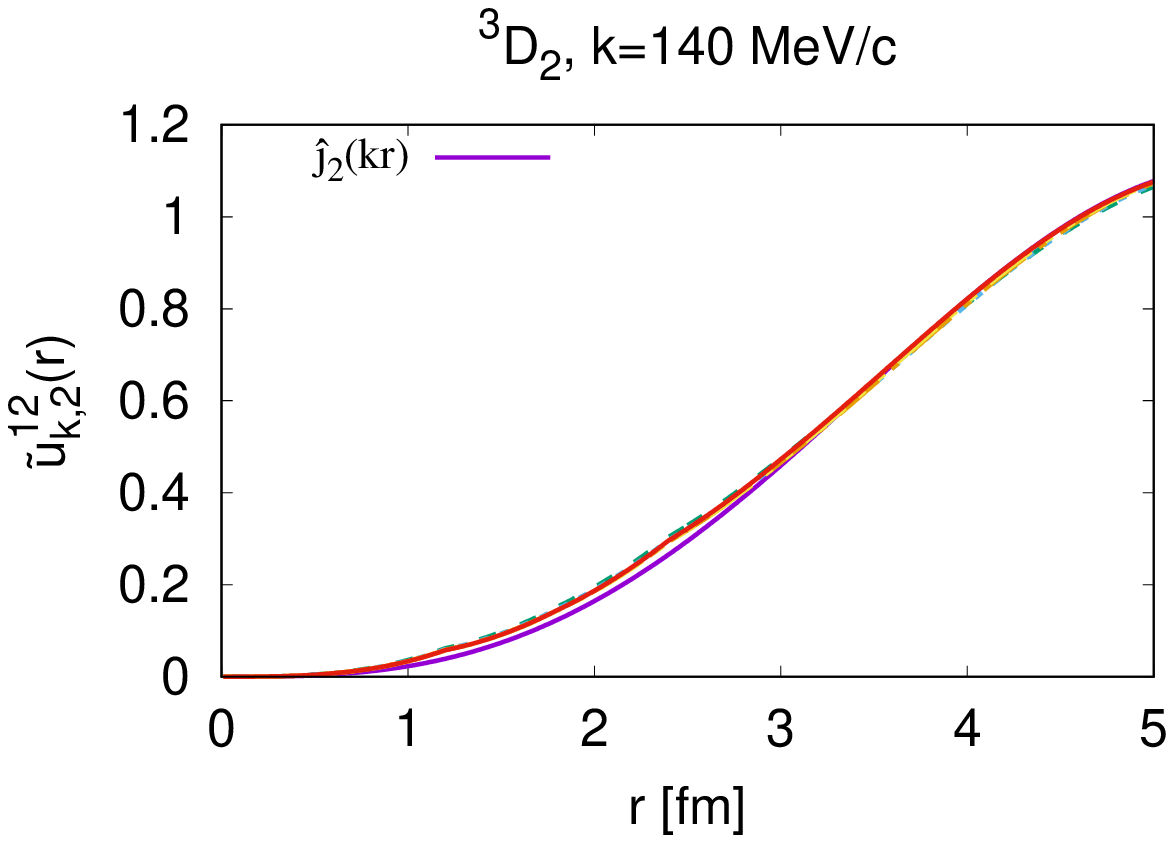}&
\includegraphics[width=7.5cm]{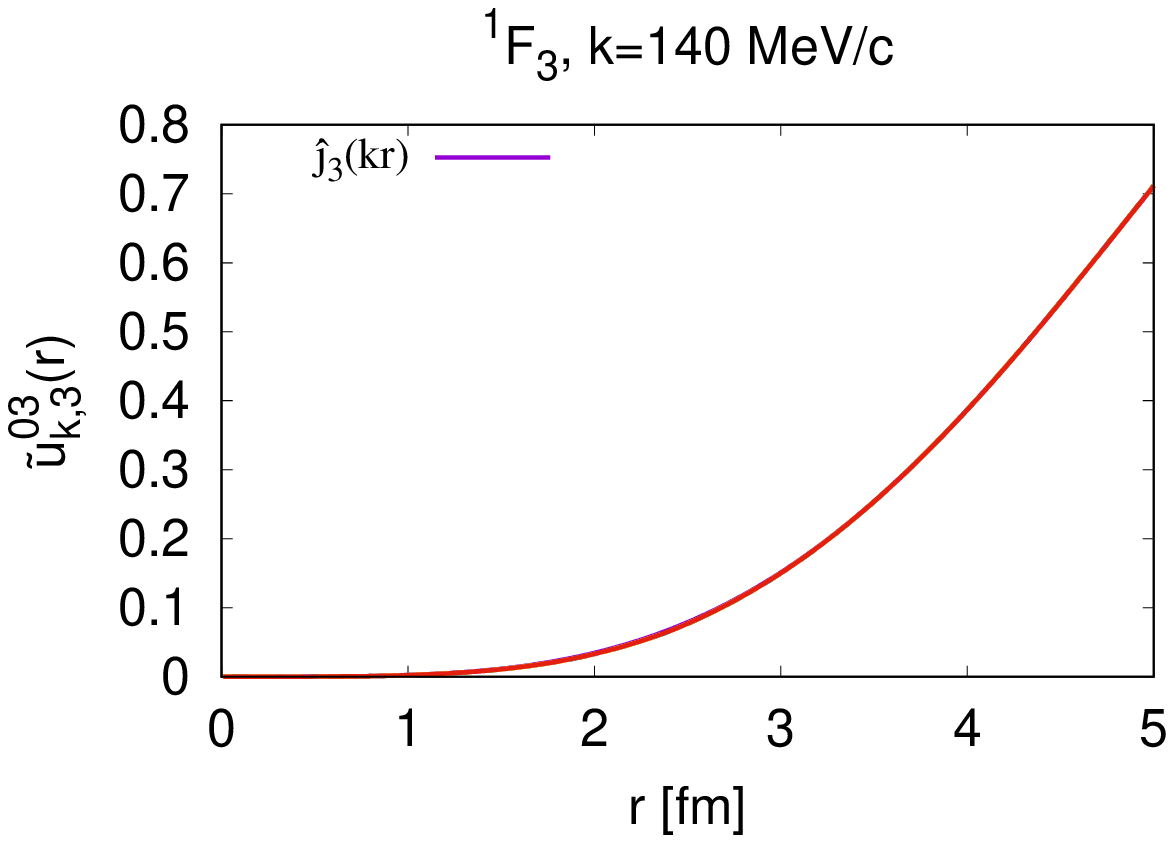}
\end{tabular}
\caption{Reduced radial wave functions $\widetilde{u}^{SJ}_{k,l}(r)$
  for the uncoupled N-N partial waves ($l=l^\prime$). The results are
  given for relative momentum $k=140$ MeV/c, and for each partial wave
  the free solution $\hat{j}_l(kr)$ as well as those for different
  values of the CM momentum are given. The results for $K_{\rm CM}=0$
  MeV/c (dashed green lines) are the same as those shown in Fig. 1 of
  Ref.~\cite{RuizSimo:2017tcb}. Although not distinguishable in all
  panels, the curves labeled in the key of the ${}^{1}$S$_0$[np]
  panel are also displayed in all the others. }
\label{Fig:uradial}
\end{figure*}

In Fig.~\ref{Fig:uradial} we show results for the radial wave
functions in coordinate representation, corresponding to the solutions
of eq.~(\ref{algebraic_radial_wf}) for the uncoupled N-N partial waves
for different total CM momenta of the nucleon pair corresponding to
$K_{\rm CM}=0,\, 100,\, 200,\, 300 \; \, {\rm and}\; \, 400$
MeV/c. All these CM momenta are compatible with having a relative
momentum of the nucleon pair $k=140$ MeV/c and both initial single
nucleon momenta fulfilling the condition of lying below the Fermi
momentum $k_F$ (in fact, the maximum allowed CM momentum for $k=140$
MeV/c under the above conditions corresponds to $K^{\rm max}_{\rm CM}
= 414.25$ MeV/c).  The values of the strength parameters
$\left(\lambda_i \right)^{SJ}_{l,l^\prime}$ of the delta-shell Granada
potential are those of Table I of Ref.~\cite{Perez:2013mwa} and they
were fitted to reproduce the N-N scattering phase-shifts of the
Granada database~\cite{NNdatabase} below the pion production
threshold.

It is evident from Fig.~\ref{Fig:uradial} that the impact of the
two-nucleon CM motion on the radial wave functions is minimal within
the scale of the figure. However, it becomes more noticeable for the
low-lying uncoupled partial waves, such as the S or P-waves.  For the
D-waves, the effect is a bit more pronounced in the triplet
${}^{3}$D$_2$ partial wave than in the singlet ${}^{1}$D$_2$ one,
because of the strength parameters of the potential at the first
delta-shell (in this case they correspond to $\lambda_2$ in Table I of
\cite{Perez:2013mwa}); the attractive behavior of the first
delta-shell parameter in the ${}^{3}$D$_2$ partial wave is much
stronger than in the ${}^{1}$D$_2$ one.

The reasons for the SRCs effects (distortions in the radial wave
functions) being more distinguishable in the low lying $l$-partial
waves have to do not only with the strength parameters of the
delta-shell potential (cf. Table I of Ref.~\cite{Perez:2013mwa}), but
also with the centrifugal barrier of each partial wave (rising with
the $l$ value), which prevents the two nucleons to approach more
closely each other. The effects of SRCs are particularly noticeable at
short inter-nucleon distances. In the case of higher partial waves,
such as D or F-waves, the probability of nucleons approaching each
other is significantly suppressed due to the presence of the
centrifugal barrier. As a result, the influence of SRCs on these
higher partial waves is less pronounced compared to the lower ones.

Nonetheless, the important point of Fig.~\ref{Fig:uradial} is that
there is little dependence on the CM momentum in the perturbed radial
wave functions at short distances, and this fact will have important
consequences in the momentum distributions for each partial wave,
$\Delta\phi^{SJ}_{k,l^\prime l}(p)$ (eq.~(\ref{high_mom_components}),
at high probed relative momenta $p$, as it will be shown later.

\begin{figure*}[!ht]
\begin{tabular}{cc}
\includegraphics[width=7.5cm]{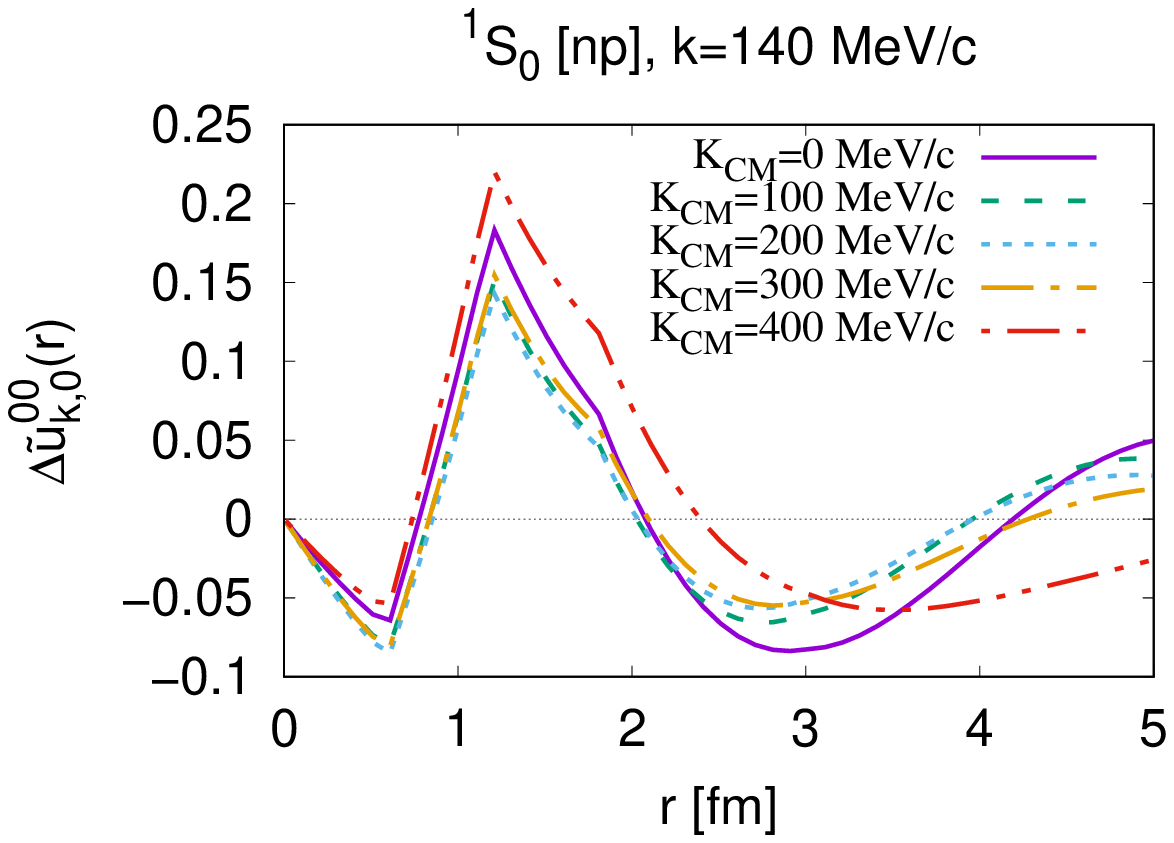}&
\includegraphics[width=7.5cm]{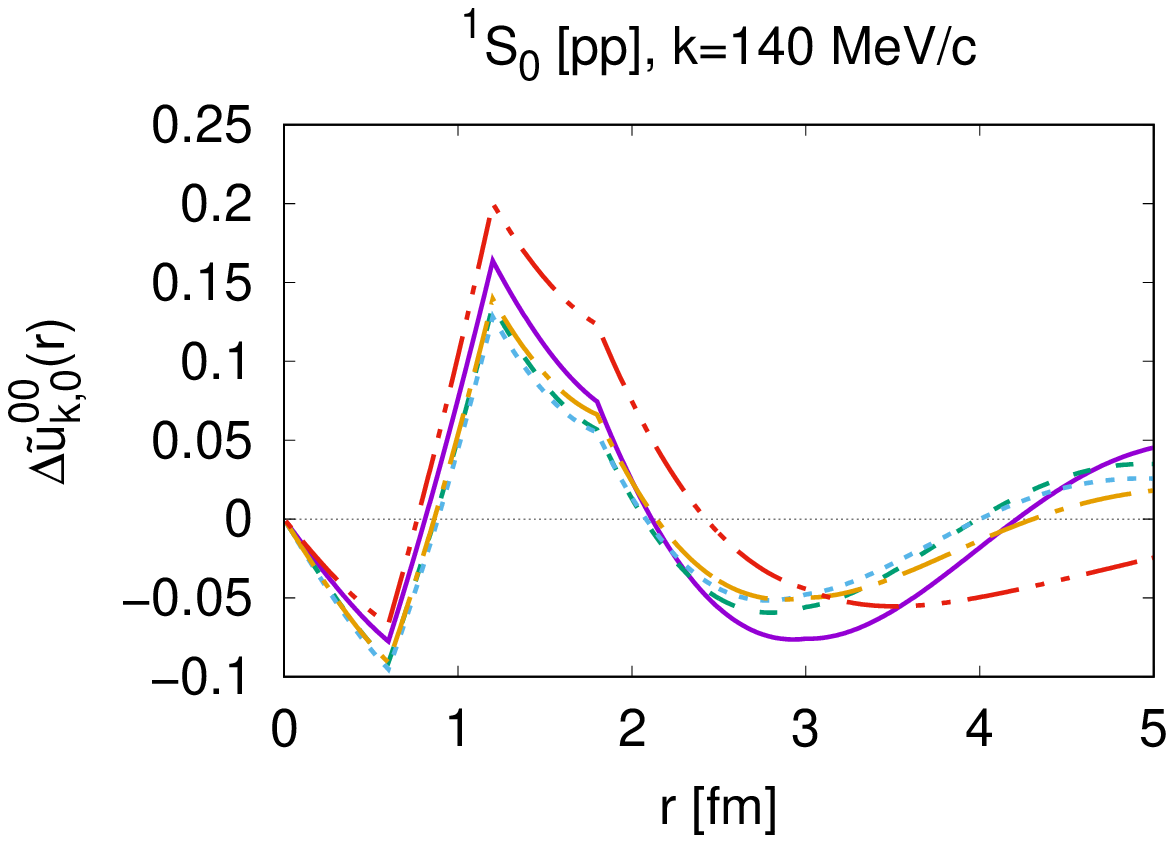}\\
\includegraphics[width=7.5cm]{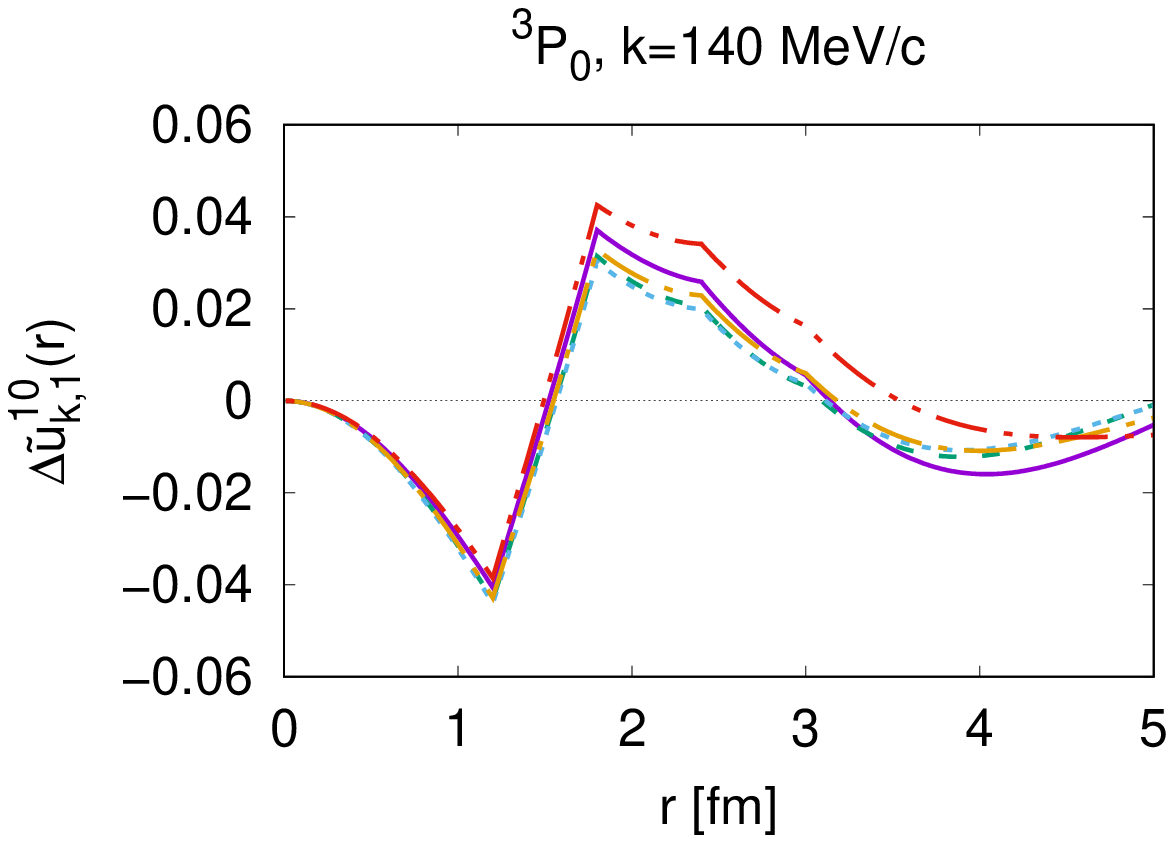}&
\includegraphics[width=7.5cm]{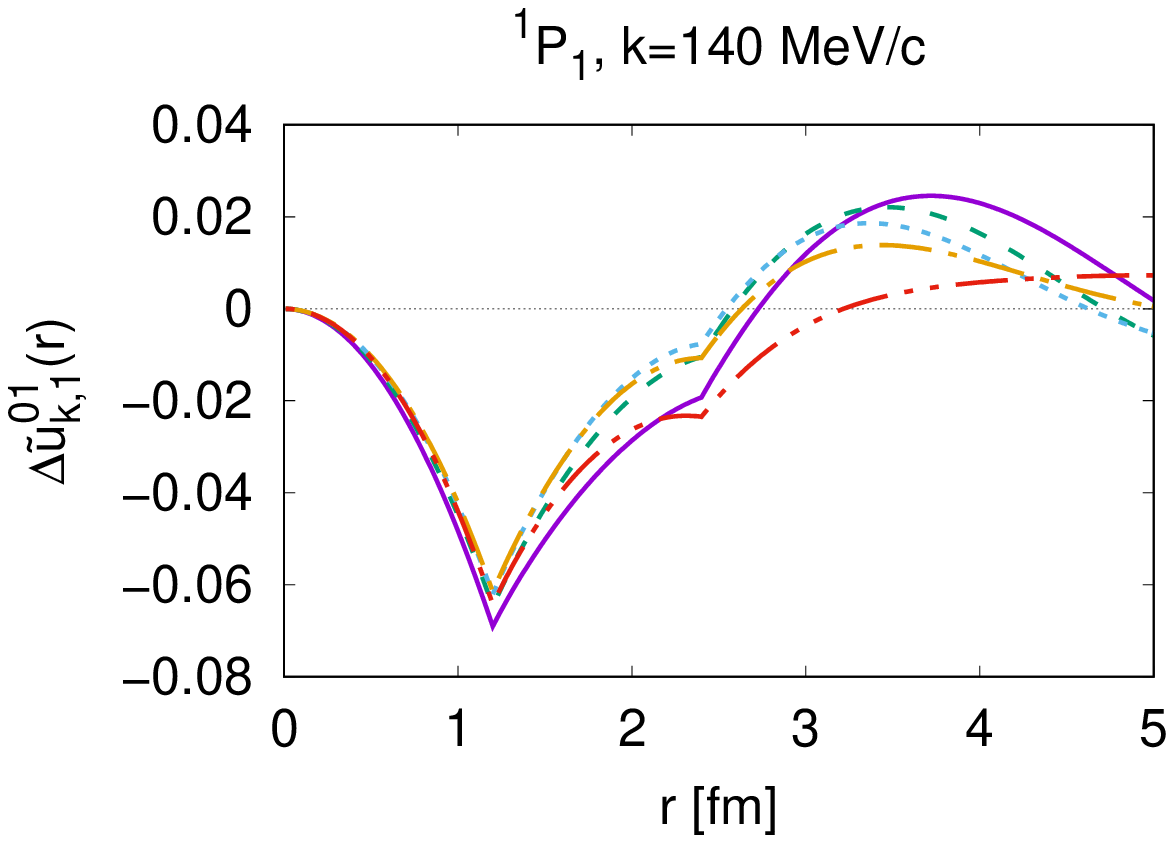}\\
\includegraphics[width=7.5cm]{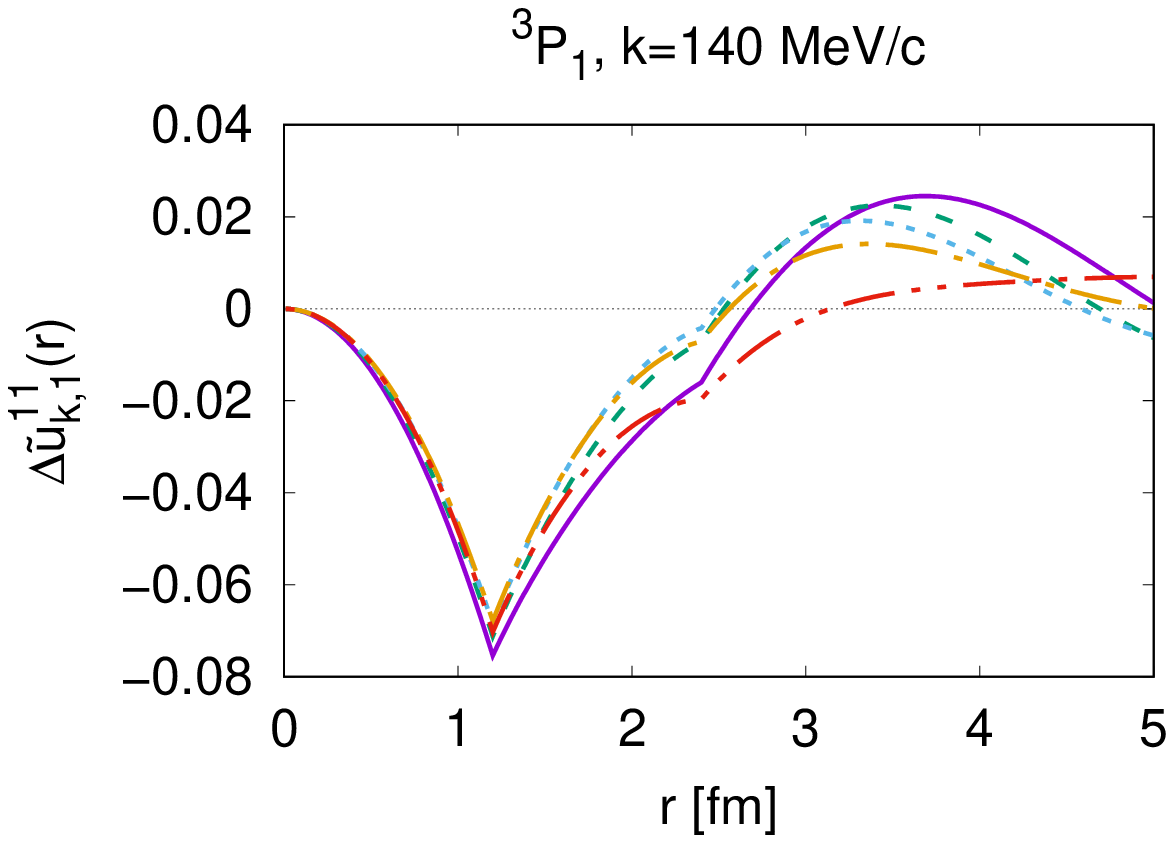}&
\includegraphics[width=7.5cm]{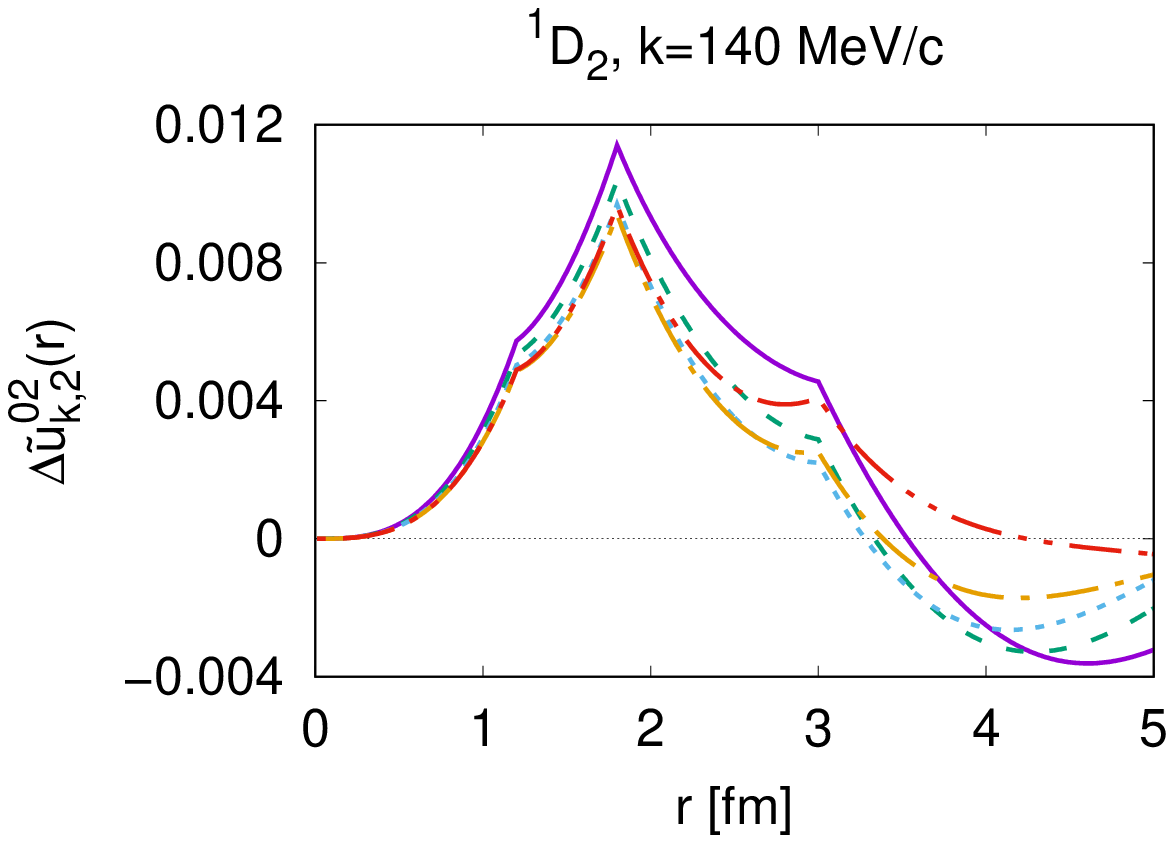}\\
\includegraphics[width=7.5cm]{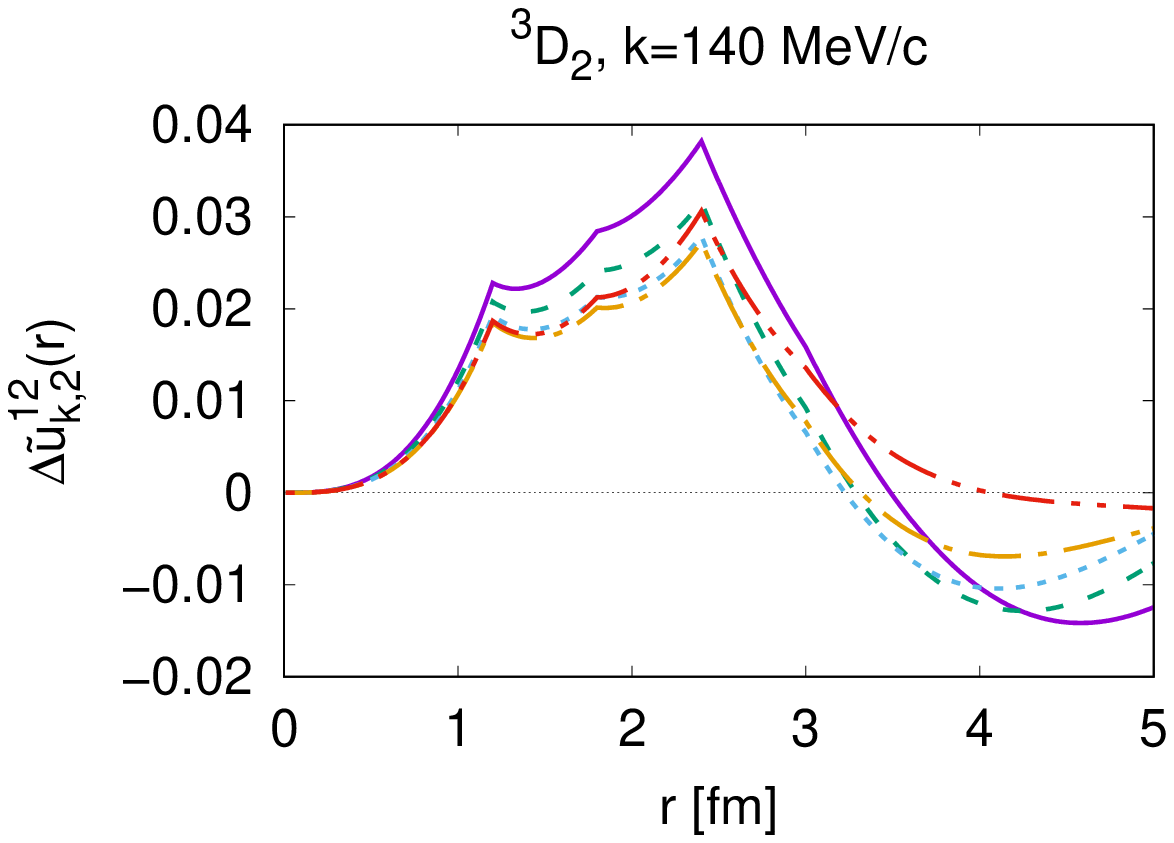}&
\includegraphics[width=7.5cm]{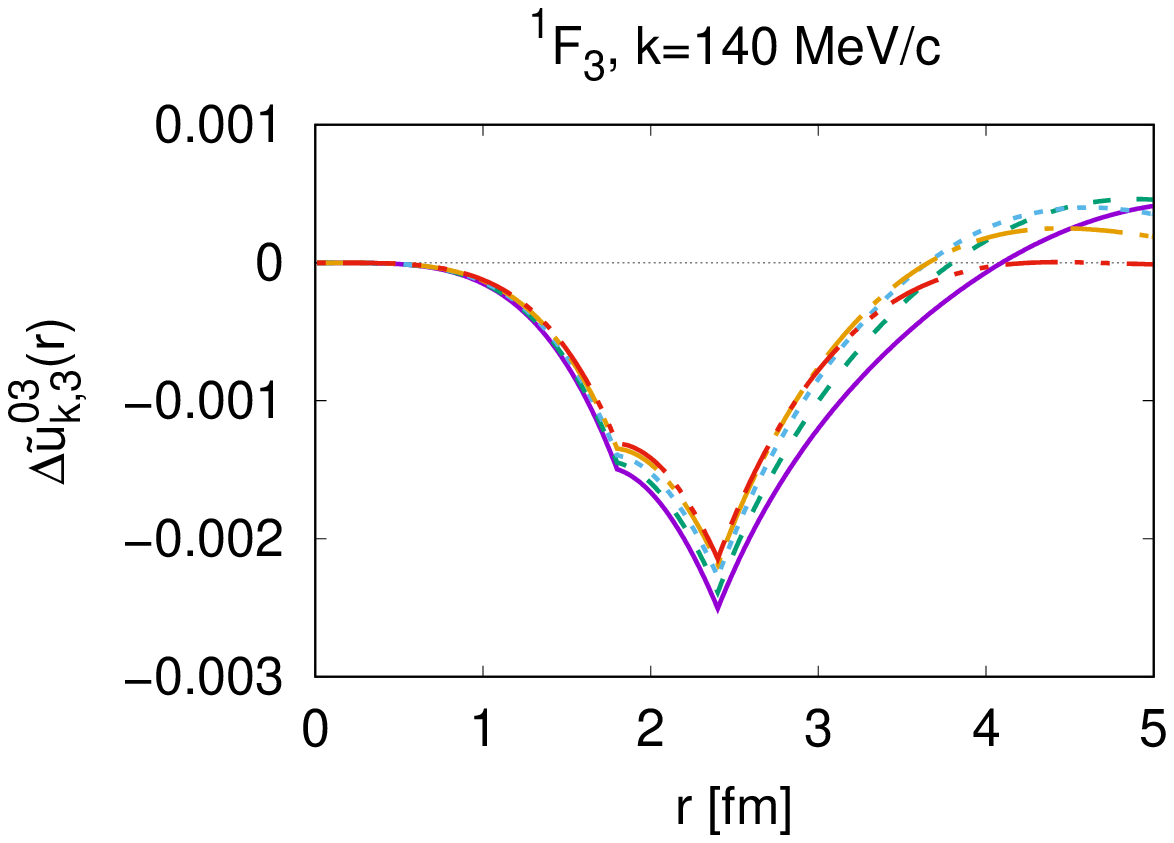}
\end{tabular}
\caption{Defect wave functions
  $\Delta\widetilde{u}^{SJ}_{k,l}(r)\equiv
  \widetilde{u}^{SJ}_{k,l}(r)-\hat{j}_l(kr)$ for the uncoupled N-N
  partial waves, $l^\prime=l$.  The results are given for relative
  momentum $k=140$ MeV/c and for the same values of the CM momentum as
  in Fig.~\ref{Fig:uradial}.  The results for $K_{\rm CM}=0$ MeV/c
  (solid purple lines) are the same as those shown as short-dashed
  green lines in Fig. 1 of Ref.~\cite{RuizSimo:2017tcb}, but on a
  different vertical scale.  Notice that the scales on the vertical
  axes are, in general, different for each partial wave as well.}
\label{Fig:defect}
\end{figure*}

In order to magnify the differences between the perturbed wave
functions and the free ones shown in Fig.~\ref{Fig:uradial}, we
present the defect wave functions in Fig.~\ref{Fig:defect}. These are
defined as the difference between the perturbed wave functions and the
free solutions
\begin{equation}\label{eq:defect}
\Delta\widetilde{u}^{SJ}_{k,l}(r)
\equiv
\widetilde{u}^{SJ}_{k,l}(r)-\hat{j}_l(kr),
\end{equation}
for the diagonal $l^\prime=l$ case.

Here we also observe the general trend discussed in
Fig.~\ref{Fig:uradial}, namely: the amplitude of the distortion in the
perturbed wave function (importance of SRCs effects), in general, gets
smaller when the value of the orbital angular momentum $l$ increases
(cf. the different scales in the vertical axes of
Fig.~\ref{Fig:defect}), thus reflecting the importance of the
centrifugal barrier that prevents the two nucleons to approach more
closely each other and to experience the short-range N-N interaction,
although the strength parameters of the delta-shell Granada potential
also play a role; and, on the other hand, the magnitude of the
distortion is quite insensitive to the state of global motion of the
two-nucleon system, i.e, the value of the CM momentum (notice that the
values at the cusps are more or less the same for the different curves
in each panel of Fig.~\ref{Fig:defect}). Therefore, we can write, in
general, a sort of hierarchy for the magnitude of the distortions in
the wave functions due to the SRCs:
  \begin{equation}\label{eq:hierarchy}
  \Delta\widetilde{u}_{l=0} >
  \Delta\widetilde{u}_{l=1} >
   \Delta\widetilde{u}_{l=2} >
    \Delta\widetilde{u}_{l=3}.
  \end{equation}
  
  It is also worth noticing the reader that the amplitudes of the
  distortions in Fig.~\ref{Fig:defect} are related to the importance
  of that partial wave in the two-nucleon relative high-momentum
  distribution $\rho^{S}_{k,K_{\rm CM}}(p)$ given by
  eq.~(\ref{high_dens_mom_dist_spinS_final}); since the different
  wavelengths overlapped (with different amplitudes of course) in the
  defect wave functions of the same figure are related with the
  corresponding high-momentum components in the relative momentum
  distribution for each partial wave $\left| \phi^{SJ}_{k,
    l^\prime\,l}(p) \right|^2$, as in any continuous harmonic Fourier
  analysis.
 
\begin{figure*}[!ht]
\begin{tabular}{cc}
\includegraphics[width=7.5cm]{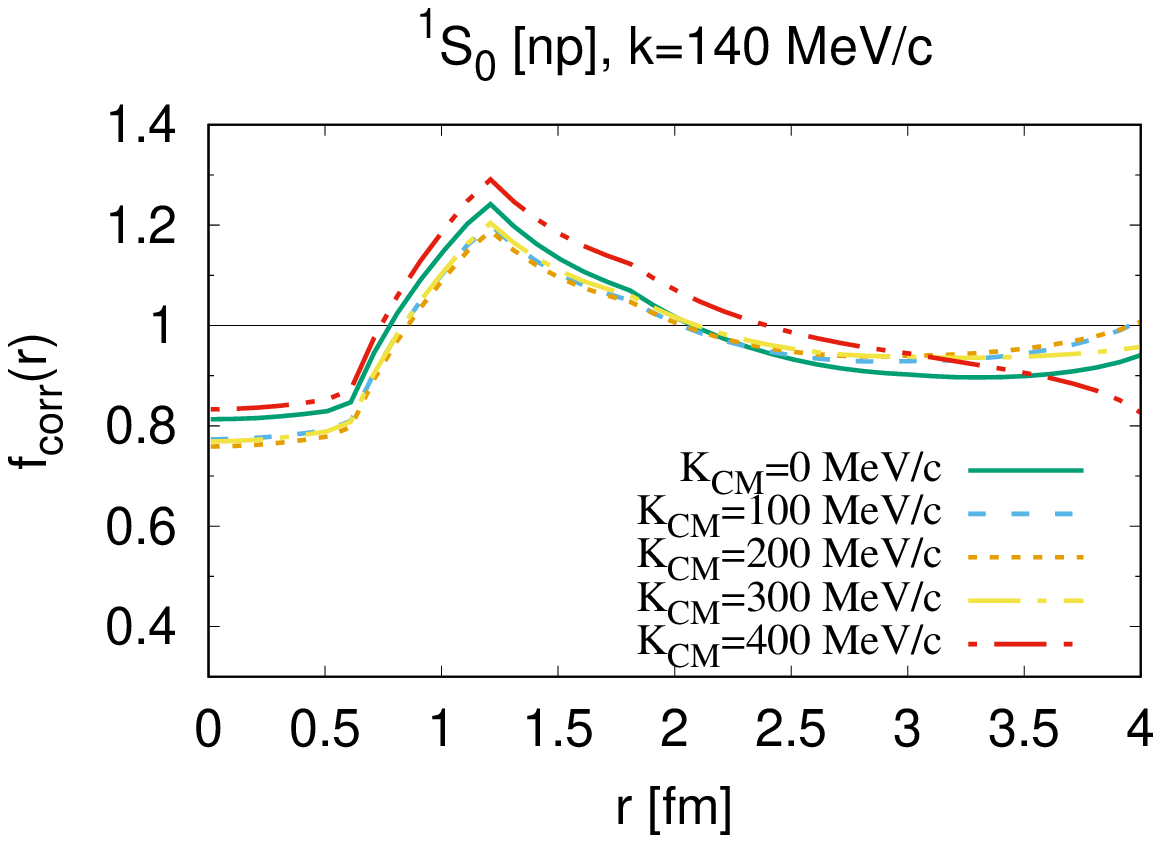}&
\includegraphics[width=7.5cm]{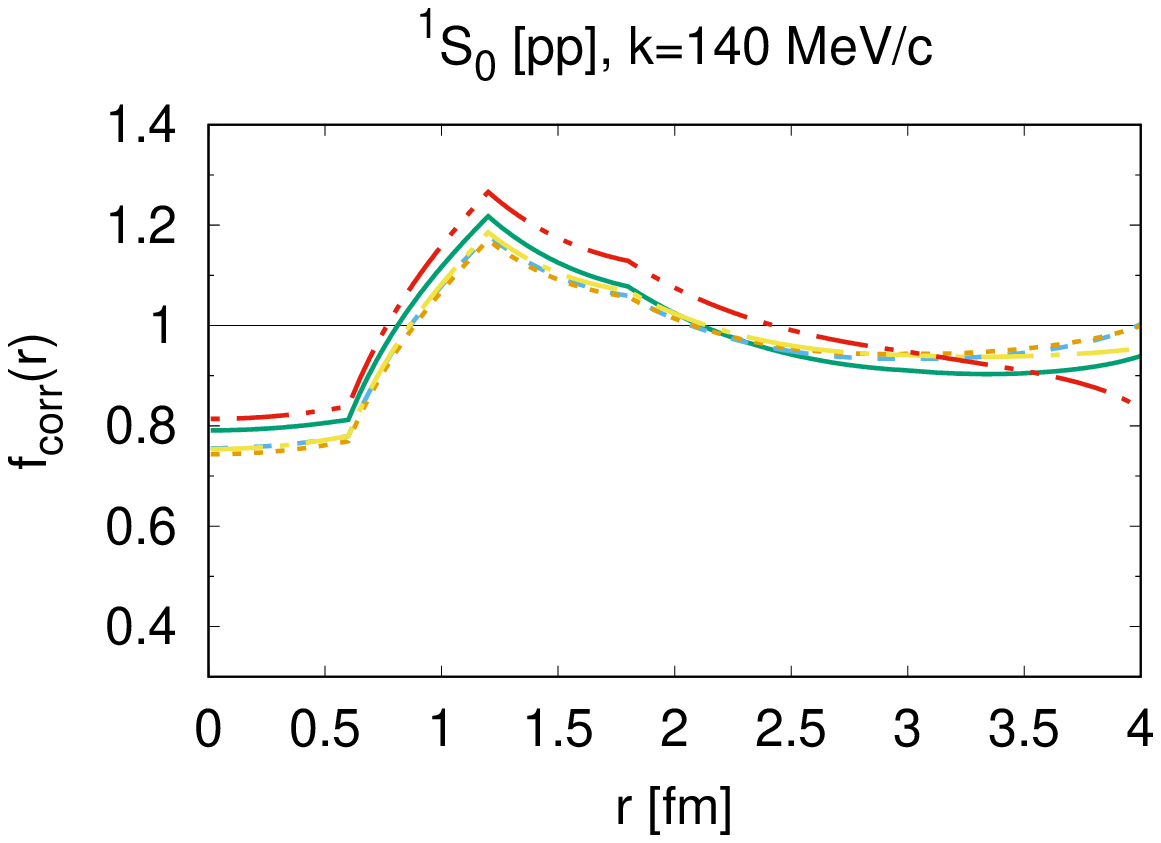}\\
\includegraphics[width=7.5cm]{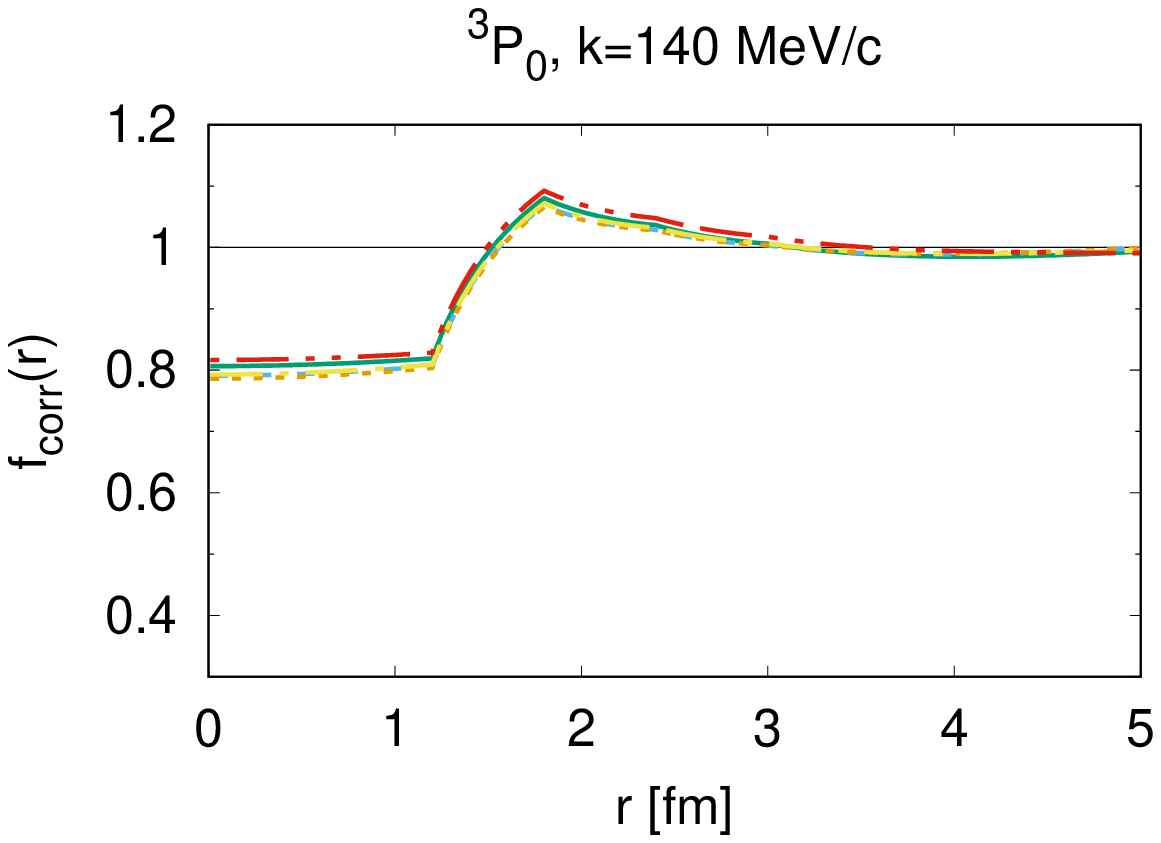}&
\includegraphics[width=7.5cm]{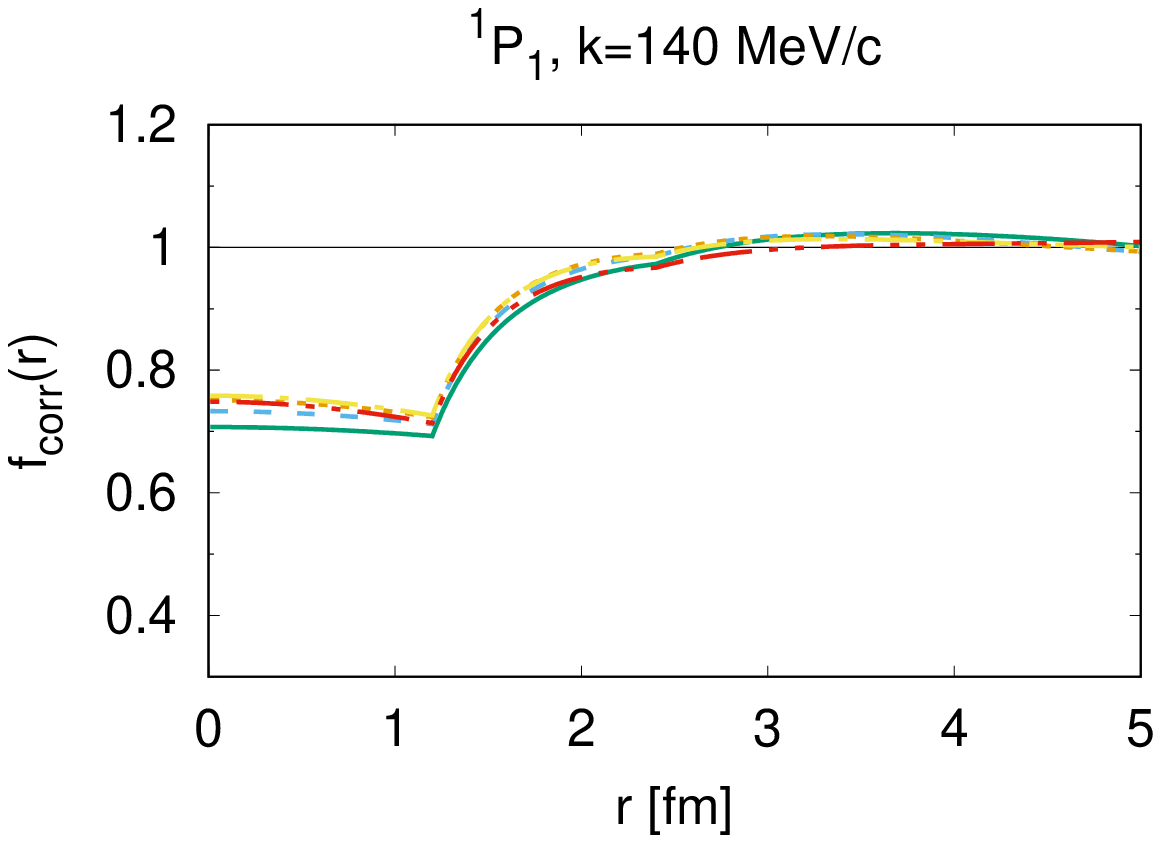}\\
\includegraphics[width=7.5cm]{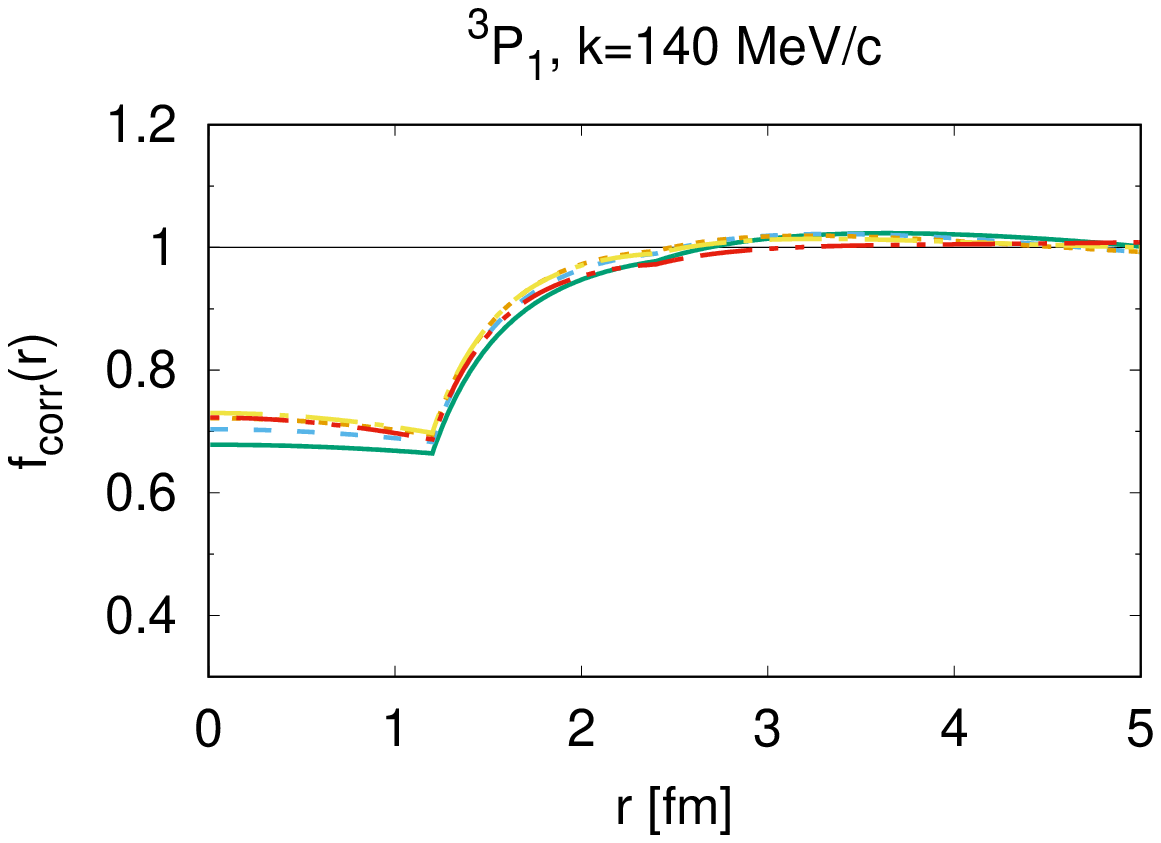}&
\includegraphics[width=7.5cm]{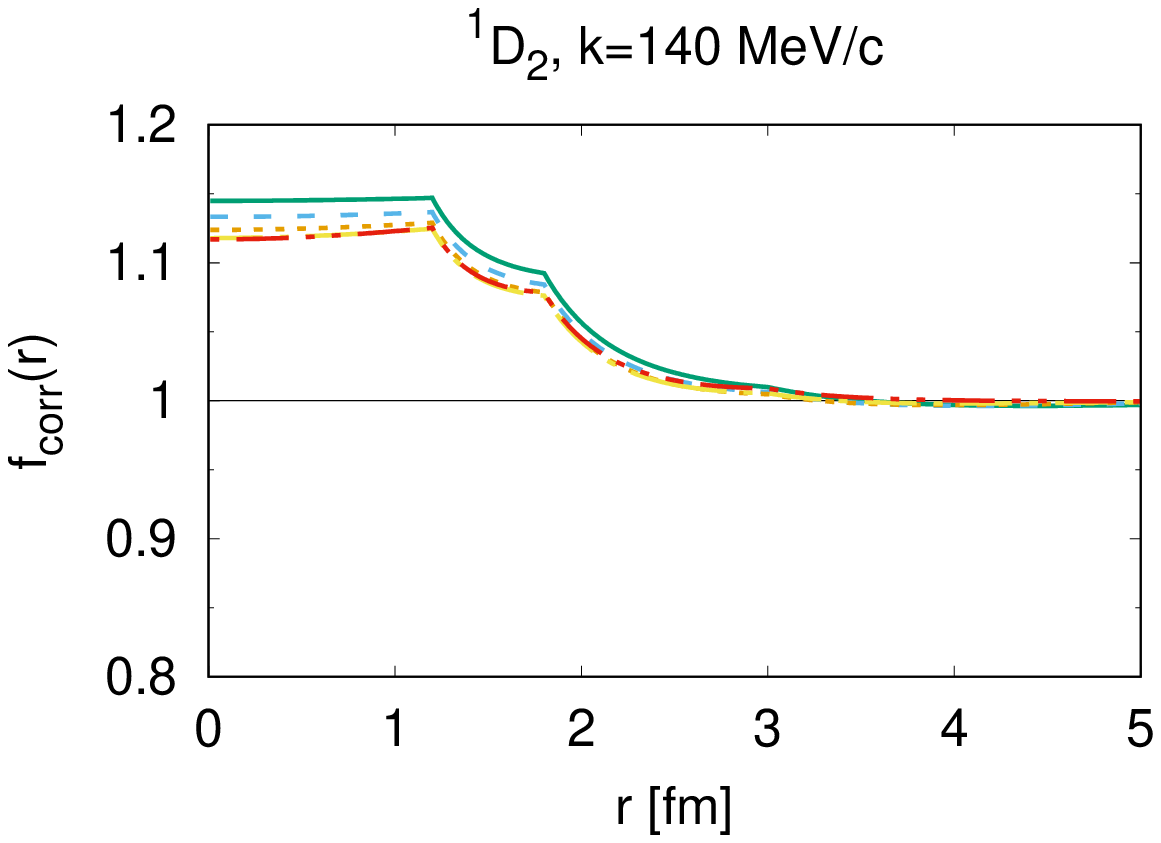}\\
\includegraphics[width=7.5cm]{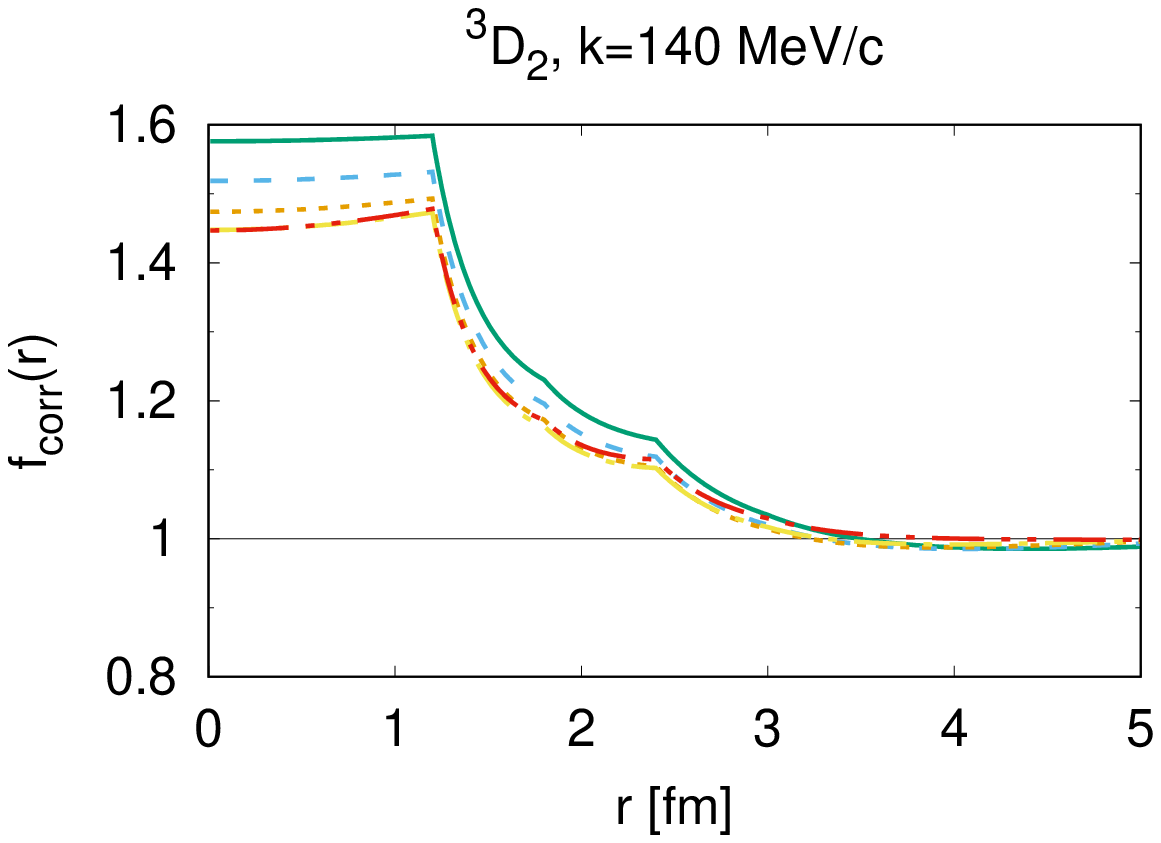}&
\includegraphics[width=7.5cm]{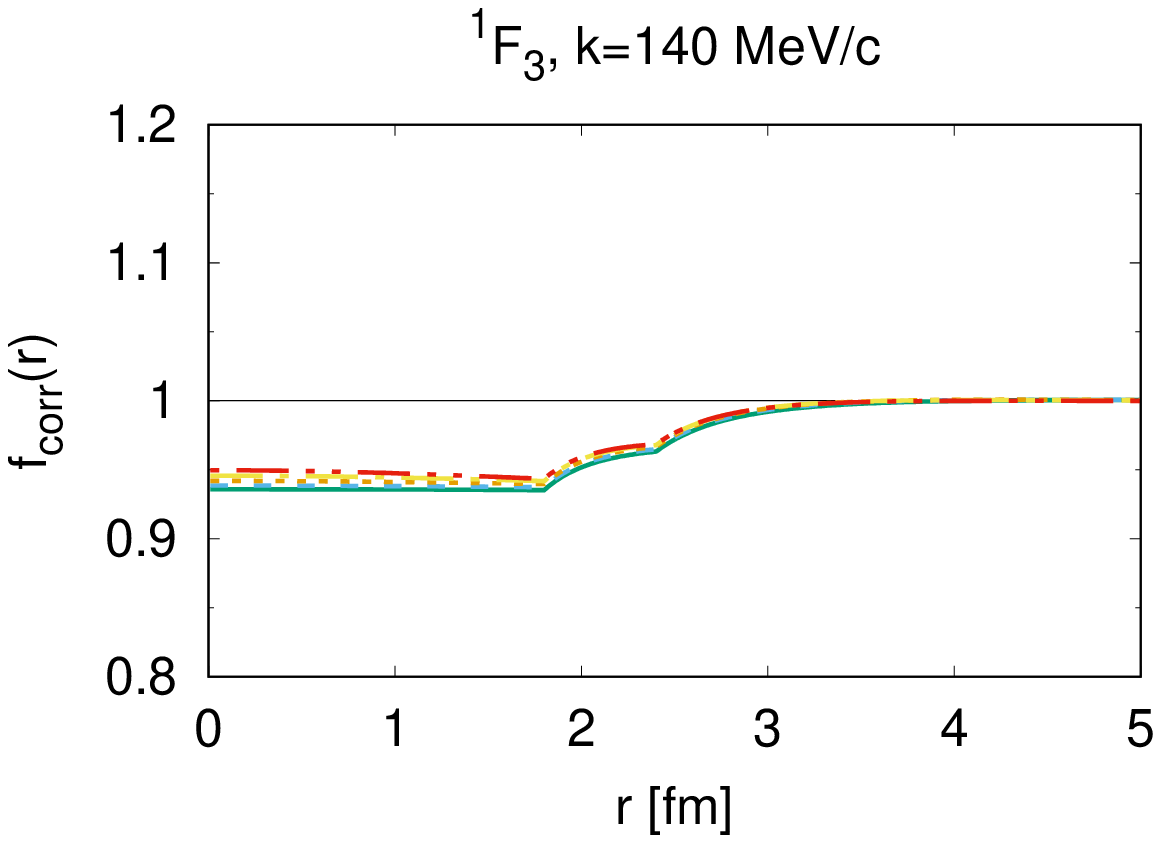}
\end{tabular}
\caption{Correlation functions $f_{\rm corr}(r)\equiv
  \frac{\widetilde{u}^{SJ}_{k,l}(r)}{\hat{j}_l(kr)}$ for the uncoupled
  N-N partial waves, $l^\prime=l$.  The results are given for relative
  momentum $k=140$ MeV/c and for the same values of the CM momentum as
  in Figs.~\ref{Fig:uradial} and \ref{Fig:defect}.  The results for
  $K_{\rm CM}=0$ MeV/c (solid green lines) are the same as those shown
  in Fig. 2 of Ref.~\cite{RuizSimo:2017tcb}.}
\label{Fig:fcorrelation}
\end{figure*}

In Fig.~\ref{Fig:fcorrelation} we display the correlation function for
each uncoupled N-N partial wave, defined by
\begin{equation}\label{eq:fcorrelation}
f_{\rm corr}(r)\equiv 
\frac{\widetilde{u}^{SJ}_{k,l}(r)}{\hat{j}_l(kr)}.
\end{equation}
Again, the most remarkable feature of these plots is the little
dependence of the correlation function on the different CM momenta of
the nucleon pair. The boldest dependence on the CM momenta, especially
close to the origin ($r=0$), occurs for the ${}^{3}$D$_2$ partial
wave. The departures of the correlation functions from unity occur
only at short distances, and these functions rapidly approach $1$ at
larger distances, which means that the perturbed solutions reach the
free ones without any phase-shift
\begin{equation}\label{limit_fcorr}
f_{\rm corr}(r) \longrightarrow 1\; \Longleftrightarrow \;
\widetilde{u}^{SJ}_{k,l}(r) \longrightarrow \hat{j}_l(kr)
\quad {\rm for} \quad r > 3 \; {\rm fm}.
\end{equation} 

Indeed, special attention must be paid to the zeros of the reduced
spherical Bessel functions $\hat{j}_l(kr)$ in the analysis. At these
zeros, the perturbed wave functions can have numerical uncertainties
that may give the impression of a non-zero phase shift at long
distances, even although the correlation functions approach unity.
This is due to two reasons: numerical uncertainties in the calculation
of $\widetilde{u}^{SJ}_{k,l}(r)$ at the nodes, that prevent an exact
cancellation of both nodes when numerically evaluating
eq.~(\ref{eq:fcorrelation}); and also to the fact that the perturbed
wave function truly converges on the free one precisely at \emph{very
long distances}, thus the quotient at the nodes is never exactly
$1$. This is, for example, the reason for plotting the correlation
function for the ${}^{1}$S$_0$ partial wave of
Fig.~\ref{Fig:fcorrelation} only up to $4$ fm, precisely because the
node for the value of $k$ considered in the same panels of
Fig.~\ref{Fig:uradial} appears between $4$ and $5$ fm.

\begin{figure*}[!ht]
\begin{tabular}{cc}
\includegraphics[width=7.5cm]{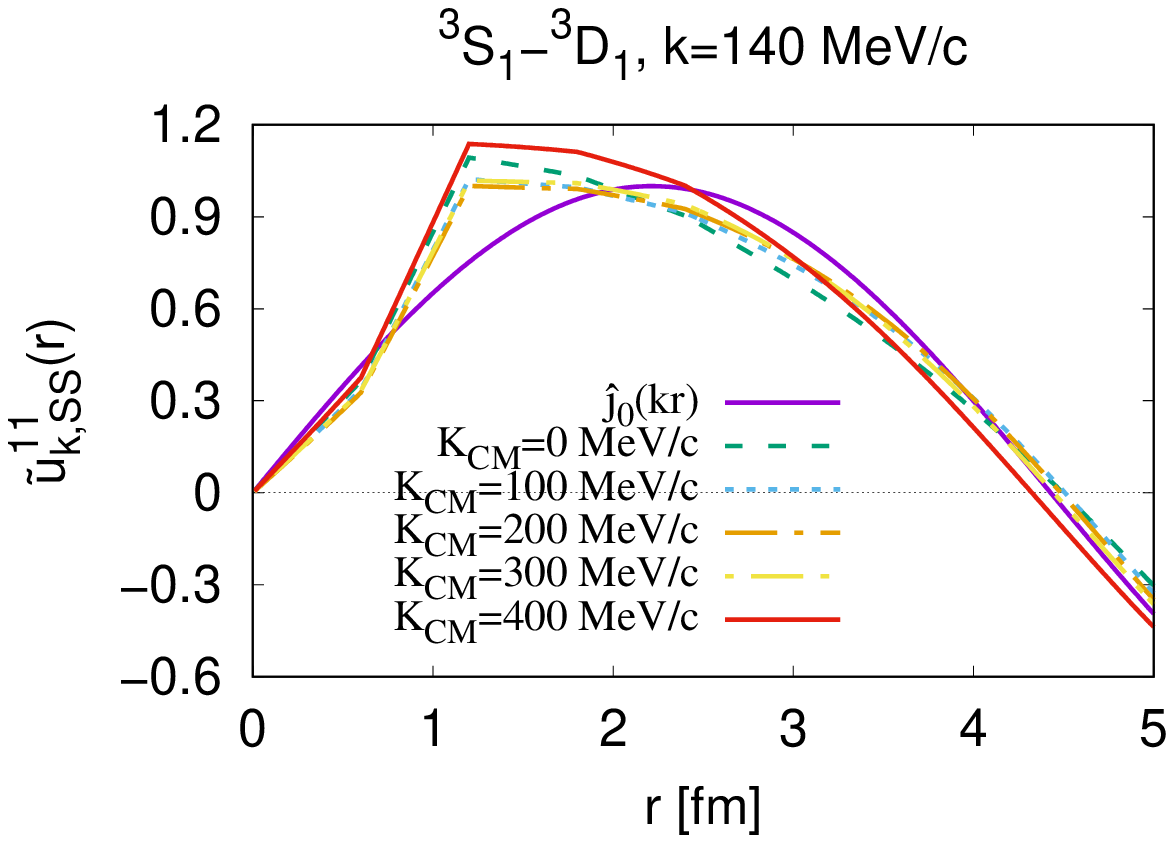}&
\includegraphics[width=7.5cm]{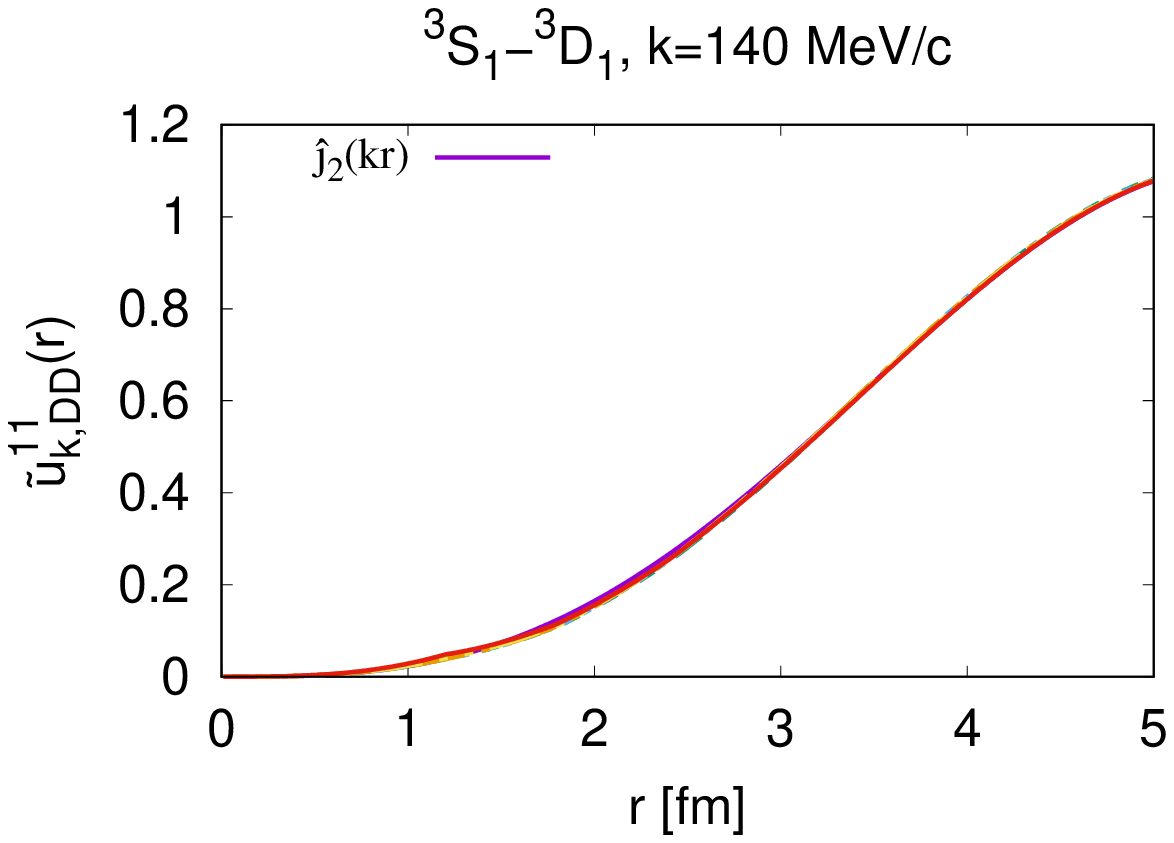}\\
\includegraphics[width=7.5cm]{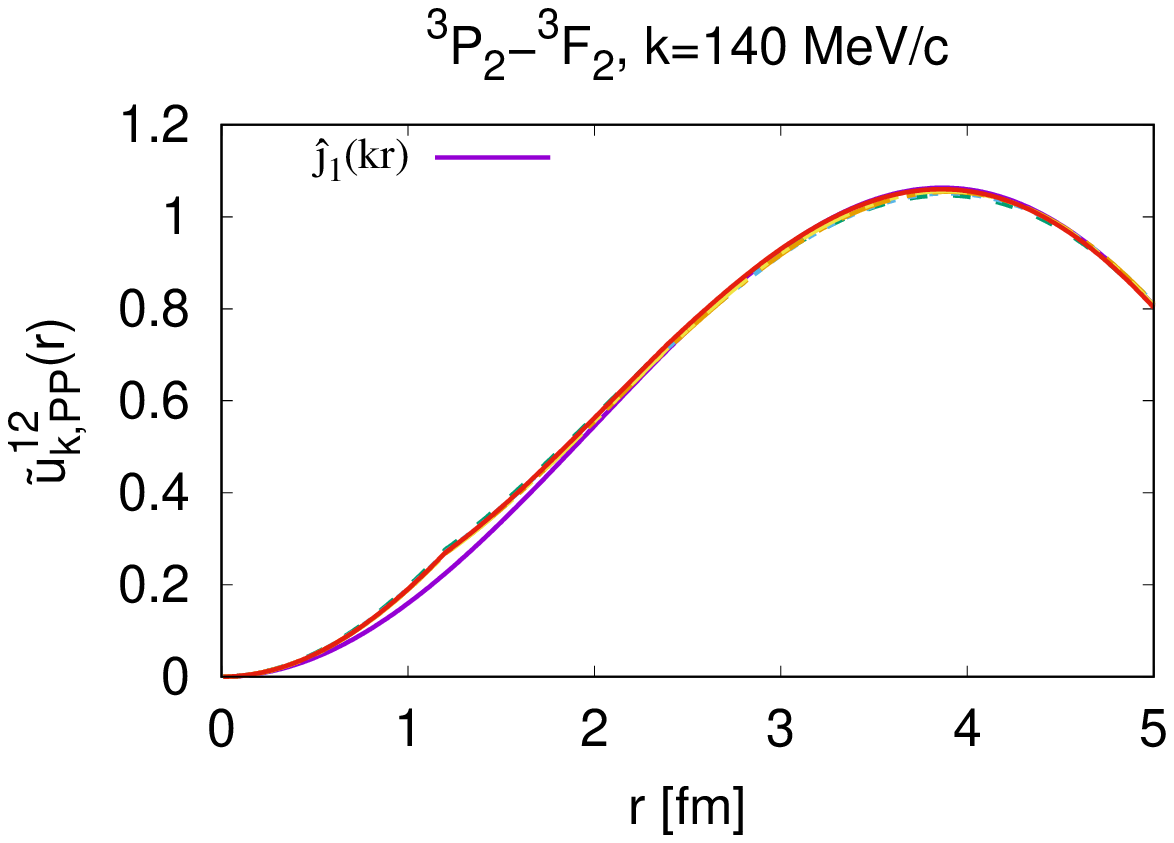}&
\includegraphics[width=7.5cm]{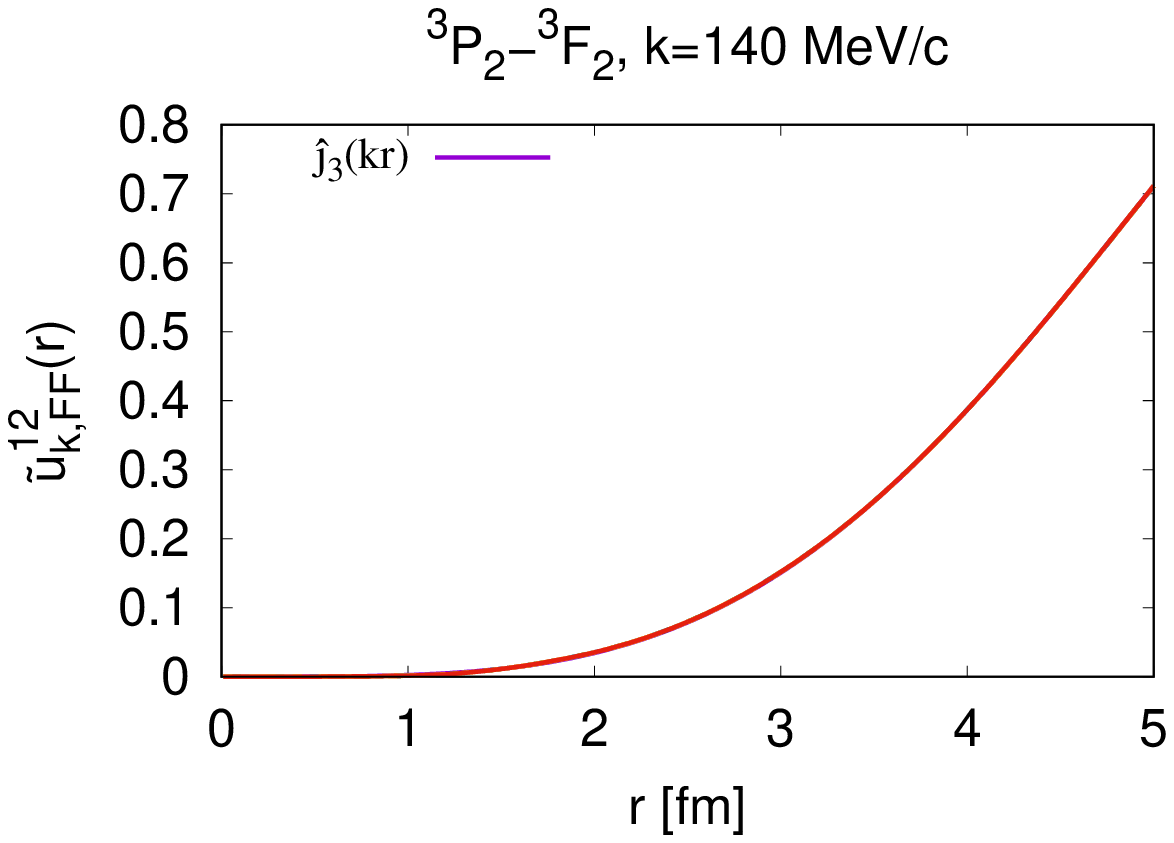}\\
\end{tabular}
\caption{Perturbed radial diagonal wave functions
  $\widetilde{u}^{SJ}_{k,l\,l}(r)$ (for $l^\prime=l$) for the coupled
  N-N partial waves ${}^{3}$S$_1$-${}^{3}$D$_1$ and
  ${}^{3}$P$_2$-${}^{3}$F$_2$. The results are given for relative
  momentum $k=140$ MeV/c, and for each partial wave the free solutions
  $\hat{j}_l(kr)$ to which they trend when $r\rightarrow\infty$ are
  shown, as well as those for different values of the CM
  momentum. Although not distinguishable in all the panels, the curves
  labeled in the key of the upper left panel are also displayed in
  all the others. }
\label{Fig:uradial_coupled}
\end{figure*}

\begin{figure*}[!ht]
\begin{tabular}{cc}
\includegraphics[width=7.5cm]{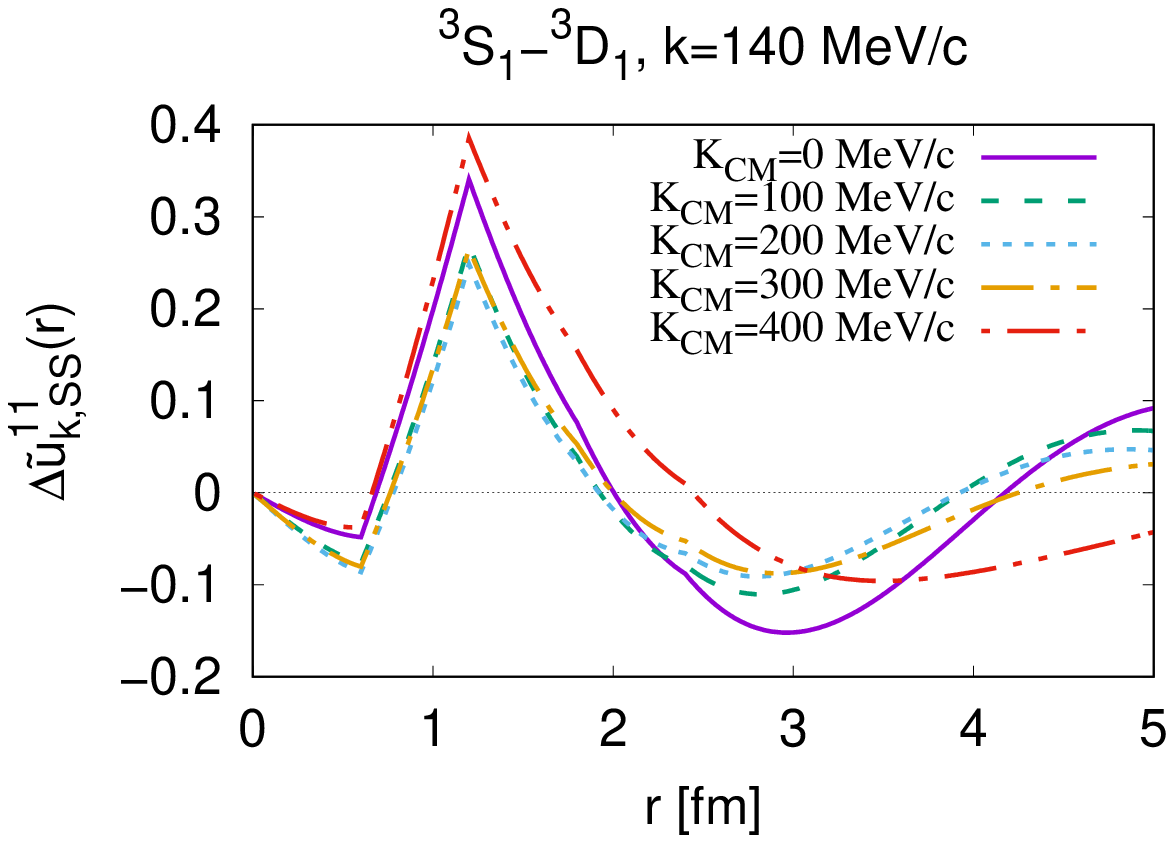}&
\includegraphics[width=7.5cm]{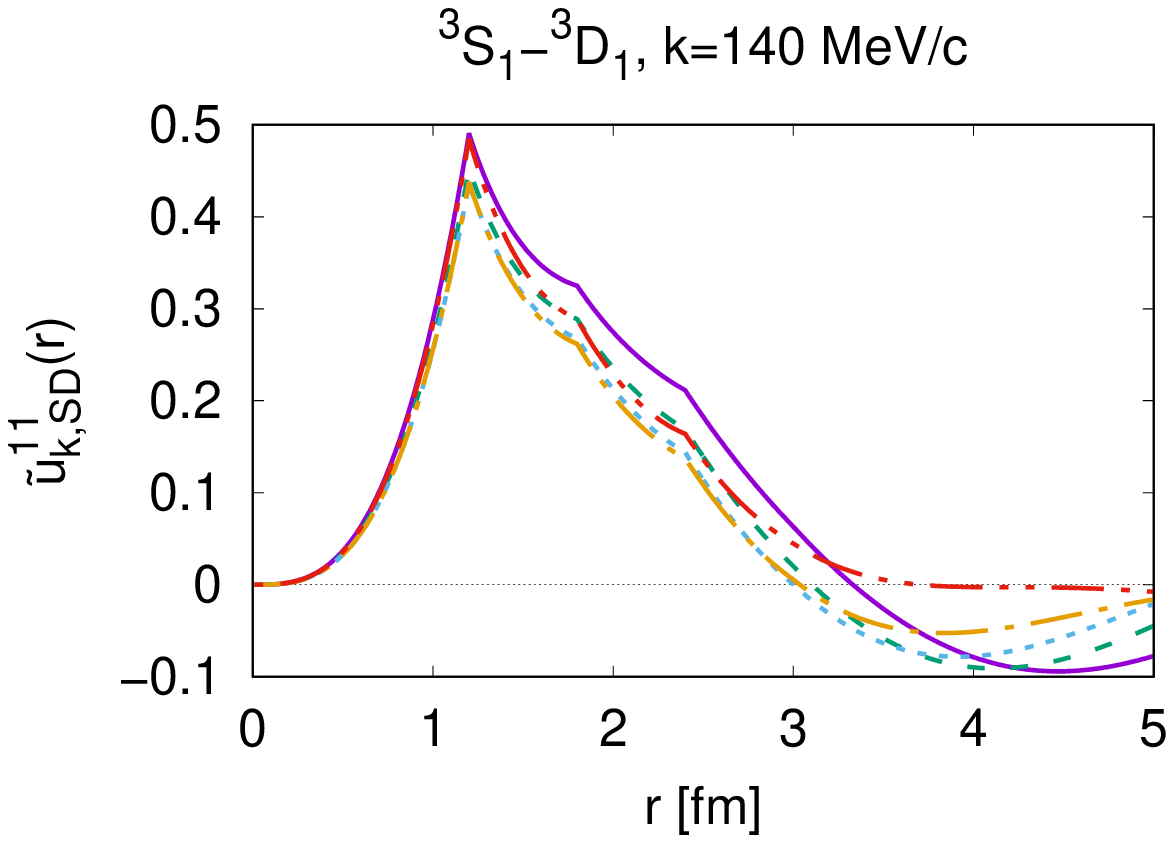}\\
\includegraphics[width=7.5cm]{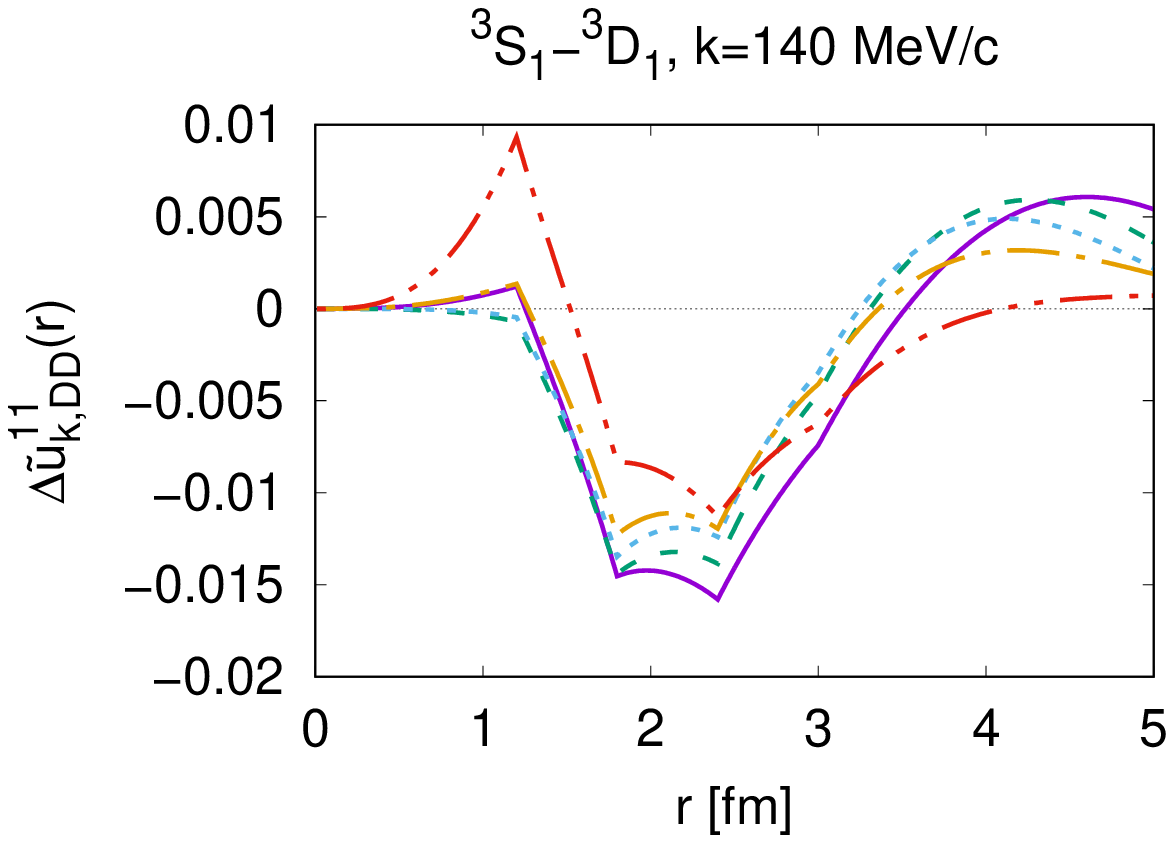}&
\includegraphics[width=7.5cm]{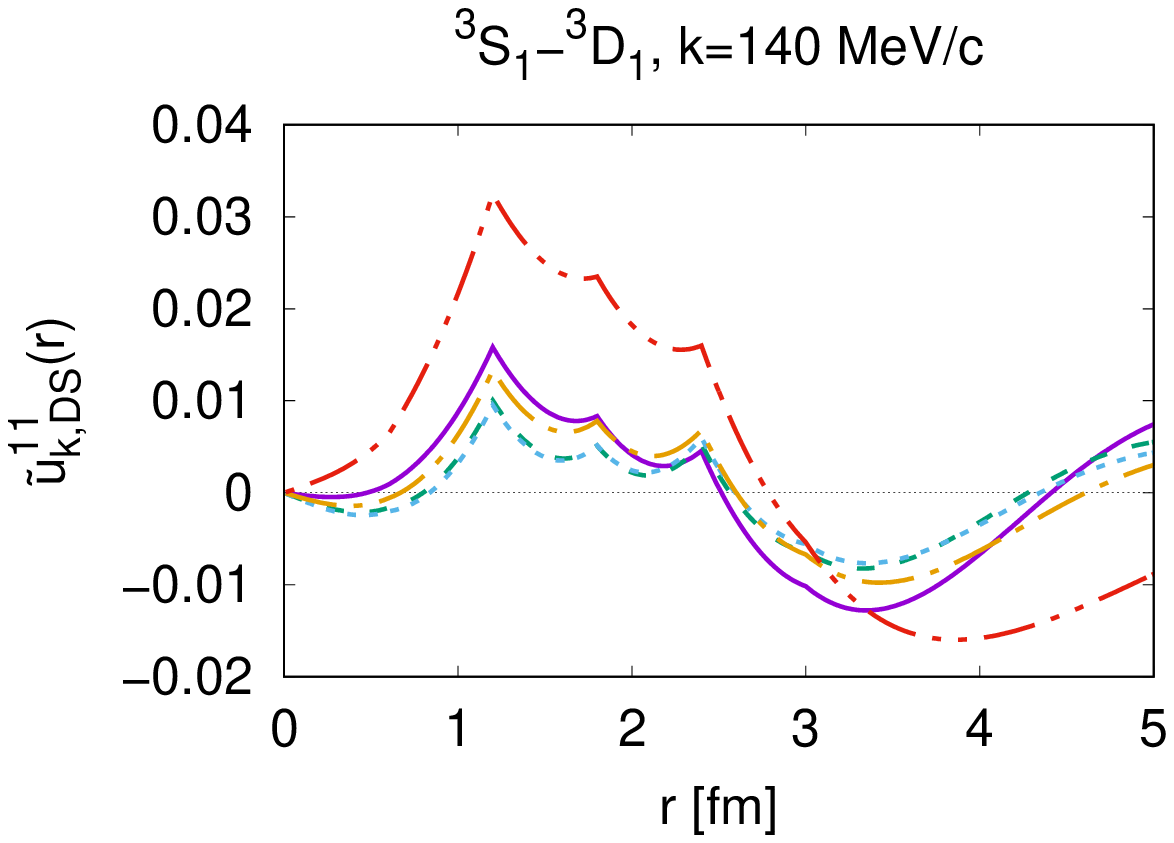}\\
\includegraphics[width=7.5cm]{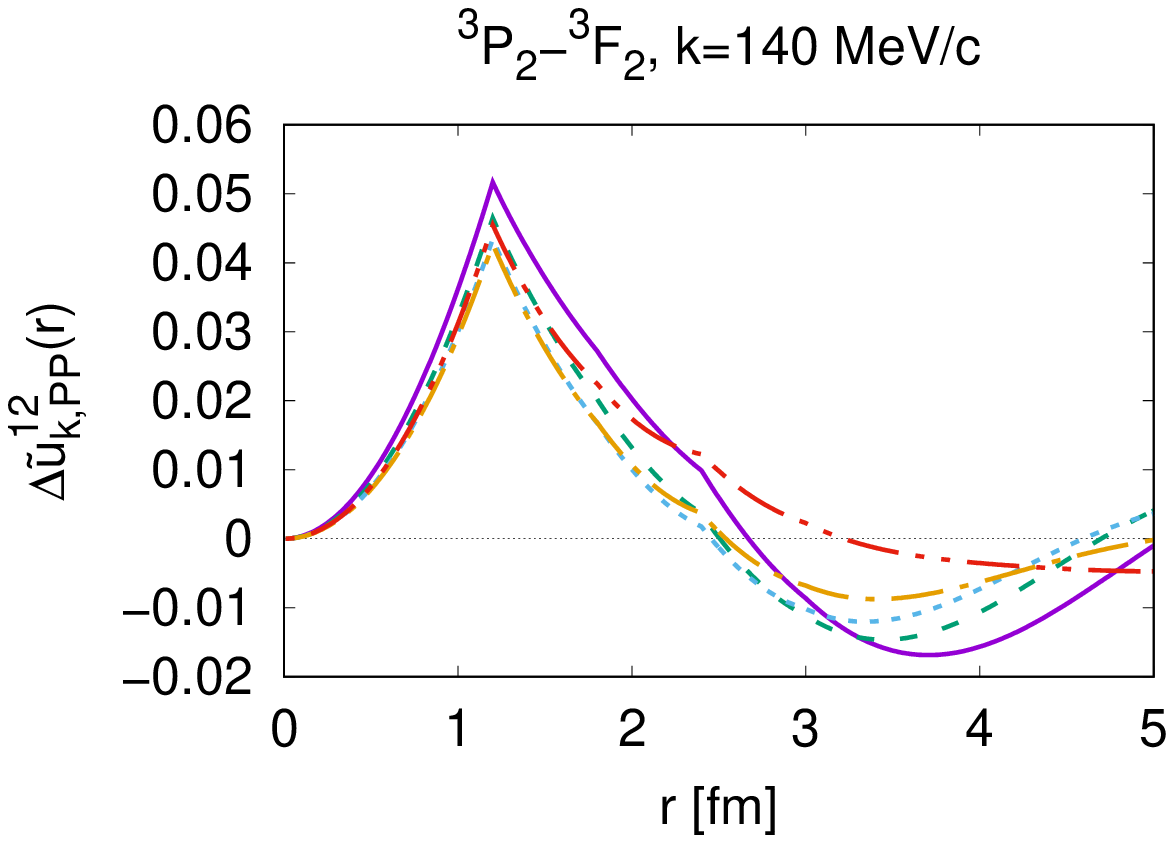}&
\includegraphics[width=7.5cm]{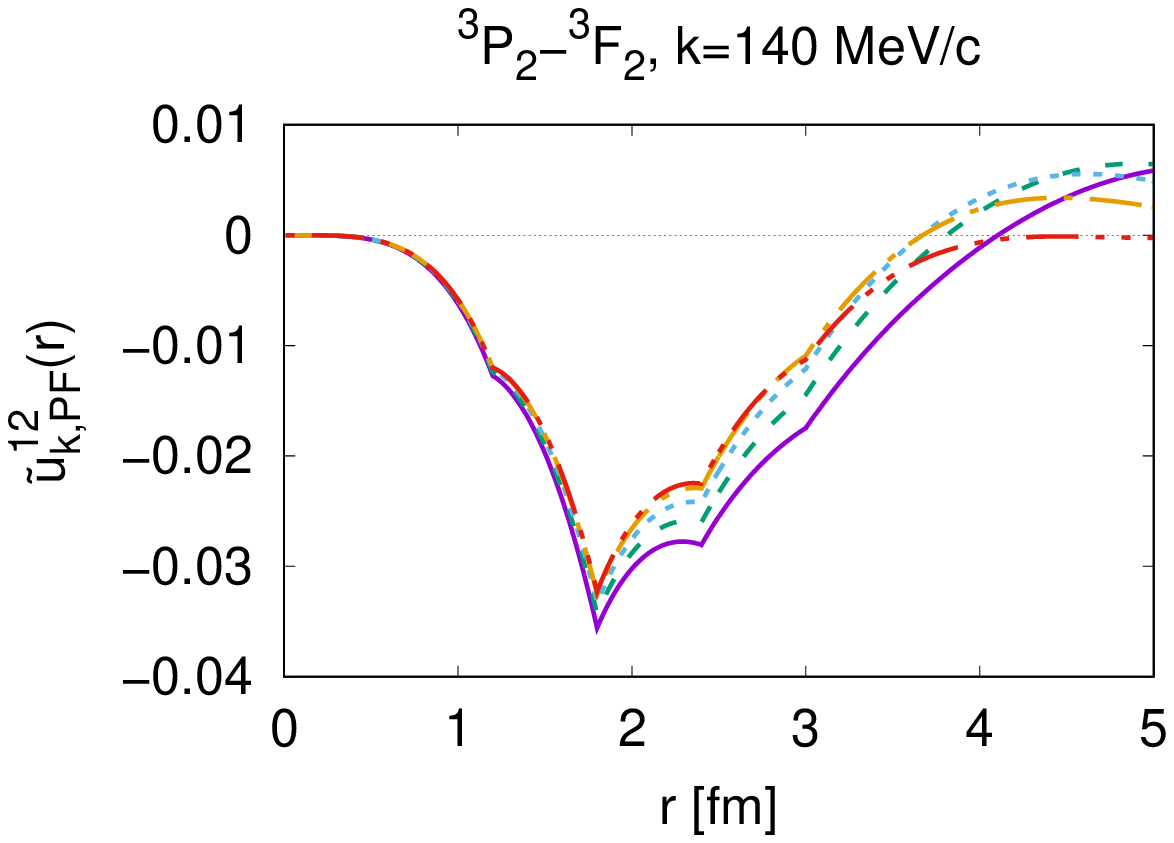}\\
\includegraphics[width=7.5cm]{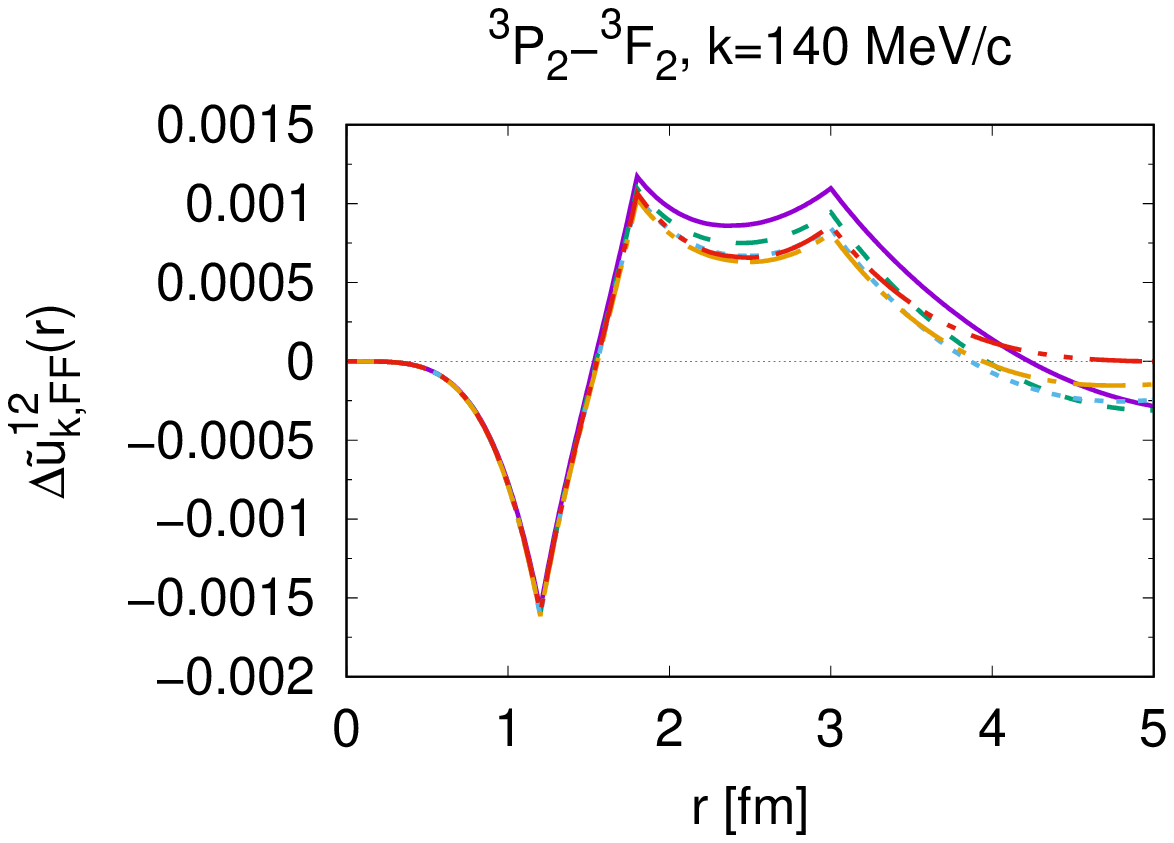}&
\includegraphics[width=7.5cm]{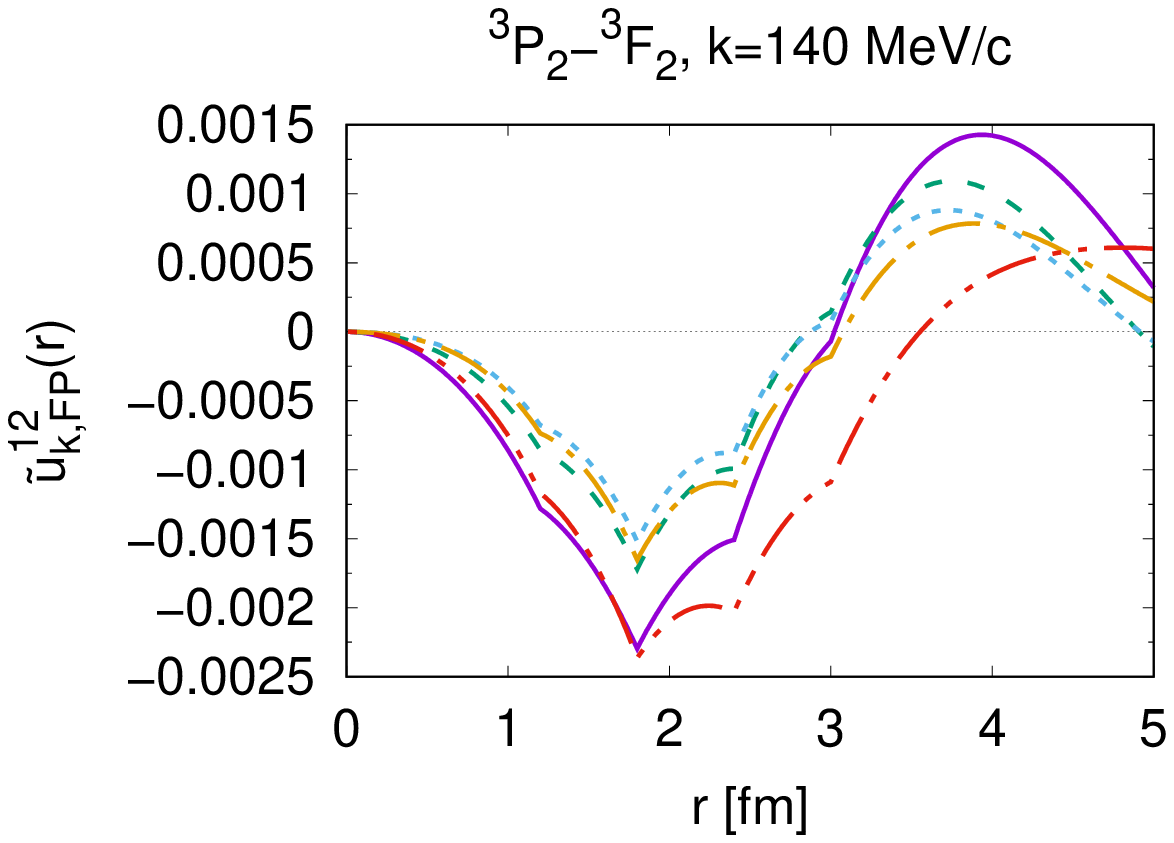}\\
\end{tabular}
\caption{Defect diagonal ($l=l^\prime$) wave functions
  $\Delta\widetilde{u}^{SJ}_{k,l\,l}(r)\equiv
  \widetilde{u}^{SJ}_{k,l\,l}(r)-\hat{j}_l(kr)$ (left panels), and
  off-diagonal wave functions $\widetilde{u}^{SJ}_{k,l\,l^\prime}(r)$
  (for $l\ne l^\prime$) (right panels) for the coupled N-N partial
  waves ${}^{3}$S$_1$-${}^{3}$D$_1$ and
  ${}^{3}$P$_2$-${}^{3}$F$_2$. The results are given for relative
  momentum $k=140$ MeV/c and for the same values of the CM momentum as
  in Fig.~\ref{Fig:uradial_coupled}.}
\label{Fig:defect_coupled}
\end{figure*}

In figure~\ref{Fig:uradial_coupled} we show the diagonal $l=l^\prime$
radial wave functions for the coupled N-N partial waves
${}^{3}$S$_1$-${}^{3}$D$_1$ and ${}^{3}$P$_2$-${}^{3}$F$_2$, for
different values of the CM momentum of the two-nucleon system and for
a relative momentum of $k=140$ MeV/c. For the coupled channels, the
notation SS refers to the partial wave with $l=l^\prime=0$, SD refers
to the partial wave with $l=0$ and $l^\prime=2$, and so on, following
the usual spectroscopic notation for the orbital angular momenta.  The
general trend with respect to the dependence of them on the CM
momentum is similar to the uncoupled partial waves, i.e, there is
little dependence on the value of the CM momentum. And the departure
from the free solution is more remarkable for the lower values of the
orbital angular momenta $l$.  The most striking dependence on the CM
momentum occurs for the SS wave at the cusp, but it is also similar to
the case of the uncoupled ${}^{1}$S$_0$ partial wave (cf. first panels
of figures~\ref{Fig:uradial} and \ref{Fig:uradial_coupled}).

In figure~\ref{Fig:defect_coupled} we show the defect diagonal
($l=l^\prime$) radial wave functions for the N-N coupled channels,
together with their coupled off-diagonal ($l\ne l^\prime$) partners on
the right panels. The most remarkable feature is that the size of the
distortion due to the short-range correlations is similar in the
partial waves which are coupled between themselves, i.e, those
corresponding to the left and right panels in each row of the
figure. Furthermore, the distortions are more sizable for the lower
$l$ partial waves, as already remarked in the discussion of
Fig.~\ref{Fig:uradial_coupled}; and, in general, there is little
dependence on the CM momentum of the nucleon pair, although this can
seem enhanced because of the scales shown in
Fig.~\ref{Fig:defect_coupled} with respect to those of
Fig.~\ref{Fig:uradial_coupled}.

\begin{figure*}[!ht]
\begin{tabular}{cc}
\includegraphics[width=7.5cm]{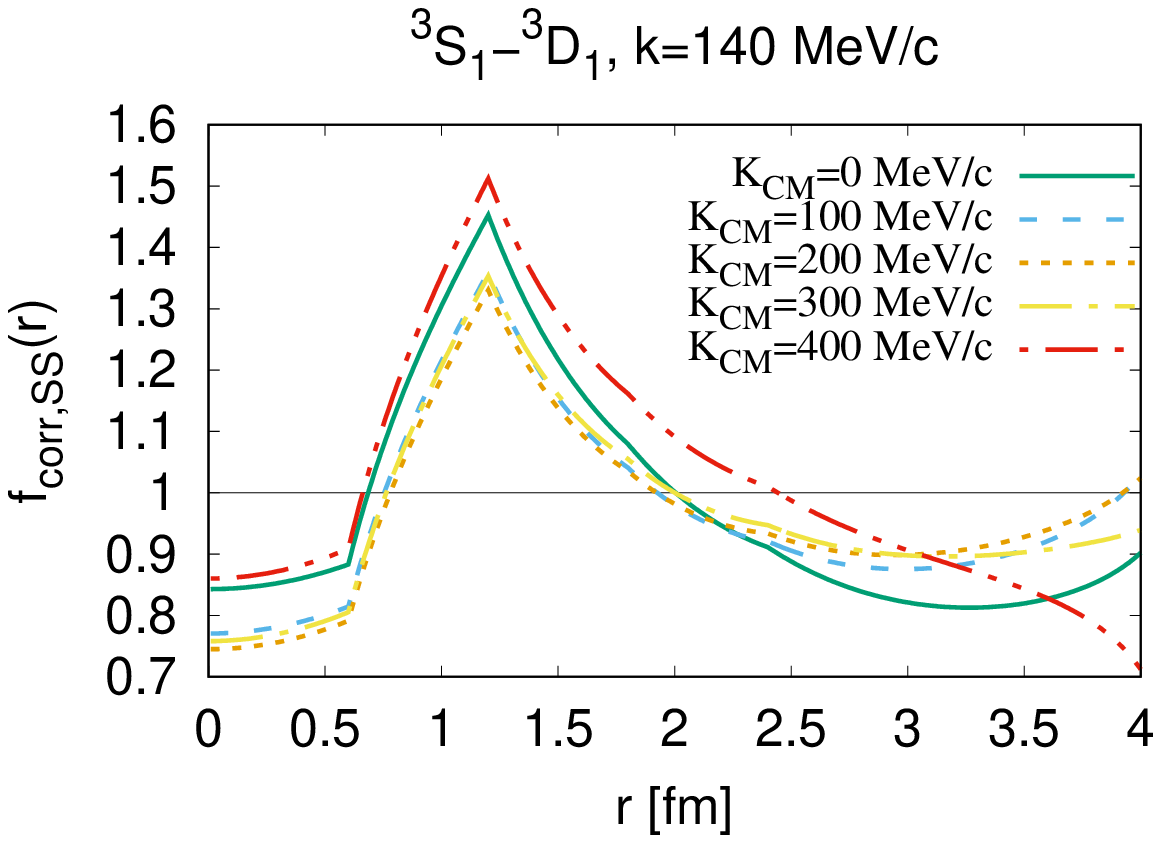}&
\includegraphics[width=7.5cm]{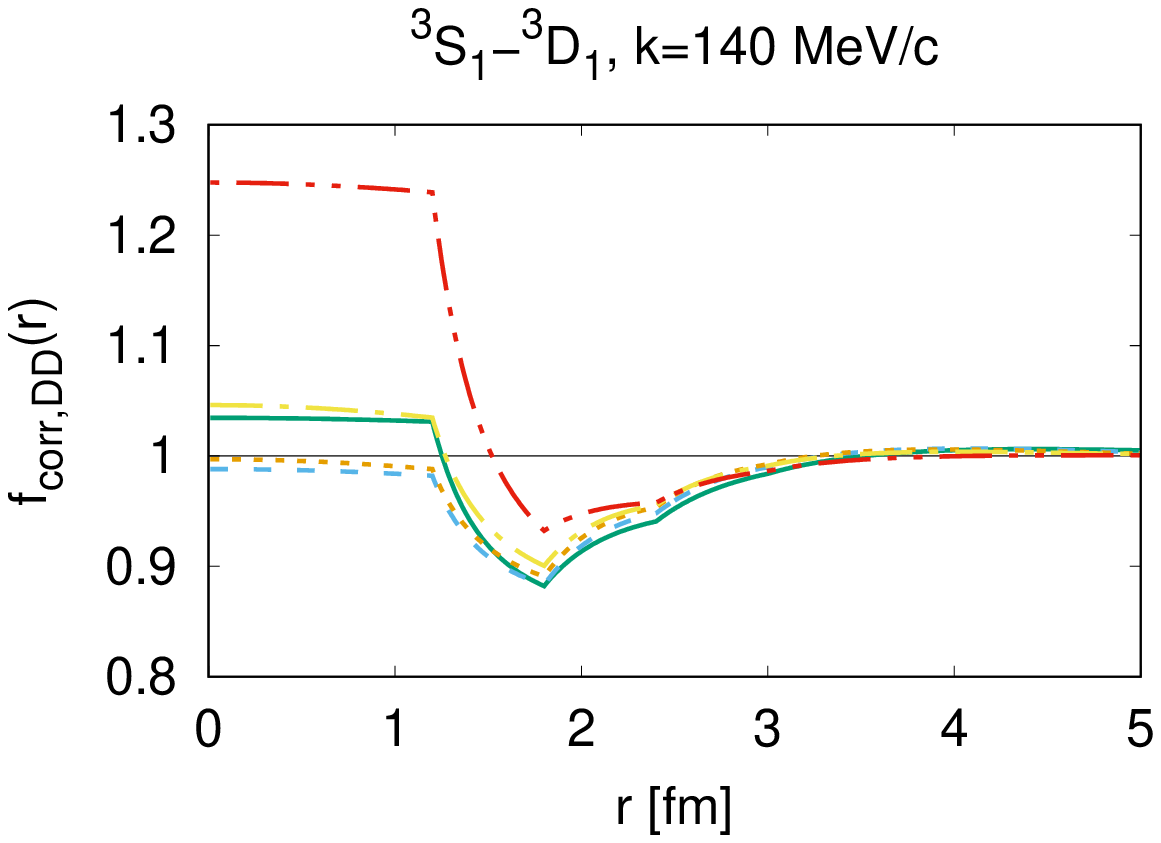}\\
\includegraphics[width=7.5cm]{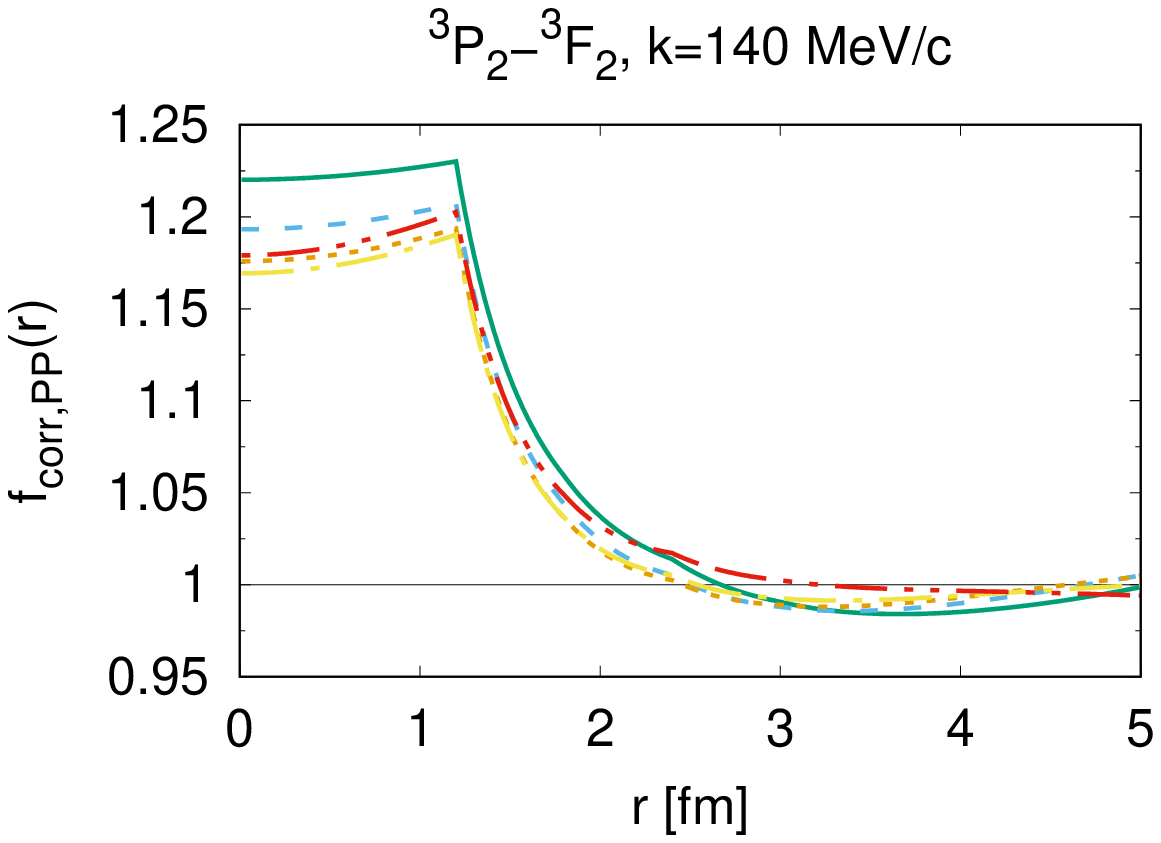}&
\includegraphics[width=7.5cm]{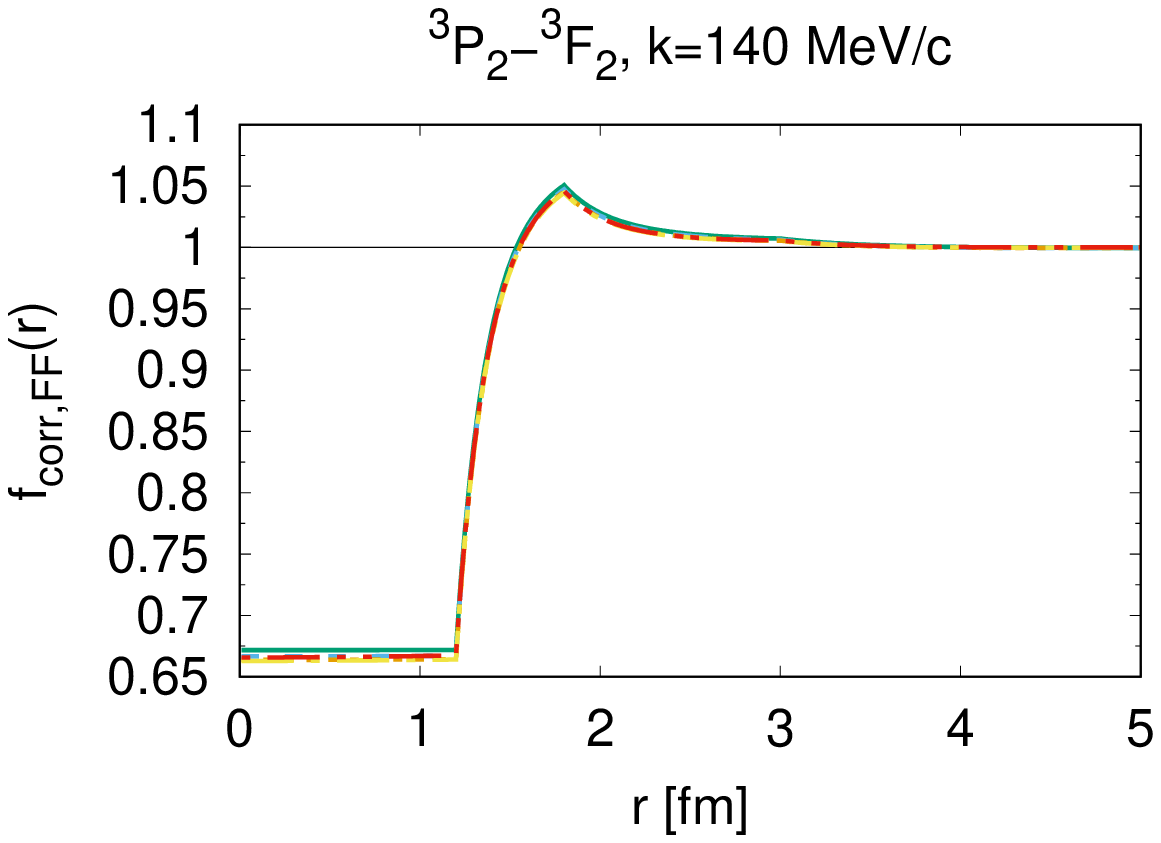}\\
\end{tabular}
\caption{Correlation functions $f_{\rm corr}(r)$ for the diagonal
  ($l=l^\prime$) wave functions, whose trend when $r\rightarrow\infty$
  should be $1$, for the coupled N-N partial waves
  ${}^{3}$S$_1$-${}^{3}$D$_1$ and ${}^{3}$P$_2$-${}^{3}$F$_2$. The
  results are given for relative momentum $k=140$ MeV/c and for the
  same values of the CM momentum as in Figs.~\ref{Fig:uradial_coupled}
  and \ref{Fig:defect_coupled}.}
\label{Fig:fcorrelation_coupled}
\end{figure*}

In Fig.~\ref{Fig:fcorrelation_coupled} we show the correlation
functions for the diagonal ($l=l^\prime$) coupled N-N partial waves.
Their behavior is, in general, similar to that of the uncoupled
partial waves, i.e, their departure from $1$ at short distances is
very similar in magnitude, and there is little dependence on the CM
momentum of the nucleon pair, except for the DD radial wave function
of the ${}^3$S$_1$-${}^3$D$_1$ coupled channel, where there is a more
pronounced dependence on the CM momentum for the highest one shown in
the upper right panel Fig.~\ref{Fig:fcorrelation_coupled}.
Nonetheless, similar behaviors can be also observed in the ${}^3$D$_2$
channel of Fig.~\ref{Fig:fcorrelation} or even, to a lesser extent, in
the PP component of the ${}^3$P$_2$-${}^3$F$_2$ channel, shown in the
bottom left panel of Fig.~\ref{Fig:fcorrelation_coupled}.

\begin{figure*}[!ht]
\begin{tabular}{cc}
\includegraphics[width=7.5cm]{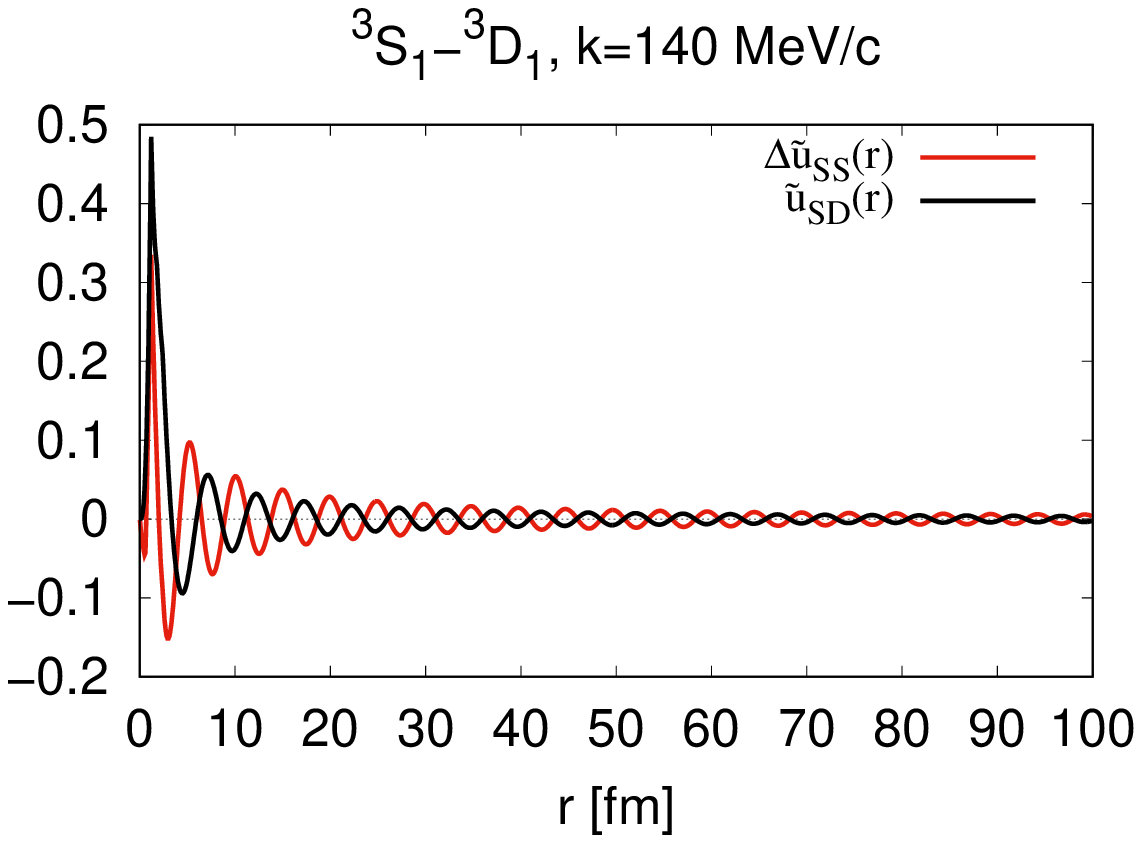}&
\includegraphics[width=7.5cm]{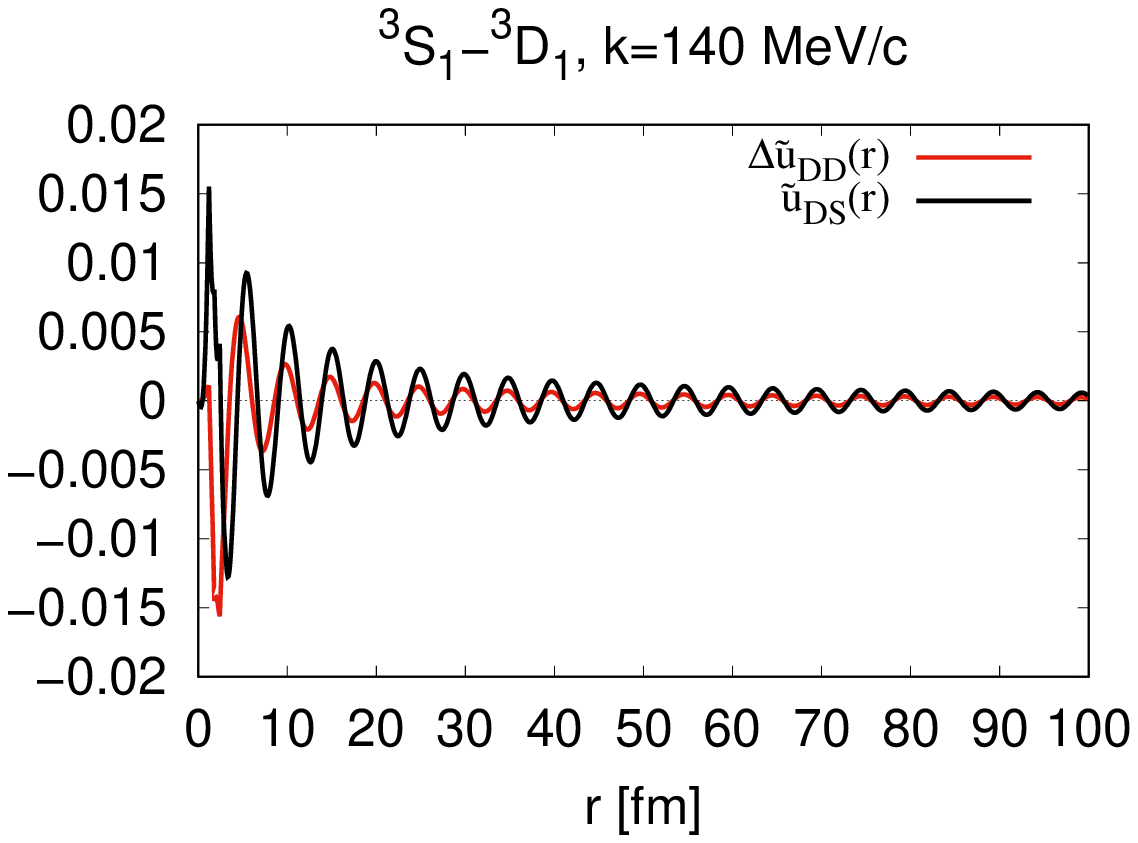}\\
\includegraphics[width=7.5cm]{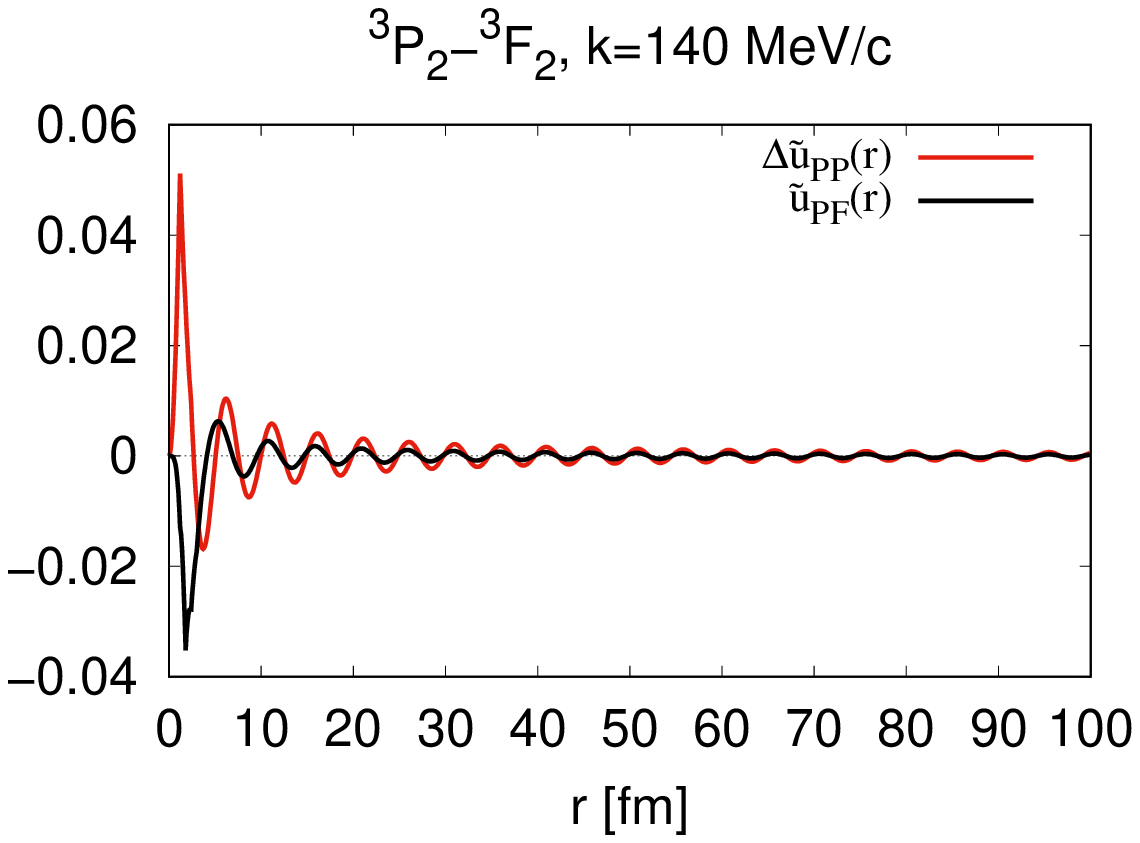}&
\includegraphics[width=7.5cm]{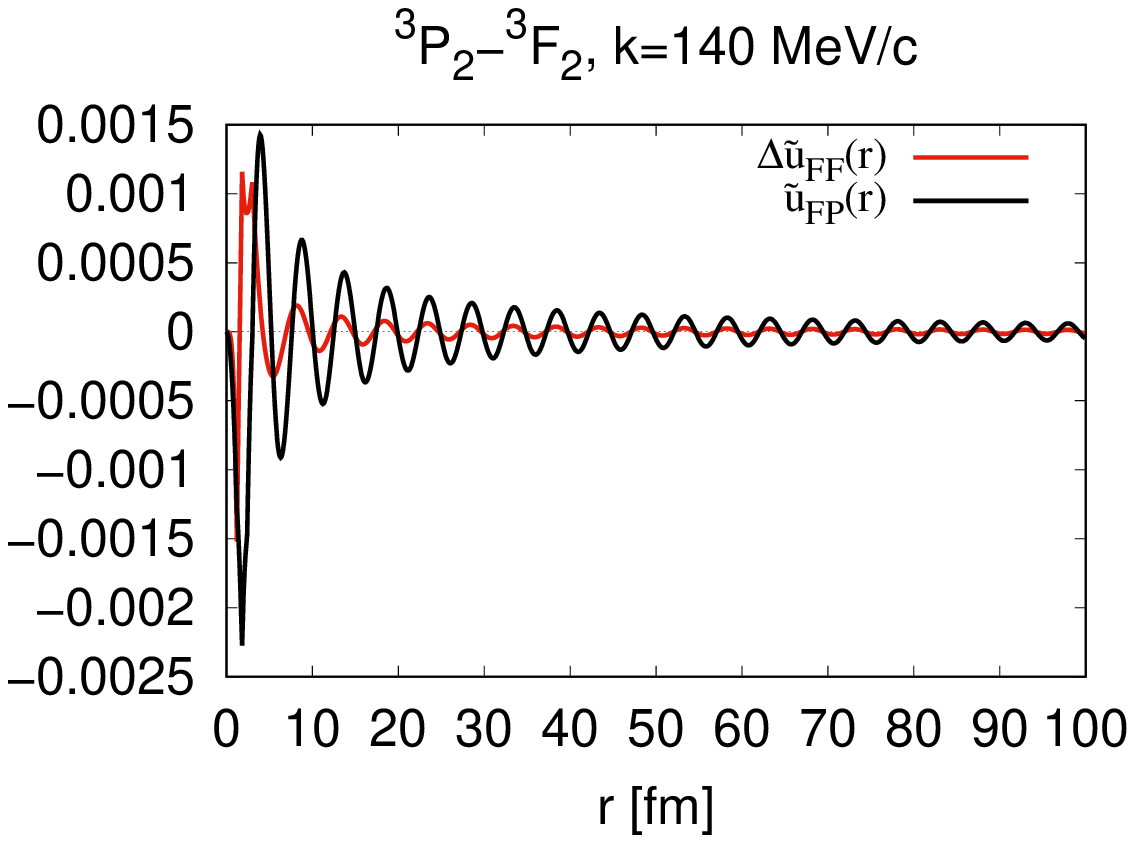}\\
\end{tabular}
\caption{Long range ($r\rightarrow\infty$) behavior of the defect
  diagonal wave functions and the off-diagonal ones for the N-N
  coupled channels, and for $K_{\rm CM}=0$.  All of them go to zero,
  but slowly converging on the scale of each figure.  This means that
  there is no phase-shift in the B-G wave functions
  $\widetilde{u}_{l\,l}(r)$ with respect to the free solution when
  $r\rightarrow\infty$.}
\label{Fig:long_range_behaviour}
\end{figure*}

In Fig.~\ref{Fig:long_range_behaviour} we observe the long-range
behavior of the defect coupled radial wave functions for $K_{\rm
  CM}=0$.  The left panels compare the $(J-1,J-1)$ waves to the
$(J-1,J+1)$ ones, while in the right panels the $(J+1,J+1)$ waves are
compared to the $(J+1,J-1)$ ones. In each panel, both functions
approach zero in an oscillatory manner, as expected. However, this
decrease occurs very slowly in the scale of each plot, indicating a
gradual decrease in amplitude as the distance increases. It is worth
noticing that the coupled waves shown in each panel have the same
order of magnitude, in concordance with the findings discussed in
Fig.~\ref{Fig:defect_coupled}.  This order of magnitude, which is a
measure of their deviation with respect to their free asymptotic
behavior, and therefore a measure of the importance of the short-range
correlations in each channel, is higher the lower the orbital angular
momentum $l$, in perfect accordance with the findings of the
discussion of Fig.~\ref{Fig:defect_coupled} as well.

\subsection{``Radial" wave functions in momentum 
representation: high-momentum components}
\label{subsec:radial_wf_momentum}

\begin{figure*}[!ht]
\begin{tabular}{cc}
\includegraphics[width=7.5cm]{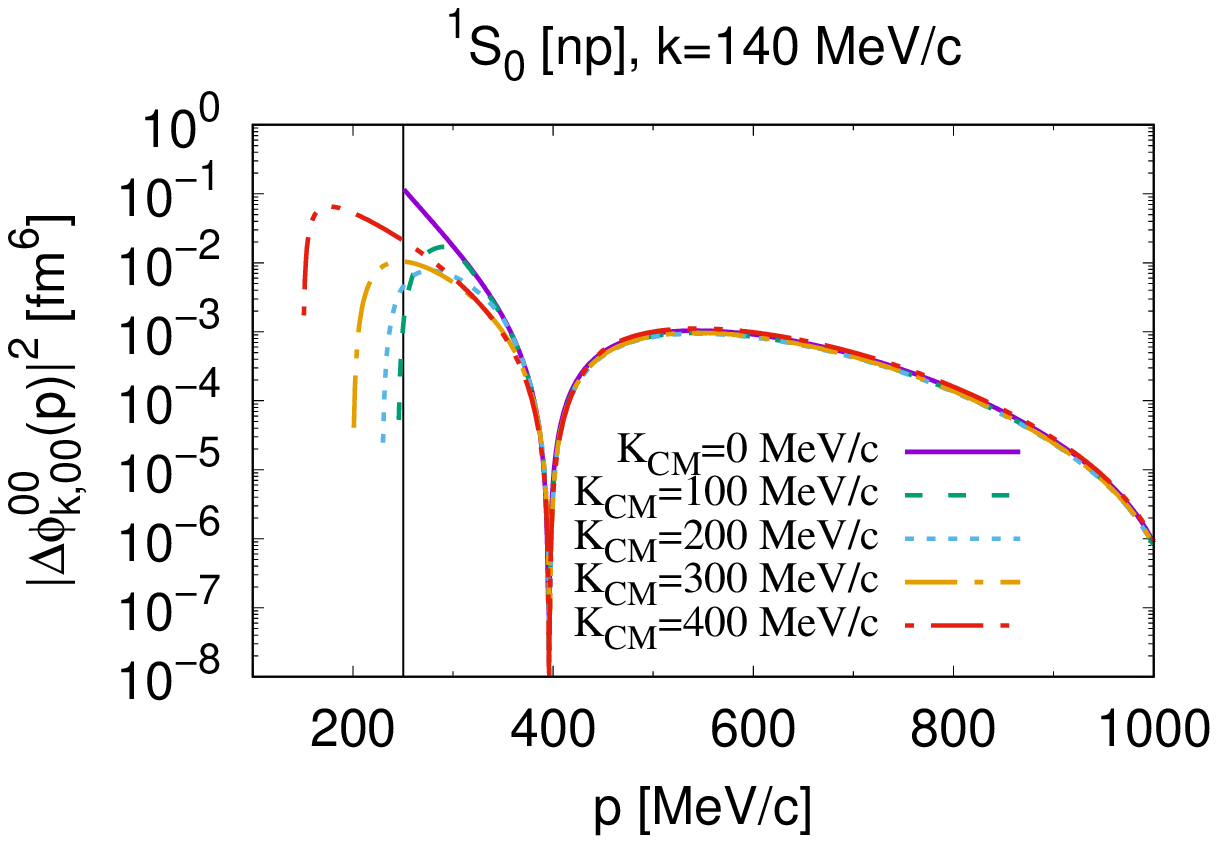}&
\includegraphics[width=7.5cm]{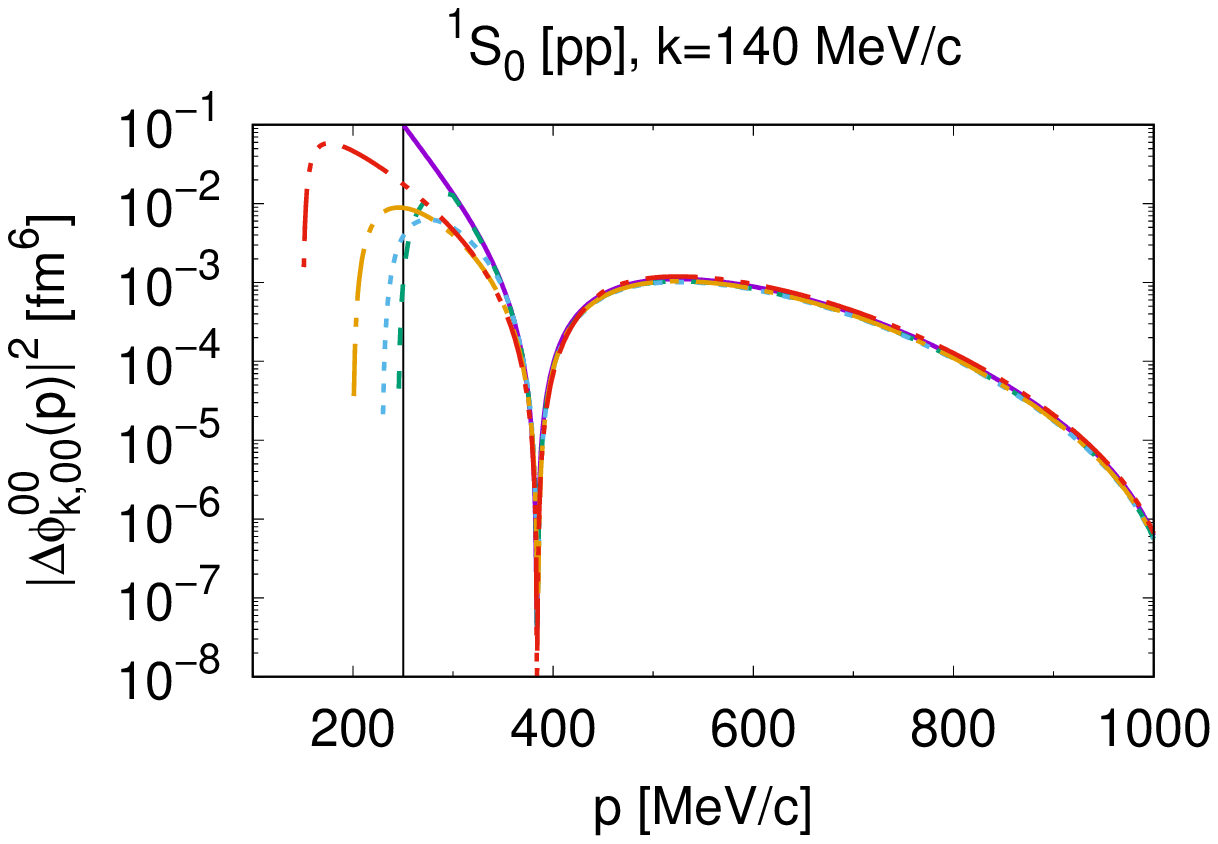}\\
\includegraphics[width=7.5cm]{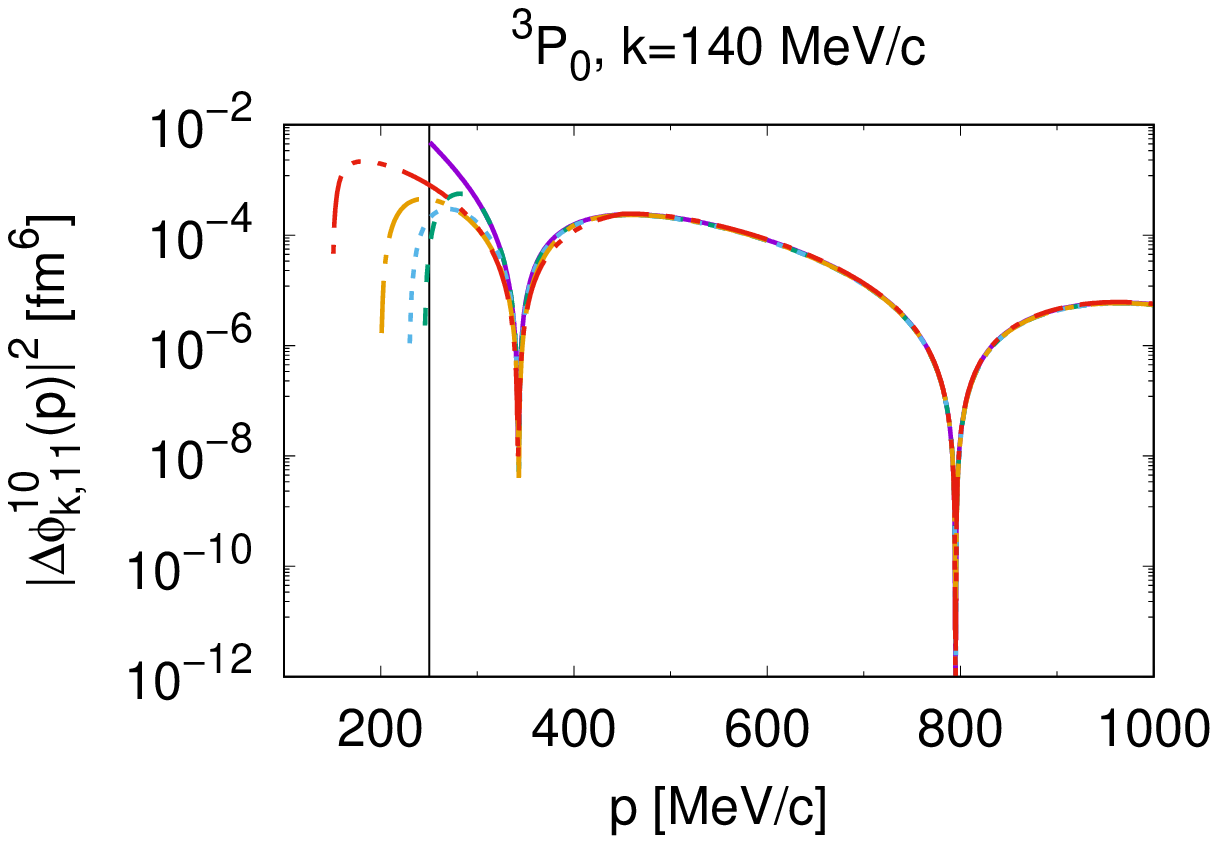}&
\includegraphics[width=7.5cm]{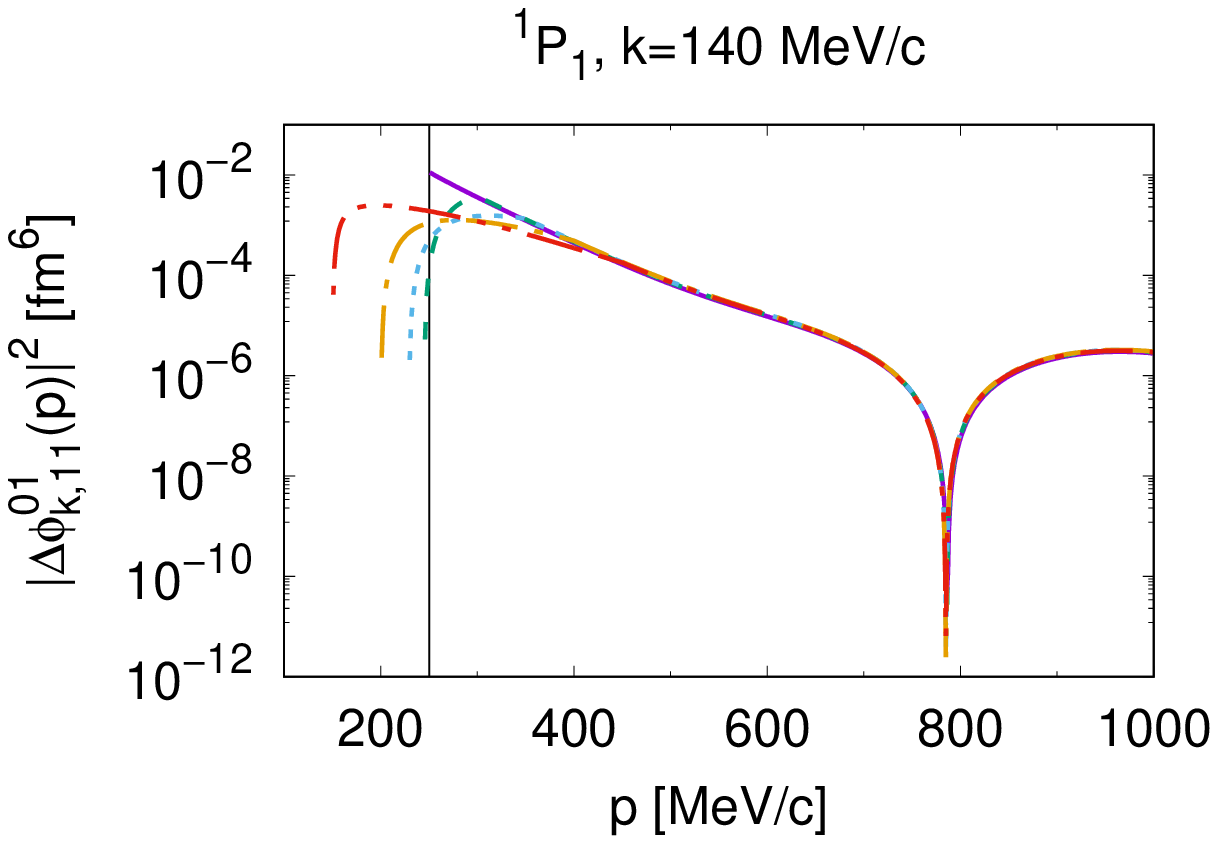}\\
\includegraphics[width=7.5cm]{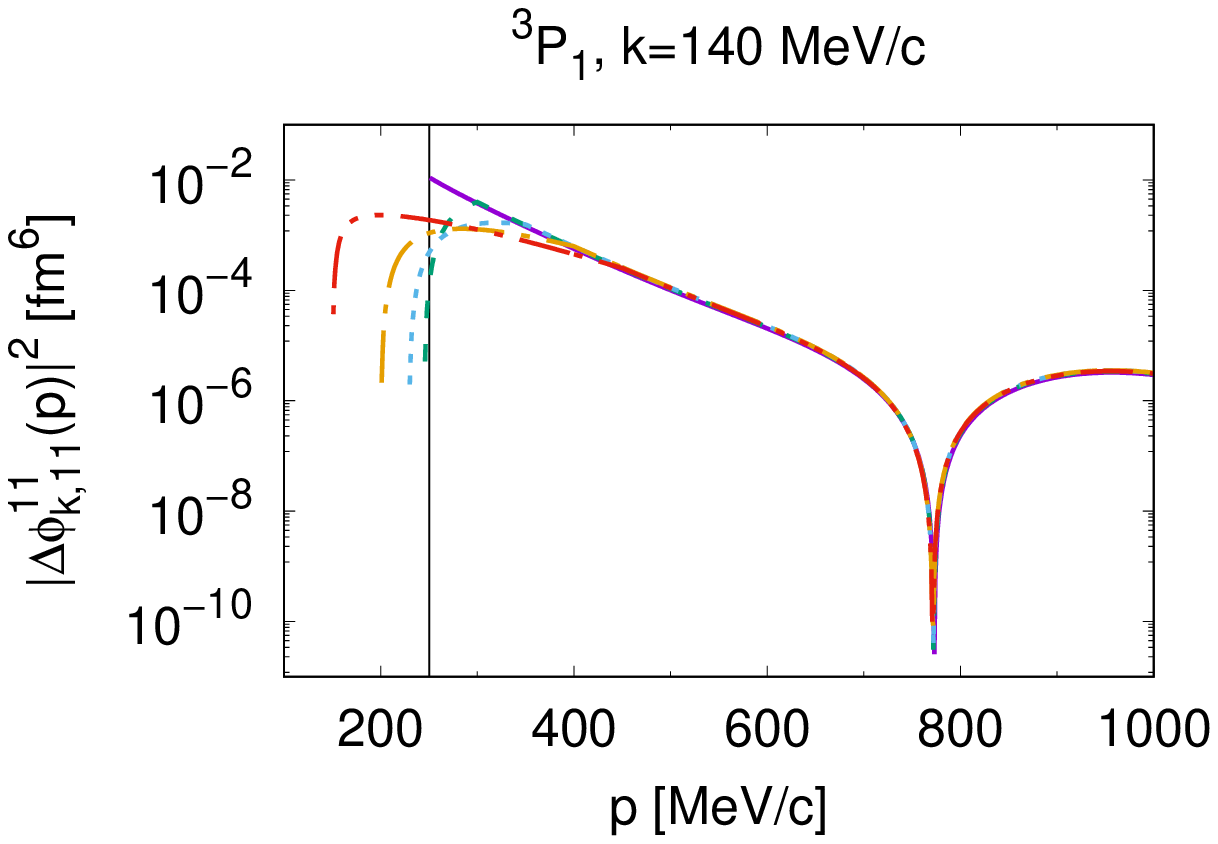}&
\includegraphics[width=7.5cm]{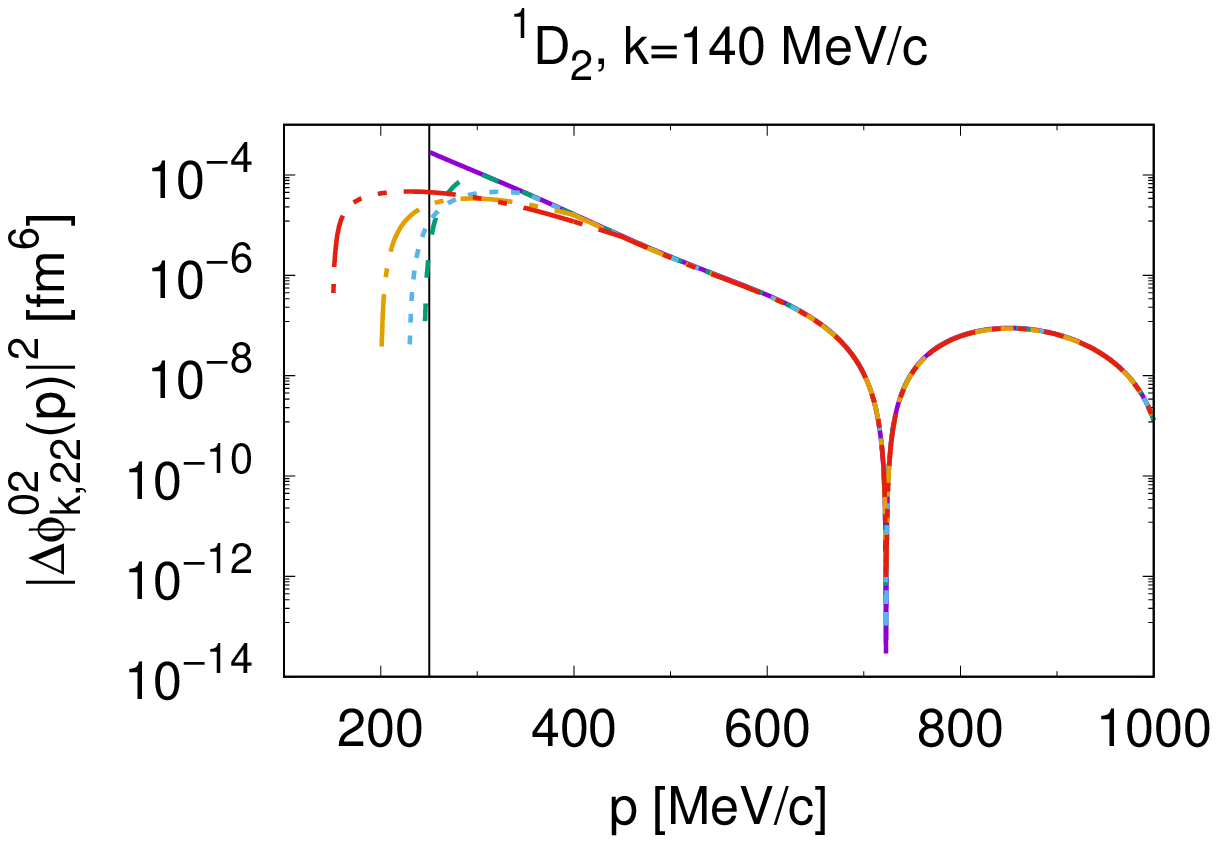}\\
\includegraphics[width=7.5cm]{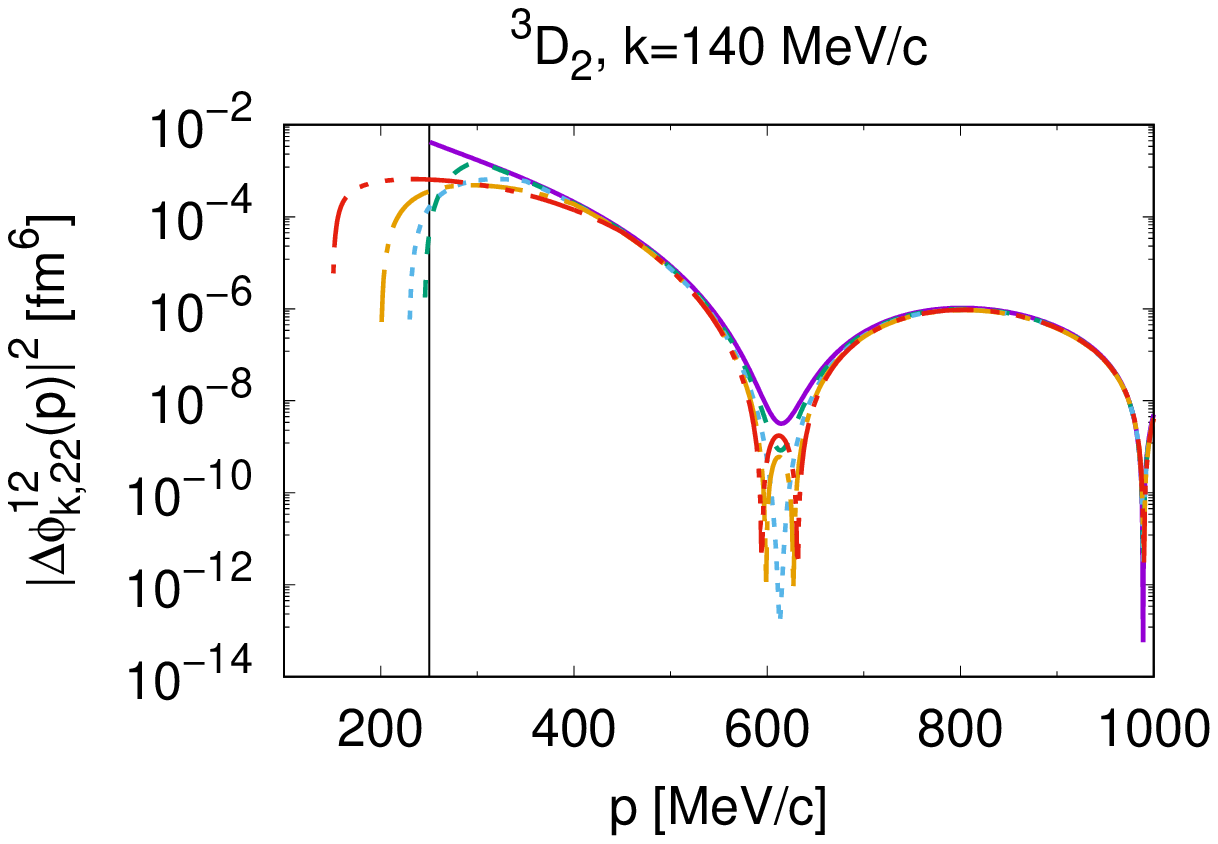}&
\includegraphics[width=7.5cm]{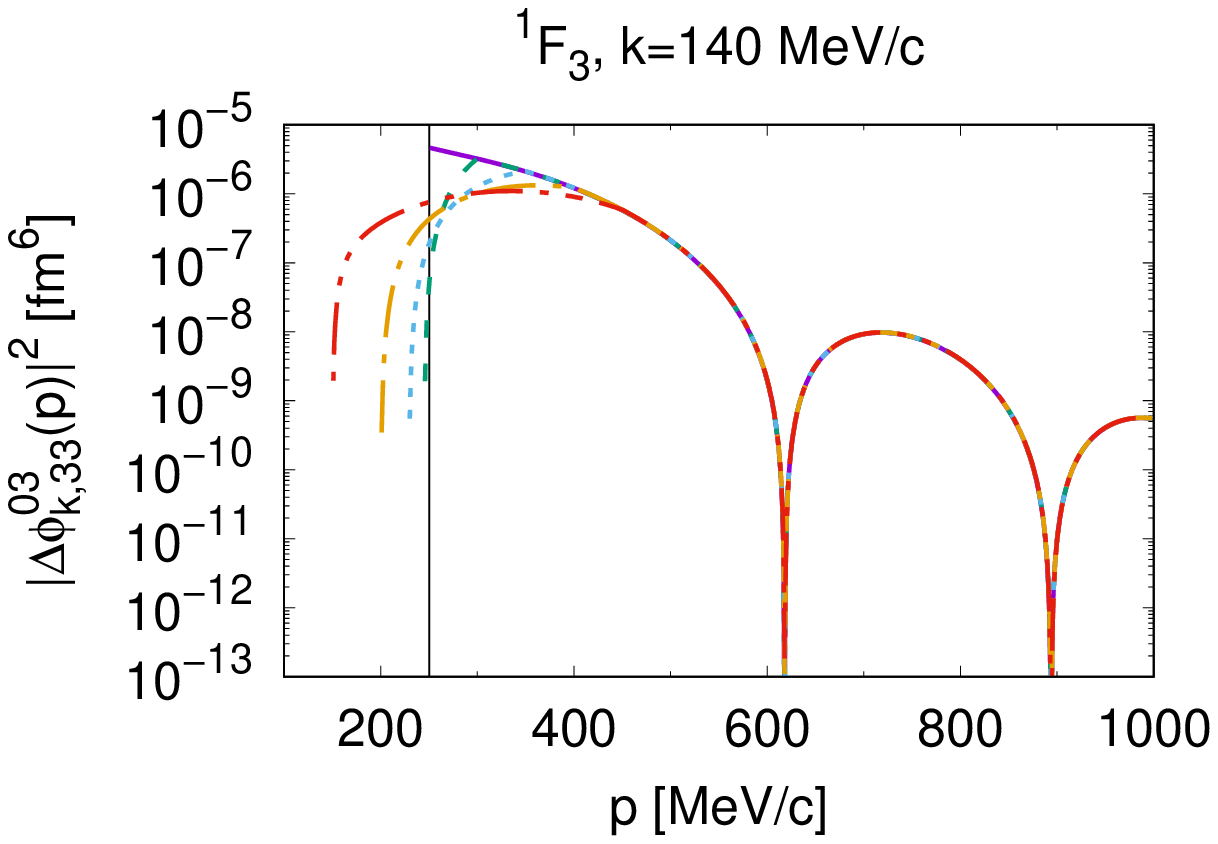}
\end{tabular}
\caption{``Radial" wave functions $\left|\Delta\phi^{SJ}_{k,
    l^\prime\,l}(p) \right|^2$ for the uncoupled N-N partial waves,
  i.e, for $l=l^\prime$.  The results are given for relative momentum
  $k=140$ MeV/c, and for different values of the CM momentum as
  labeled in the key of the first panel.  The results for $K_{\rm
    CM}=0$ MeV/c (solid purple lines) are the same as those shown in
  the upper panel of Fig. 6 of Ref.~\cite{RuizSimo:2017tcb}.}
\label{Fig:phiradial}
\end{figure*}

\begin{figure*}[!ht]
\begin{tabular}{cc}
\includegraphics[width=7.5cm]{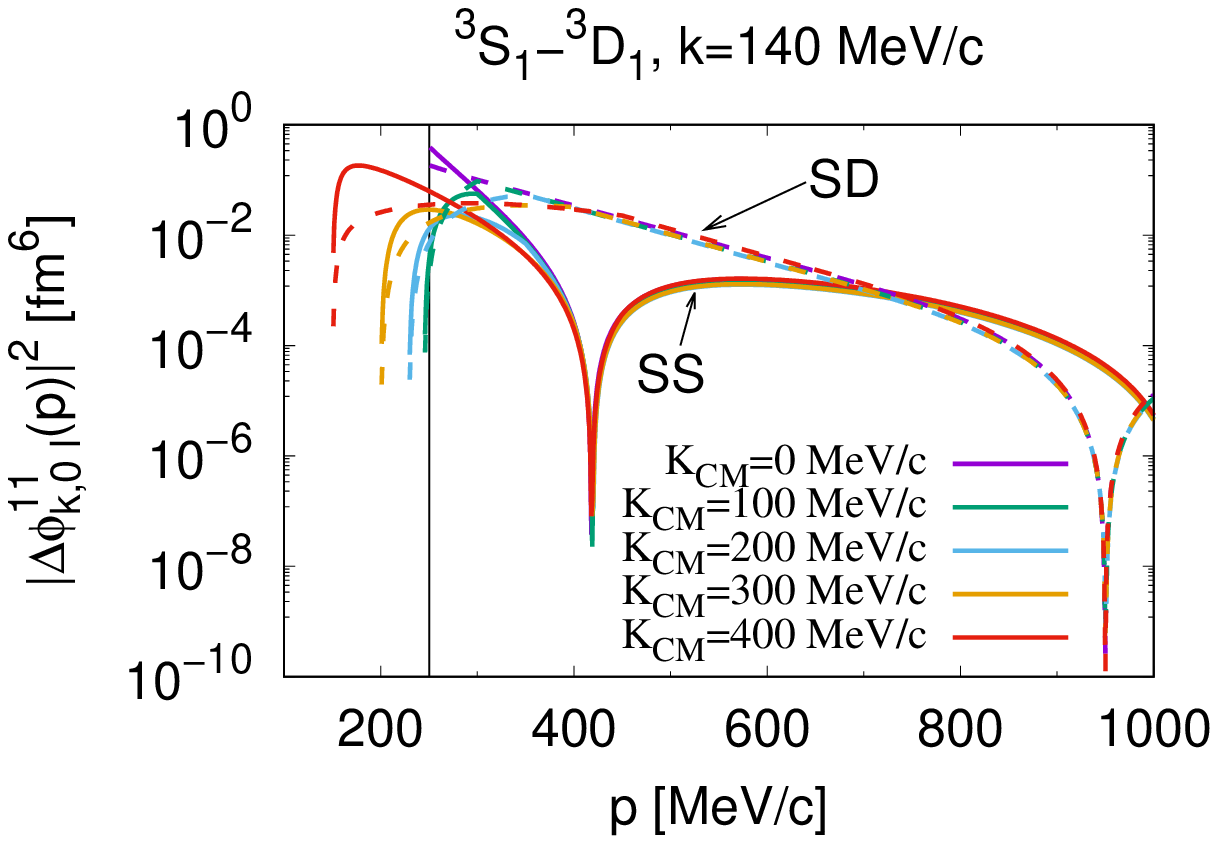}&
\includegraphics[width=7.5cm]{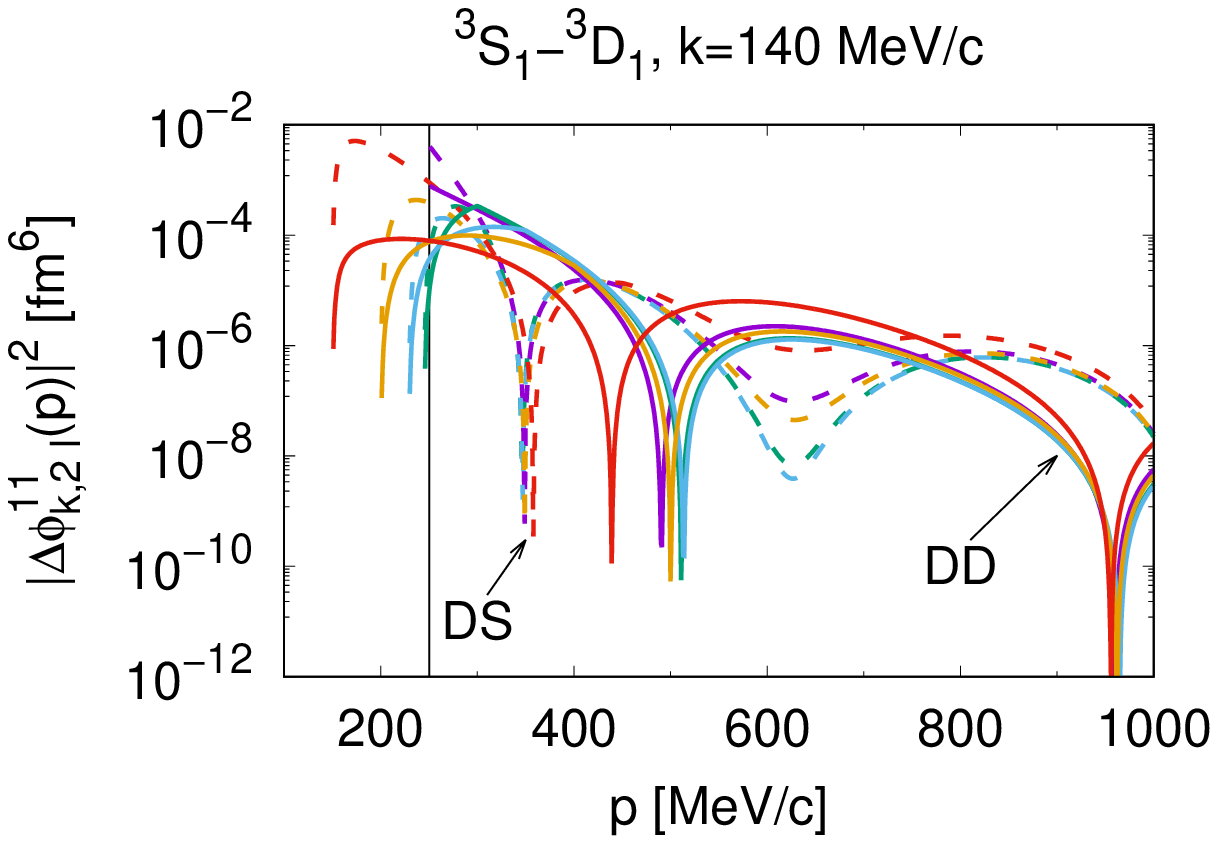}\\
\includegraphics[width=7.5cm]{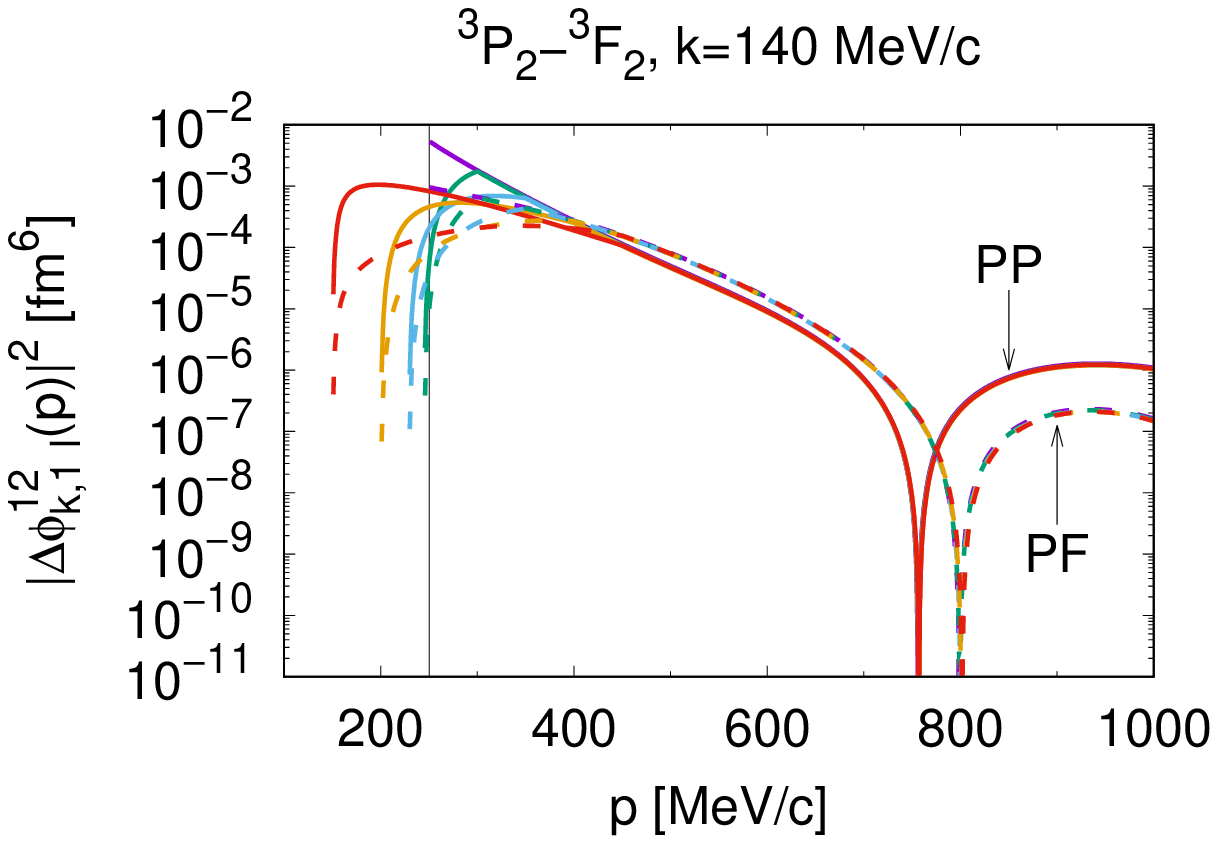}&
\includegraphics[width=7.5cm]{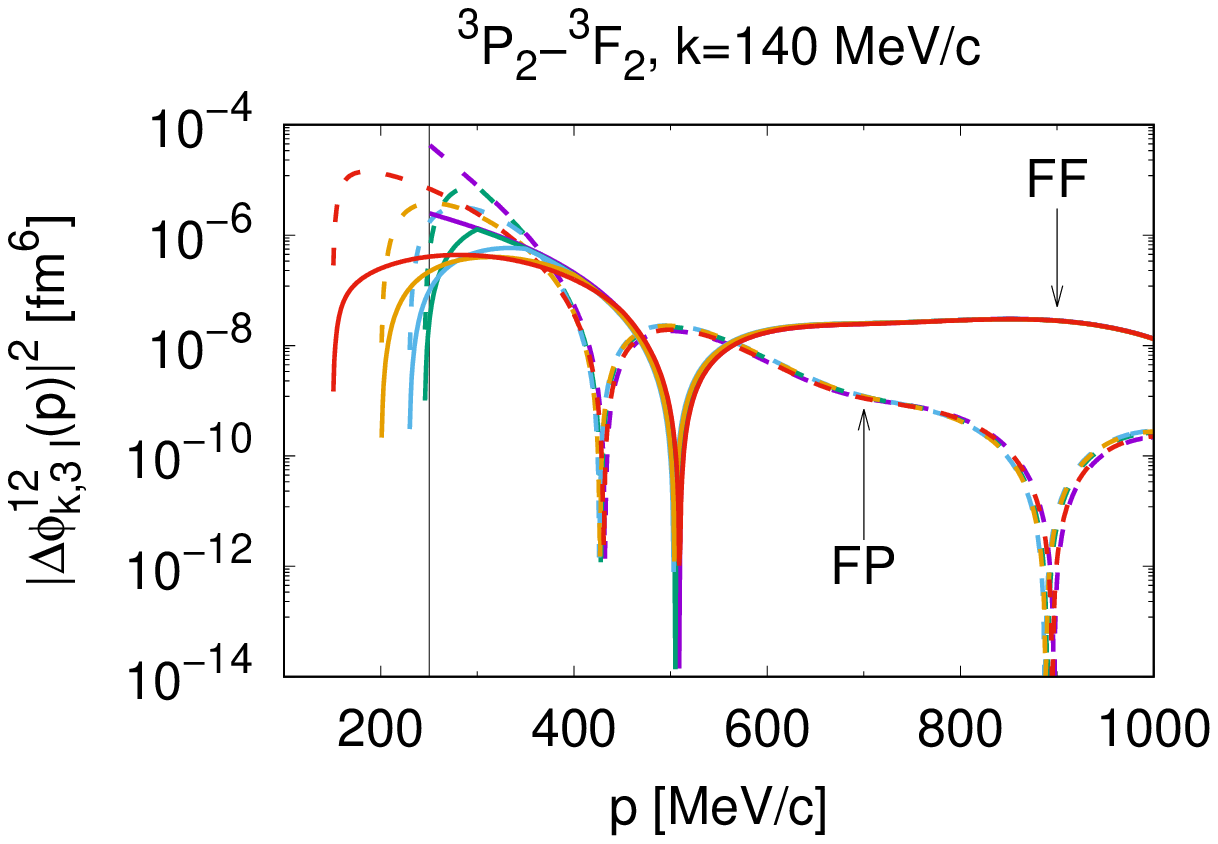}
\end{tabular}
\caption{``Radial" wave functions $\left| \Delta\phi^{SJ}_{k,
    l^\prime\,l}(p) \right|^2$ for the coupled N-N partial waves. The
  results are given for relative momentum $k=140$ MeV/c, and for
  different values of the CM momentum as labeled in different colors
  in the key of the first panel. The solid lines correspond to the
  diagonal $l^\prime=l$ waves, while the dashed lines are the results
  corresponding to the off-diagonal ($l^\prime\ne l$) coupled waves.}
\label{Fig:phiradial_coupled}
\end{figure*}

In figures~\ref{Fig:phiradial} and \ref{Fig:phiradial_coupled} we show
the square of the high-momentum component of the ``radial" wave
function $\left| \Delta\phi^{SJ}_{k, l^\prime\,l}(p) \right|^2$ in
momentum space (see eq.~\ref{high_mom_components}), for different CM
momenta of the nucleon pair, of the uncoupled and coupled N-N partial
waves, respectively. The vertical black lines in the different panels
at $p=250$ MeV/c mark the position of the Fermi momentum. These show
the high-momentum components in the relative wave function due to the
short-range correlations.

As it has been already discussed in Fig.~\ref{Fig:Qbarra}, the
angle-averaged Pauli-blocking function appearing in
eq.~(\ref{high_mom_components}) is always a piece-wise function
bounded between $0$ and $1$, depending on the values of the CM
momentum $K_{\rm CM}$ and the probed relative one $p$ of the nucleon
pair. Therefore, if one wants to look at the effects produced in the
high-momentum components by the influence of the CM momentum on the
radial wave functions at the grid points, one should look at regions
of $p$ in Figs.~\ref{Fig:phiradial} and \ref{Fig:phiradial_coupled}
where the angle-averaged Pauli blocking operator does not depend at
all on $K_{\rm CM}$.  These regions of $p$ where $\overline{Q}(K_{\rm
  CM},p)=1$ correspond to zone (c) in Fig.~\ref{Fig:zones}, i.e, when
$p > k_F + \frac{K_{\rm CM}}{2}$. In the most unfavorable situation,
we should look at $p > 2 k_F = 500$ MeV/c in Figs.~\ref{Fig:phiradial}
and \ref{Fig:phiradial_coupled}.

In this region of values of $p$, the only observable differences
between the curves for distinct CM momenta can only come from the
differences in the radial wave functions at the grid points $r_i$,
shown in Figs.~\ref{Fig:uradial}, \ref{Fig:uradial_coupled} and the
right panels of Fig.~\ref{Fig:defect_coupled}.  However, if we look in
Figs.~\ref{Fig:phiradial} and \ref{Fig:phiradial_coupled} at $p
\gtrsim 500$ MeV/c, we do not observe almost any difference between
the curves for distinct CM momenta, except for the particular cases of
the ${}^{3}$D$_2$ and the DS-DD coupled components of the
${}^3$S$_1$-${}^3$D$_1$ partial waves, which will be explained later.
Therefore, the conclusion is that the differences due to the CM
momentum dependence on the radial wave functions (even the largest
ones for the low $l$ partial waves) observed in
Figs.~\ref{Fig:uradial}, \ref{Fig:defect}, \ref{Fig:uradial_coupled}
and \ref{Fig:defect_coupled} are mostly completely irrelevant for the
tail of high-momentum components ($p \gtrsim 2 k_F$) in the relative
wave function of the nucleon pair. This points out the universality of
SRCs or, \emph{at least, that the global motion state of the nucleon
pair has negligible influence in the tail of high momentum
components.}

However, if we observe Figs.~\ref{Fig:phiradial} and
\ref{Fig:phiradial_coupled} for $p\lesssim 2k_F = 500$ MeV/c, the
angle-averaged Pauli-blocking function, $\overline{Q}(K_{\rm CM},p)$
of eq.~(\ref{high_mom_components}), starts to play a significant
role. For a fixed value of the CM momentum of the pair, when
diminishing the probed relative momentum $p$, we are entering into
region (b) of Fig.~\ref{Fig:zones} from region (c) of the same figure.
And in the region (b) of Fig.~\ref{Fig:zones}, the value of the
$\overline{Q}(K_{\rm CM},p)$ function starts to get reduced from $1$
at the right line $p=k_F + \frac{K_{\rm CM}}{2}$ to $0$ at the ellipse
$p=\sqrt{k^2_F - \frac{K^2_{\rm CM}}{4}}$.

This reduction in the value of the angle-averaged Pauli-blocking
function gets reflected in the departures from the purple lines of
almost all the curves for $K_{\rm CM} > 0$ MeV/c in
Figs.~\ref{Fig:phiradial} and \ref{Fig:phiradial_coupled} at different
values of $p$.  The smaller the value of $K_{\rm CM}$, the smaller the
value of $p$ is at the point where the deviation from the purple
curves occurs.  This fact can be easily understood looking again at
Fig.~\ref{Fig:zones}.  Indeed, if we plot imaginary vertical lines in
Fig.~\ref{Fig:zones} at the CM momenta depicted in
Figs.~\ref{Fig:phiradial} and \ref{Fig:phiradial_coupled}, we observe
that region (b) along these imaginary vertical lines starts to become
larger when the CM momentum increases. This is so because the right
line is growing and the ellipse is diminishing.  This causes the point
of deviation from the purple curves (corresponding to $K_{\rm CM}=0$)
in Figs.~\ref{Fig:phiradial} and \ref{Fig:phiradial_coupled} to be
larger in the $p$ variable when the CM momentum is also larger.  In
fact, the exact point of deviation from the purple curves occurs at
$p_{\rm dev}(K_{\rm CM})=k_F + \frac{K_{\rm CM}}{2}$, which of course
depends on the value of the CM momentum.

Other interesting features that can be observed in
Figs.~\ref{Fig:phiradial} and \ref{Fig:phiradial_coupled} is that for
$K_{\rm CM} > 0$ MeV/c, the high-momentum distributions intrude below
the Fermi momentum marked by the vertical lines on the same
figures. This, again, can be easily understood by looking at
Fig.~\ref{Fig:zones}: for $K_{\rm CM} > 0$ MeV/c, the points $p$ below
which the high-momentum distributions are zero correspond to the
ellipse points, and these are always below the Fermi momentum when
$K_{\rm CM} > 0$.  The particular case when $K_{\rm CM} = 0$
corresponds to the purple curves in Figs.~\ref{Fig:phiradial} and
\ref{Fig:phiradial_coupled}, and for this CM momentum the
high-momentum distribution is zero exactly at $p=k_F$. However, there
are never high-momentum distributions for $p < k$. The only low
momentum component is the unperturbed component for $p=k$, represented
by the Dirac delta function in eq.~(\ref{analytical_radia_wf_mom}).

Another interesting point that is worth being remarked is that the
high-momentum distributions $\left| \Delta\phi^{SJ}_{k, l^\prime\,
  l}(p) \right|^2$ are continuous at the deviation points from the
purple lines, $p_{\rm dev}(K_{\rm CM})$, but not their derivatives
with respect to $p$ at these points. This fact is completely related
to the discontinuity in the derivative of the angle-averaged
Pauli-blocking function (see Fig.~\ref{Fig:Qbarra}) at the joining
point between regions (b) and (c) of Fig.~\ref{Fig:zones} along a
vertical line for a constant value of $K_{\rm CM}$. This feature was
already discussed in point 5 at the end of Sect.~\ref{ang_average}.
In particular, this effect is very clearly observable for the curves
of Figs.~\ref{Fig:phiradial} and \ref{Fig:phiradial_coupled}
corresponding to $K_{\rm CM}=100$ MeV/c, whose angle-averaged
Pauli-blocking function has a behavior very similar to that of the
second panel ($K_{\rm CM}=0.5\, k_F$) in Fig.~\ref{Fig:Qbarra}.
Indeed, in these cases of low CM momenta, the discontinuities in the
derivative of the $\overline{Q}(K_{\rm CM},p)$ function at $p_{\rm
  dev}=k_F + \frac{K_{\rm CM}}{2}$ are much more pronounced than for
larger CM momenta, as it can be observed on the different panels of
Fig.~\ref{Fig:Qbarra} (notice that for the first panels the slope of
the transition curve between $0$ and $1$ is much steeper than for the
last panels).

The case of the ${}^{3}$D$_2$ partial wave high-momentum distribution
shown in Fig.~\ref{Fig:phiradial} deserves a separate explanation for
its behavior at $p\sim 600$ MeV/c.  This is the region of $p$ values
where the angle-averaged Pauli-blocking function is equal to $1$, and
therefore, any difference between the curves for distinct CM momenta
can be solely ascribed to differences in the perturbed radial wave
functions at the grid points $(\widetilde{u}^{12}_{k,2}(r_i))$ for the
different CM momenta of the nucleon pair, as appearing in
eq.~(\ref{high_mom_components}).  The explanation is as follows: when
$K_{\rm CM}=0$ MeV/c, the corresponding curve (purple line) in
Fig.~\ref{Fig:phiradial} does not have a node at $p\sim 600$ MeV/c,
but a local minimum very close to $0$; however, the differences in the
perturbed radial wave functions at the grid points when varying the CM
momentum make this local minimum to become also a node for $K_{\rm
  CM}\approx 200$ MeV/c (short-dashed blue line).  Finally, if one
increases the value of the CM momentum above $200$ MeV/c, the minimum
of the function $\Delta\phi^{12}_{k,22}(p)$ at $p\sim 600$ MeV/c
starts to have negative values and cuts the $p$-axis at two nodes very
close to $p\sim 600$ MeV/c, but each one of them at one side of the
negative minimum. This gives the particular pattern shown in the
${}^{3}$D$_2$ panel of Fig.~\ref{Fig:phiradial} for the square of the
function $\Delta\phi^{12}_{k,22}(p)$ around $p\sim 600$ MeV/c,
presenting two very close nodes for $K_{\rm CM}=300$ and $400$ MeV/c.
 
 The differences observed in the DS-DD coupled waves of the
 ${}^3$S$_1$-${}^3$D$_1$ (top right panel of
 Fig.~\ref{Fig:phiradial_coupled}) channel, particularly between the
 case of $K_{\rm CM}=400$ MeV/c and the other CM momenta, can be
 explained by looking at the defect radial wave functions for that
 channel in Fig.~\ref{Fig:defect_coupled} (second line plots of that
 figure). In this case, on the contrary to the others of the same
 figure, one can observe a relevant difference at short distances
 between the curve for $K_{\rm CM}=400$ MeV/c and for the other CM
 momenta.  This is particularly evident for the DS relative wave
 function. Notice that, despite being basically D-waves, the magnitude
 of the distortion at short distances is quite similar to that of the
 PP-PF ($l=1$) coupled waves. However, for this latter case, all the
 distortions depend little on the CM momentum, while in the DS-DD case
 there is a significant difference between the case with $K_{\rm
   CM}=400$ MeV/c and the other CM momenta configurations, whose
 curves show a softer dependence on the total momentum.  This makes
 the high-momentum components for this DS-DD coupled channel to depend
 substantially more on the CM momentum, especially for the DS
 component at $p\simeq 625$ MeV/c (note that the vertical scales in
 Fig.~\ref{Fig:phiradial_coupled} are logarithmic).
 
\begin{figure*}[!ht]
\begin{tabular}{cc}
\includegraphics[width=9cm]{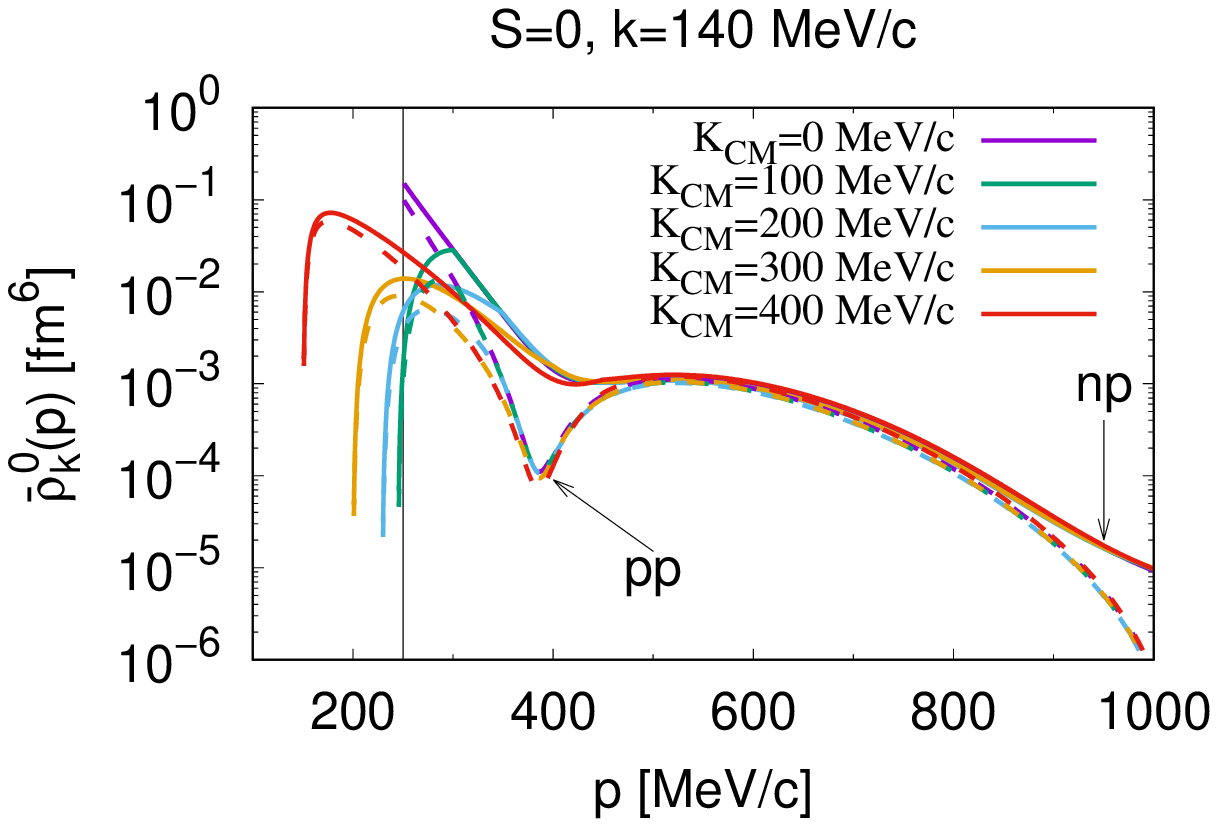}&
\includegraphics[width=9cm]{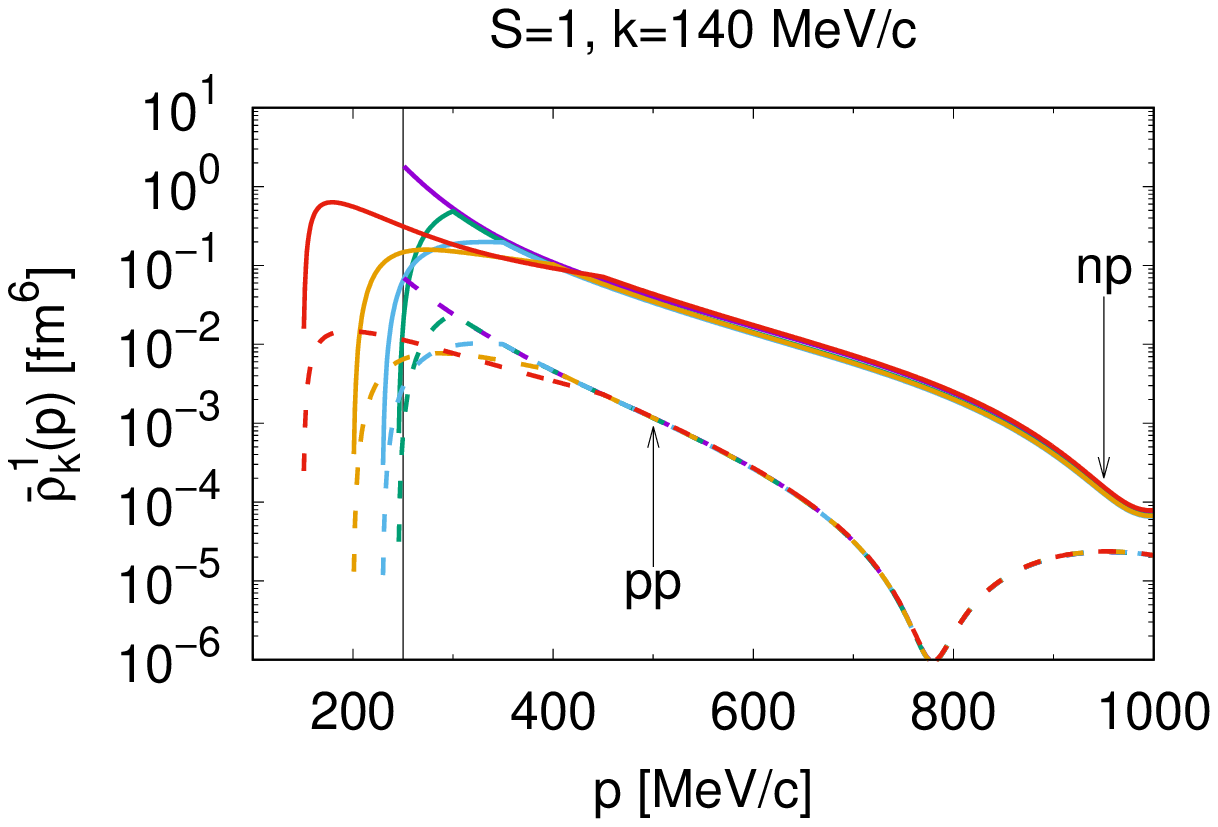}\\
\includegraphics[width=9cm]{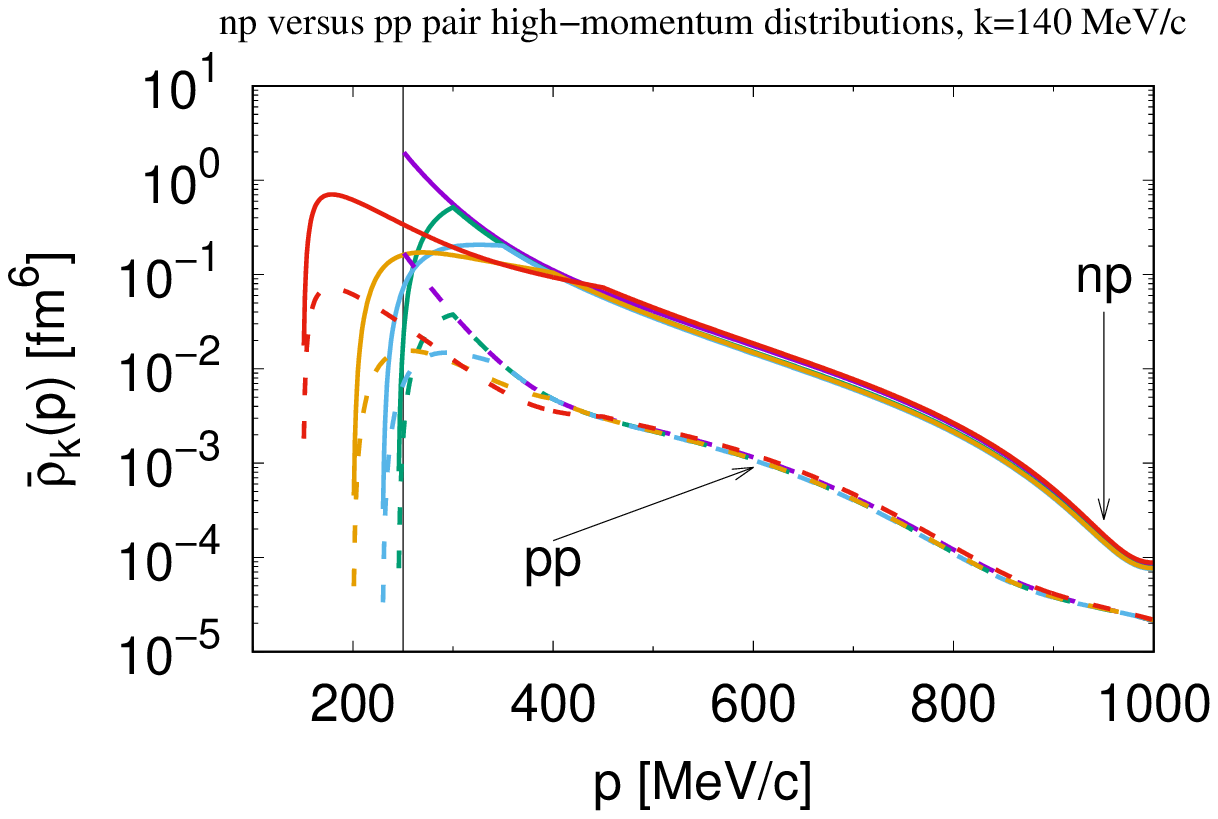}&
\includegraphics[width=9cm]{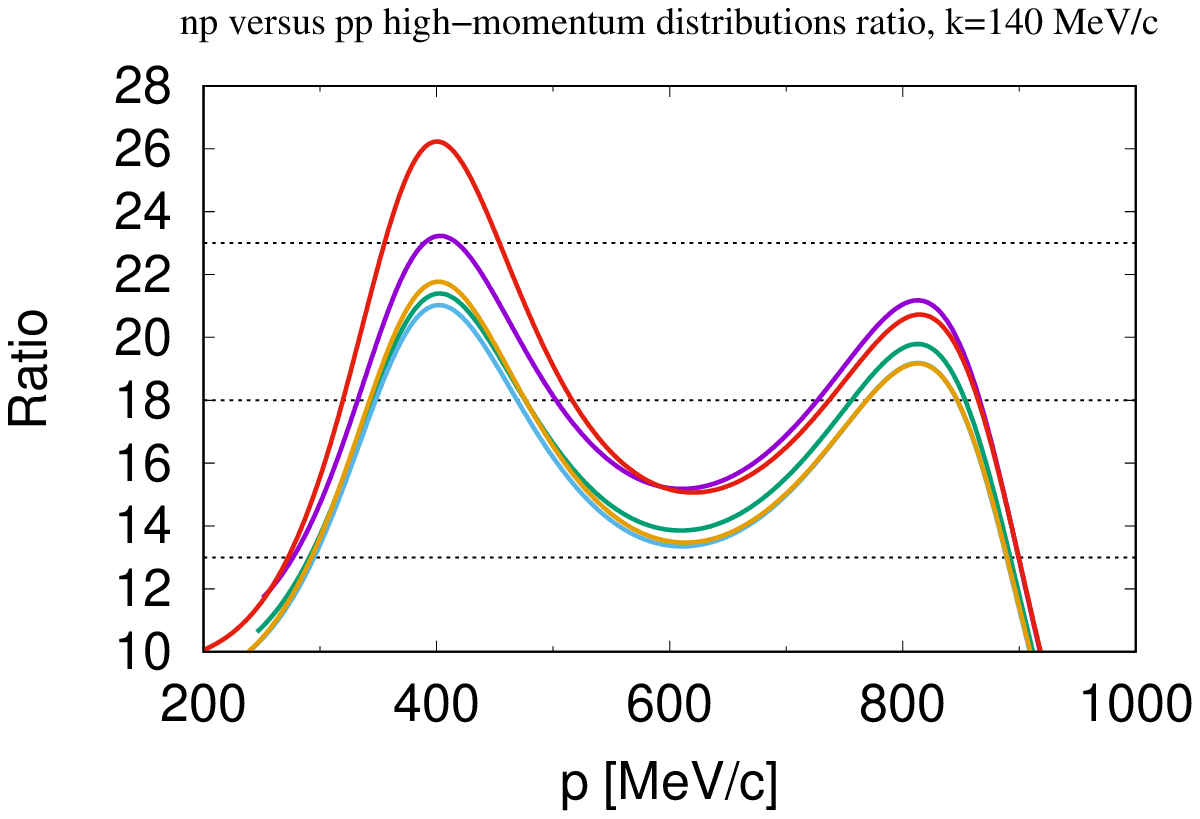}
\end{tabular}
\caption{Probability densities of high-momentum components per nucleon
  pair $\bar{\rho}^{S}_k(p)$ (see definition given in
  eq.~(\ref{high_density_mom_dist_spin})) for $S=0$ (upper left
  panel), $S=1$ (upper right panel) and summing the contributions from
  the two spin configurations (lower left panel).  The results are
  given for relative momentum $k=140$ MeV/c, and for different values
  of the CM momentum as labeled in the key of the first panel. The
  solid lines refer to momentum distributions of np pairs, while the
  dashed lines correspond to pp pairs.  The results for $K_{\rm CM}=0$
  MeV/c (purple lines) are essentially the same as those shown in
  Fig. 8 and in the upper panel of Fig. 9 of
  Ref.~\cite{RuizSimo:2017tcb}. Therefore, these curves are directly
  comparable with the quantity defined in eq. (43) of
  Ref.~\cite{RuizSimo:2017tcb}.  On the other hand, the lower right
  panel corresponds to the ratio of high-momentum density distribution
  of np over pp pairs, for the same CM momenta displayed in the other
  panels. This is straightforward comparable with the lower panel of
  Fig. 9 in Ref.~\cite{RuizSimo:2017tcb}.  The ratios are, generally,
  in between $18\pm5$ for a wide range of high momentum components and
  are almost insensitive to the CM momenta of the nucleon pair.}
\label{Fig:density_high_mom}
\end{figure*}

In Fig.~\ref{Fig:density_high_mom} we display the total high-momentum
distributions of a nucleon pair with initial relative momentum $k$ and
spin $S$, for $p>k$, which we define as:
\begin{eqnarray}
&&\bar{\rho}^{S}_{k}(p)=
4\pi\,(2S+1)
\rho^{S}_{k,K_{\rm CM}}(p)
\nonumber\\
&&=\sum_{l,l^\prime,J}
\Delta(JSl^\prime)\,(2J+1)\,
\left|\Delta\phi^{SJ}_{k,l^\prime\,l}(p)\right|^2,
\label{high_density_mom_dist_spin}
\end{eqnarray} 
where $\rho^{S}_{k,K_{\rm CM}}(p)$ is given by
eq.~(\ref{high_dens_mom_dist_spinS_final}) but replacing
$\phi^{SJ}_{k,l^\prime\,l}$ by $\Delta\phi^{SJ}_{k,l^\prime\,l}$.  We
show results for np (solid lines) and pp pairs (dashed lines) at
relative momentum of $k=140$ MeV/c for different CM momenta of the
pair, as labeled in the key of the top left panel of
Fig.~\ref{Fig:density_high_mom}.  The purpose of giving the quantity
$\bar{\rho}^{S}_{k}(p)$ is because it is directly comparable with
eq. (43) of Ref.~\cite{RuizSimo:2017tcb} and with figures 8 and 9
(upper panel) of the same reference.

The upper left panel of Fig.~\ref{Fig:density_high_mom} shows the
total high-momentum density distribution (summed over the different
partial waves) for correlated np and pp pairs with total spin $S=0$.
Both np and pp momentum distributions are very similar in the
intermediate region of probed relative momentum $450\lesssim p
\lesssim 850$ MeV/c. This is due to the fact that the ${}^{1}$P$_1$
contribution, which is present in np pairs with $S=0$ but not in pp
pairs with the same total spin, is quite irrelevant in this region of
$p$ if compared with the dominant component coming from the
${}^{1}$S$_0$ partial wave. However, the differences between np and pp
pairs are patent at $p\sim 400$ MeV/c, where the dominant
${}^{1}$S$_0$ partial wave has a node, and then the ${}^{1}$P$_1$
contribution makes the difference between np and pp pairs, because for
the latter only the ${}^{1}$D$_2$ partial wave can be added (due to
antisymmetry considerations of the relative wave function for two
identical fermions), and its contribution is far less important than
that of the ${}^{1}$P$_1$ channel.

In the upper right panel of Fig.~\ref{Fig:density_high_mom} we show
the same total high-momentum distributions for correlated np and pp
pairs in triplet spin state, $S=1$. In this case the differences
between np and pp pairs are much clearer in the whole range of
$p$. This is because for pp pairs in triplet state only odd $l$
partial waves contribute (P and F-waves), while for np pairs all
triplet partial waves are summed, especially the most relevant ones,
such as the ${}^{3}$S$_1$-${}^{3}$D$_1$ coupled channel. Basically,
the presence of the ${}^{3}$S$_1$-${}^{3}$D$_1$ channel in the np
high-momentum distribution, while not in the pp one, makes the former
to be much larger, in general, by several orders of magnitude.

In the lower left panel of Fig.~\ref{Fig:density_high_mom}, we can
observe the sum of both singlet and triplet contributions for the
high-momentum distributions of np and pp pairs. This panel represents
the high-momentum density distribution of a nucleon pair regardless of
its total spin state. It is evident that the np distribution is
approximately an order of magnitude larger than the pp distribution.
This observation is consistent with the findings of
Ref.~\cite{RuizSimo:2017tcb}, which also reported a similar trend in
its Figs. 9 and 10.

It is also worth pointing out that the so far discussed three panels
shown in Fig.~\ref{Fig:density_high_mom} share several common features
with those of Figs.~\ref{Fig:phiradial} and
\ref{Fig:phiradial_coupled}, namely: very little dependence of the
high-momentum distributions on the CM momentum of the pair for $p
\gtrsim k_F + \frac{K_{\rm CM}}{2}$; intrusion of the momentum
distributions below the Fermi momentum for $K_{\rm CM} > 0$; and,
finally, it is very clear that the pair momentum distributions are
continuous at the deviation points $p_{\rm dev}(K_{\rm
  CM})=k_F+\frac{K_{\rm CM}}{2}$, but not their derivatives at these
points.  This latter fact has been already discussed in relation with
Fig.~\ref{Fig:Qbarra}.

Finally, in the lower right panel of Fig.~\ref{Fig:density_high_mom}
we show the ratio $\frac{\bar{\rho}^{\rm np}_k(p)}{\bar{\rho}^{\rm
    pp}_k(p)}$ for a relative momentum of the pair of $k=140$ MeV/c
and for the five different CM momenta displayed in the other panels of
the same figure, as a function of the probed high momentum $p$.
Again, the ratio is quite insensitive to the CM momenta of the nucleon
pair, and, for a wide range of probed high momentum $p$, the ratio is
in between $18\pm 5$, which is the claimed averaged ratio measured in
Ref.~\cite{Subedi:2008zz} for the ground state of the ${}^{12}$C
nucleus.

\begin{figure}[!ht]
\centering
\includegraphics[width=9cm]{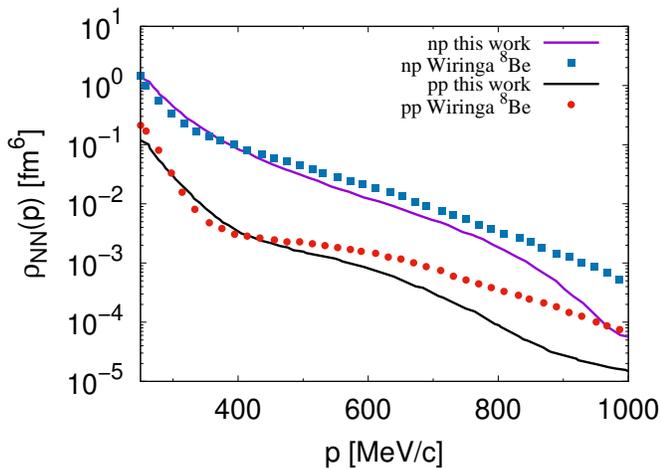}
\caption{Comparison of the np and pp pairs high-momentum distributions
  of this work, for $K_{\rm CM}=0$ and $k=140$ MeV/c, with those for
  ${}^8$Be nucleus found in Fig. 15 of
  reference~\cite{Wiringa:2013ala}, calculated within the variational
  Monte Carlo (VMC) approach. The two-nucleon momentum densities are
  plotted as functions of the high relative momentum $p$.  The high
  momentum np distributions have been normalized to $1$ fm${}^3$.}
\label{Fig:comparison_wiringa}
\end{figure}

To finish, in Fig.~\ref{Fig:comparison_wiringa} we present a
comparative analysis of our results with those obtained from a
realistic calculation in a finite nucleus using the Variational Monte
Carlo (VMC) approach, as reported in Ref.~\cite{Wiringa:2013ala}.  The
comparison is focused on the relative-momentum densities for
neutron-proton (np) and proton-proton (pp) pairs, which were
determined from the solution of the Bethe-Goldstone equation for a
center-of-mass momentum, $K_{\rm CM}$, of $0$ and an initial relative
momentum, $k$, of $140$ MeV/c.  To provide a comprehensive assessment,
we compared our findings with the high momentum pair distribution of
the nucleus ${}^8$Be, specifically for $K_{\rm CM}=0$, as depicted in
Fig. 15 of Ref.~\cite{Wiringa:2013ala}.

It is important to note that the pair momentum distribution presented
in reference~\cite{Wiringa:2013ala} involves the utilization of the
full nuclear wave function. Establishing a direct and straightforward
relationship between this distribution and the solution of the
Bethe-Goldstone equation for a specific pair with an initial relative
momentum, $k$, is not a trivial task which deserves further
investigation.

Some remarks to understand how this comparison has been carried out
must be carefully explained. As our two- nucleon densities have, in
general, a weak dependence (see figure 10 of
Ref.~\cite{RuizSimo:2017tcb}) on the initial relative momentum $k$ of
the pair, we have chosen again $k=140$ MeV/c to perform the
comparison, which is an intermediate value for the initial relative
momentum of the pair when its total momentum is zero. In the
comparison with the nucleon-pair momentum distributions in ${}^{8}$Be
for back-to-back ($K_{\rm CM}=0$) pairs found in figure 15 of
Ref.~\cite{Wiringa:2013ala}, we have disregarded the part of the
distributions below $q<k_F=\frac{250}{197.33}=1.267$ fm${}^{-1}$, for
normalization purposes, in order to focus on the high momentum
contribution.

However, given that our pair momentum distributions exhibit only weak
dependence on the precise value of $k$, and under the reasonable
assumption that the contribution from high momenta primarily reflects
short-distance behavior which is relatively independent on the nuclear
size, the trends provided by Fig.~\ref{Fig:comparison_wiringa} offer a
valuable insight on the comparison between np and pp pair
distributions and what is expected in finite nuclei.  Indeed, we
observe that the relationship between the np and pp distributions in
our study approximates the same trend observed in
reference~\cite{Wiringa:2013ala}.

What we have done is to normalize our np-pair distribution, for
$K_{\rm CM}=0$, taken from the lower left panel of
Fig.~\ref{Fig:density_high_mom}, in such a way that the integral $\int
\bar{\rho}_{\rm np}(p)\,p^2\,dp = 1$ fm${}^3$. And we have done
exactly the same for the np-pair distribution of figure 15 of
Ref.~\cite{Wiringa:2013ala} starting from $q\simeq 1.267$ fm${}^{-1}$.
Later, we have scaled the pp-pair momentum distributions in both
models accordingly with the normalization factors found in the
previous procedure, in order to keep the proportionality between the
nucleon-pair momentum densities. Although in figure 15 of
Ref.~\cite{Wiringa:2013ala} there are no units in the Y-axis for the
momentum distribution, accordingly to eq. (6) of the same reference,
the right units for the nucleon-pair distribution of that figure are
fm$^{6}$~\cite{Wiringa_comm}.

The result of the comparison can be observed in
Fig.~\ref{Fig:comparison_wiringa}, where our results are shown as
solid lines, while the results of Ref.~\cite{Wiringa:2013ala} are
displayed as filled squares for the np-pair distribution, and as
filled circles for the pp one, respectively. For relative intermediate
momentum, $p\le500$ MeV/c, the respective momentum distributions look
quite similar in size and shape.  However, for larger relative
momenta, they start to differ significantly: the nucleon-pair momentum
distributions calculated in ${}^8$Be start to be larger by almost one
order of magnitude with respect to those of this work for nuclear
matter. However, recent calculations~\cite{Piarulli:2022ulk,
  Wiringa_online} with the same methods based on chiral interactions,
clearly show a similar trend in qualitative agreement with that
presented in Fig.~\ref{Fig:comparison_wiringa}.  Presumably, this
feature is related to the comparatively harder core of the AV18
potential, as compared to the current chiral interactions and our
coarse grained potential.

Finally, this approximation suggests that, despite the inherent
complexities associated with a full nuclear wave function, and
differences in nuclear matter and finite nuclei, our theoretical
framework captures important aspects of neutron-proton and
proton-proton pair interactions. These results are in themselves
remarkable and suggest a quantitative connection between nuclear
matter and finite nuclei.

\section{Conclusions}\label{Sec:conclusions}
In this work we have extended our previous studies
\cite{RuizSimo:2016vsh,RuizSimo:2017tcb} about the effects of SRCs on
the high-momentum components of the relative wave function for a
nucleon pair in nuclear matter. The extension amounts to take the
angular average of the Pauli-blocking operator for the case with
$K_{\rm CM}\neq0$, and observing its effects on the tail of relative
high-momentum components.

Our findings indicate minimal dependence on the CM momentum of the
nucleon pair in the majority of plots presented in this paper.  This
consistency is observed in various aspects, including the relative
wave function at short distances, correlation functions in the
proximity of the origin, and defect wave functions. Furthermore, we
also observe limited sensitivity to the overall CM momentum of the
pair when examining the higher-momentum components of the relative
wave function. This holds true as long as the probed relative momentum
$p$ exceeds $2k_F$, where $k_F$ represents the Fermi momentum. This
last finding is consistent with the universality of SRCs and with the
factorization ansatzs used in the literature to express the pair
momentum distribution as a product of the momentum distribution of the
CM times the momentum distribution of the relative motion, being the
last an universal function \cite{Cohen:2018gzh, Vanhalst:2014cqa,
  Weiss:2016obx, Alvioli:2016wwp, CiofidegliAtti:2017tnm}.

In our case, the momentum distribution of the CM motion can be
described by a three-dimensional Dirac delta function. This is due to
the conservation of total momentum in the Bethe-Goldstone equation. In
finite nuclei, the momentum distribution of the CM is broadened. This
broadening has been observed and modeled using a three-dimensional
Gaussian function in Ref.~\cite{Cohen:2018gzh}.  This approach is
reasonable as the Dirac delta function can be considered as a limiting
case of a Gaussian function with an infinitesimally small width.

Therefore, it is reasonable to assume that our findings regarding the
independence of the relative high-momentum distribution of a pair on
the CM momentum could be extrapolated to finite nuclei. This
assumption holds true as long as the distribution is generated by
universal short-range correlations, considering the inherent
limitations in obtaining an exact resolution of the problem in such
systems.

\section{Acknowledgments}
This work has been partially supported by grant PID2020-114767GB-I00
funded by MCIN/AEI/10.13039/501100011033, by FEDER/Junta de
Andaluc\'ia-Consejer\'ia de Transformaci\'on Econ\'omica, Industria,
Conocimiento y Universidades/A-FQM-390-UGR20, and by Junta de
Andaluc\'ia (grant FQM-225). The authors of this work are deeply
indebted to Prof. Robert B. Wiringa for many clarifications about the
interpretation of his results, specially in connection with the
discussion of the last figure of the present manuscript.

\appendix
\section{Formal derivation of B-G equation for
total and relative perturbed states}\label{formal_derivation}

In this appendix we give the formal derivation of eqs.
(\ref{corr_2nucleon_state}) and (\ref{corr_relative_ket_equation})
starting from eq. (\ref{G-operator}).  If we apply
eq. (\ref{G-operator}) in operator form to a ket in the CM and
relative momenta representation $\left| \nK_{\rm CM},\nk\right\rangle$
which represents a two-nucleon state with definite total CM momentum
$\nK_{\rm CM}$ and relative one $\nk$, we obtain
\begin{equation}\label{G_matrix_ket_equation}
G\left| \nK_{\rm CM},\nk\right\rangle =
V\left| \nK_{\rm CM},\nk\right\rangle+
V\;\frac{Q}{E-H_0}\,G\left| \nK_{\rm CM},\nk\right\rangle.
\end{equation}
If we use now the definition of the G-matrix or effective interaction
\begin{eqnarray}
G \underbrace{\left| \nK_{\rm CM},\nk\right\rangle} &\equiv&
V \underbrace{\left| \Psi_{\nK_{\rm CM},\nk}\right\rangle},
\label{definition_effective_interaction}\\
{\rm unperturbed}&&\quad {\rm perturbed} \nonumber\\
{\rm state}&&\quad \quad {\rm state}\nonumber
\end{eqnarray}
which means that the action of the effective interaction over the
unperturbed state is the same as the action of the potential over the
corresponding perturbed state, then eq. (\ref{G_matrix_ket_equation})
transforms into
\begin{equation}\label{V_matrix_ket_equation}
V\left| \Psi_{\nK_{\rm CM},\nk}\right\rangle =
V\left| \nK_{\rm CM},\nk\right\rangle +
V\;\frac{Q}{E-H_0}\,V\left| \Psi_{\nK_{\rm CM},\nk}\right\rangle,
\end{equation}
where $E$ is the energy eigenvalue of the perturbed two-nucleon state
$\left| \Psi_{\nK_{\rm CM},\nk}\right\rangle$, and $H_0= T_1+T_2$ is
the unperturbed Hamiltonian containing only the one-body kinetic
energy operators.

Formally, in eq.~(\ref{V_matrix_ket_equation}), assuming that $V$ is
invertible, we can act from the left by the inverse potential operator
$V^{-1}$, thus eliminating the first appearance of the potential
operator in all the terms of the equation. Additionally, we can also
introduce a resolution of the identity operator in terms of the direct
product of two single-particle momentum eigenstates, $\mathbf{I}=\int
d^3k_1 \, d^3k_2 \left| \nk_1,\nk_2 \right\rangle \left\langle
\nk_1,\nk_2\right|$, in between the $\frac{Q}{E-H_0}$ and $V$
operators. With this, eq.~(\ref{V_matrix_ket_equation}) becomes
\begin{eqnarray}
\left| \Psi_{\nK_{\rm CM},\nk}\right\rangle&=&
\left| \nK_{\rm CM},\nk\right\rangle +
\int d^3k_1 \, d^3k_2\, \frac{\theta(\left|\nk_1 \right|-k_F)\,
\theta(\left|\nk_2 \right|-k_F)}{E-\left(T_{\nk_1}+T_{\nk_2} \right)}
\nonumber\\
&\times& \left| \nk_1,\nk_2 \right\rangle \left\langle \nk_1,\nk_2\right|
V \left| \Psi_{\nK_{\rm CM},\nk}\right\rangle, \label{perturbed_state}
\end{eqnarray}
where the two step functions come from the action of the
Pauli-blocking operator $Q$ over the two-particle momentum eigenstates
$ \left| \nk_1,\nk_2 \right\rangle$, and $T_{\nk_i}=
\frac{\nk^2_i}{2M_N}$ (with $i=1,2$) are the kinetic energy
eigenvalues.

Again, in eq.~(\ref{perturbed_state}), we can introduce another
resolution of the identity operator in terms of the CM and relative
momenta eigenstates representation, $\mathbf{I}=\int d^3K^\prime_{\rm
  CM} \, d^3k^\prime \left| \nK^\prime_{\rm CM},\nk^\prime
\right\rangle \left\langle \nK^\prime_{\rm CM},\nk^\prime \right|$, in
between the bra $\left\langle \nk_1,\nk_2\right|$ and the potential
$V$ operator.  In this way, we obtain
\begin{eqnarray}
\left| \Psi_{\nK_{\rm CM},\nk}\right\rangle&=&
\left| \nK_{\rm CM},\nk\right\rangle + 
\int d^3k_1 \, d^3k_2\, \frac{\theta(\left|\nk_1 \right|-k_F)\,
\theta(\left|\nk_2 \right|-k_F)}{E-\left(T_{\nk_1}+T_{\nk_2} \right)}
\nonumber\\
&\times& \left| \nk_1,\nk_2 \right\rangle 
\int d^3K^\prime_{\rm CM} \, d^3k^\prime
\left\langle \nk_1,\nk_2\right| \left. 
\nK^\prime_{\rm CM},\nk^\prime \right\rangle \nonumber\\
&\times&\left\langle \nK^\prime_{\rm CM},\nk^\prime \right| V 
\left| \Psi_{\nK_{\rm CM},\nk}\right\rangle. \label{perturbed_state_2}
\end{eqnarray}

Finally, using the first line for the bra-ket product $\left\langle
\nk_1,\nk_2\right| \left.  \nK^\prime_{\rm CM},\nk^\prime
\right\rangle$ given in eqs.~(\ref{CM_single_part_represent}), we can
easily perform in eq.~(\ref{perturbed_state_2}) the integrations over
$\nk_1$ and $\nk_2$ with the aid of the two Dirac delta functions,
obtaining
\begin{eqnarray}
\left| \Psi_{\nK_{\rm CM},\nk}\right\rangle&=&
\left| \nK_{\rm CM},\nk\right\rangle + 
\int d^3K^\prime_{\rm CM} \, d^3k^\prime
\frac{Q(\nK^\prime_{\rm CM}, \nk^\prime)}{E- \left( 
\frac{\nK^{\prime\,2}_{\rm CM}}{4M_N} + 
\frac{\nk^{\prime\,2}}{M_N}\right)} \nonumber\\
&\times& \left| \frac{\nK^\prime_{\rm CM}}{2} + \nk^\prime,
\frac{\nK^\prime_{\rm CM}}{2} - \nk^\prime \right\rangle
\left\langle \nK^\prime_{\rm CM},\nk^\prime \right| V 
\left| \Psi_{\nK_{\rm CM},\nk}\right\rangle. \nonumber\\ \label{perturbed_state_3}
\end{eqnarray}
In eq.~(\ref{perturbed_state_3}), $Q(\nK^\prime_{\rm CM}, \nk^\prime)$
stands for the two step functions written in terms of the CM and
relative momenta (see eq.~(\ref{Q_CM_rel_momenta})).

The final step to get eq.~(\ref{corr_2nucleon_state}) of
Sect.~\ref{sect:general} is to assume that the true energy eigenvalue
$E$ of the perturbed state does not change too much from the energy
eigenvalue of the unperturbed initial state, i.e, $E\simeq
\frac{\nK^{2}_{\rm CM}}{4M_N} + \frac{\nk^{2}}{M_N}$, and to write the
energy denominator in eq.~(\ref{perturbed_state_3}) in terms of the
total and reduced masses of the two-nucleon system. This approximation
for the true energy eigenvalue has also been done by other authors,
such as Ref.~\cite{Walecka1995} in the context of the independent pair
approximation.

The next step to obtain eq.~(\ref{corr_relative_ket_equation}) of
Sect.~\ref{sect:general} consists in trying to remove as much as
possible the dependence on the CM momentum in
eq.~(\ref{perturbed_state_3}). For this to be possible it is
completely necessary to assume that the potential does not depend at
all on the CM coordinate; we will further assume that it is also local
in the relative coordinate as well, as given in
eq.~(\ref{pot_mat_elem}).  To this end, we may introduce two
resolutions of the identity operator: one in terms of the CM and
relative momenta eigenkets representation $\mathbf{I}=\int
d^3K^{\prime\prime}_{\rm CM} \, d^3k^{\prime\prime} \left|
\nK^{\prime\prime}_{\rm CM},\nk^{\prime\prime} \right\rangle
\left\langle \nK^{\prime\prime}_{\rm CM},\nk^{\prime\prime} \right|$
inside the term with the integrals in eq.~(\ref{perturbed_state_3})
and acting from the left on the $ \left| \frac{\nK^\prime_{\rm CM}}{2}
+ \nk^\prime,\frac{\nK^\prime_{\rm CM}}{2} - \nk^\prime \right\rangle$
ket; and the other one in terms of the CM and relative position
eigenkets, $\mathbf{I}=\int d^3R^{\prime}_{\rm CM} \, d^3r^{\prime}
\left| \nR^{\prime}_{\rm CM},\nr^{\prime} \right\rangle \left\langle
\nR^{\prime}_{\rm CM},\nr^{\prime} \right|$, and acting in between the
potential $V$ operator and the perturbed state in
eq.~(\ref{perturbed_state_3}) as well:
\begin{eqnarray}
&&\left| \Psi_{\nK_{\rm CM},\nk}\right\rangle=
\left| \nK_{\rm CM},\nk\right\rangle +
\int d^3K^\prime_{\rm CM} \, d^3k^\prime
\frac{Q(\nK^\prime_{\rm CM}, \nk^\prime)}{
\frac{\left( \nK^2_{\rm CM}-\nK^{\prime\,2}_{\rm CM}\right)}{2M_T} 
 + \frac{\left(\nk^2-\nk^{\prime\,2}\right)}{2\mu}}\nonumber\\
 &\times& \int d^3K^{\prime\prime}_{\rm CM} \, d^3k^{\prime\prime}
\left| \nK^{\prime\prime}_{\rm CM},\nk^{\prime\prime} \right\rangle 
\; \delta^3\left(\frac{\nK^\prime_{\rm CM}-
\nK^{\prime\prime}_{\rm CM}}{2}+ \nk^{\prime\prime}-\nk^\prime \right)
\nonumber\\
&\times&\delta^3\left(\frac{\nK^\prime_{\rm CM}-
\nK^{\prime\prime}_{\rm CM}}{2}+ \nk^{\prime}-\nk^{\prime\prime} 
\right)\, \int d^3R^{\prime}_{\rm CM} \, 
d^3r^{\prime}\; \; V(\nr^\prime) \nonumber\\
&\times& \left\langle \nK^\prime_{\rm CM}, \nk^\prime \right| \left.
\nR^\prime_{\rm CM}, \nr^\prime \right\rangle
\left\langle \nR^\prime_{\rm CM}, \nr^\prime \right| \left.
\Psi_{\nK_{\rm CM},\nk}\right\rangle. \label{perturbed_state_4}
\end{eqnarray}
The two Dirac delta functions in eq.~(\ref{perturbed_state_4}) come
from the bra-ket product $\left\langle \nK^{\prime\prime}_{\rm
  CM},\nk^{\prime\prime} \right| \left. \frac{\nK^\prime_{\rm CM}}{2}
+ \nk^\prime, \frac{\nK^\prime_{\rm CM}}{2} - \nk^\prime
\right\rangle$, where the first line of the bra-ket product
$\left\langle \nK_{\rm CM}, \nk \right| \left. \nk_1, \nk_2
\right\rangle$ given in eqs.~(\ref{CM_single_part_represent}) has been
used with $\nk_1= \frac{\nK^\prime_{\rm CM}}{2} + \nk^\prime$ and
$\nk_2= \frac{\nK^\prime_{\rm CM}}{2} - \nk^\prime$, and accordingly
for the doubly primed CM and relative momenta variables of the bra.

To further proceed with eq.~(\ref{perturbed_state_4}), it is necessary
to pass from ket notation to wave function notation, in order to
remove totally all the integrals over CM coordinates and momenta. To
this end, we multiply both sides of eq.~(\ref{perturbed_state_4}) from
the left by the bra $\left\langle \nR_{\rm CM},\nr \right|$; we also
use the final line of eqs.~(\ref{CM_single_part_represent}) for the
unperturbed or plane wave states and
\begin{equation}\label{perturbed_wf_appendix}
\left\langle \nR_{\rm CM},\nr \right|
\left. \Psi_{\nK_{\rm CM},\nk}\right\rangle = 
\frac{e^{i\, \nK_{\rm CM} \cdot \nR_{\rm CM}}}{(2\pi)^{\frac32}}\;
\frac{\psi_{\nK_{\rm CM},\nk}(\nr)}{(2\pi)^{\frac32}}
\end{equation}
for the perturbed wave functions in coordinate representation.  It is
worth noting that the plane wave for the CM motion in
eq.~(\ref{perturbed_wf_appendix}) appears because the potential does
not depend on the CM coordinate and it is, therefore, a constant of
motion in our problem. Then, substituting the plane waves and
eq.~(\ref{perturbed_wf_appendix}) into eq.~(\ref{perturbed_state_4}),
we can straightforwardly carry out the integrals over
$\nk^{\prime\prime}$, $\nR^{\prime}_{\rm CM}$, $\nK^\prime_{\rm CM}$
and $\nK^{\prime\prime}_{\rm CM}$ in
eq.~(\ref{perturbed_state_4}). The final result is
\begin{eqnarray}
&&\frac{e^{i\, \nK_{\rm CM} \cdot \nR_{\rm CM}}}{(2\pi)^{\frac32}}\;
\frac{\psi_{\nK_{\rm CM},\nk}(\nr)}{(2\pi)^{\frac32}}=
\frac{e^{i\, \nK_{\rm CM} \cdot \nR_{\rm CM}}}{(2\pi)^{\frac32}}\;
\frac{e^{i\, \nk \cdot \nr}}{(2\pi)^{\frac32}} \nonumber\\
&+&\frac{e^{i\, \nK_{\rm CM} \cdot \nR_{\rm CM}}}{(2\pi)^{\frac32}}
\int d^3k^\prime \; \frac{Q(\nK_{\rm CM},\nk^\prime)}{k^2-k^{\prime\,2}}\;
\frac{e^{i\, \nk^\prime \cdot \nr}}{(2\pi)^{\frac32}}\nonumber\\
&\times& \int d^3r^\prime \; \;
\frac{e^{-i\, \nk^\prime \cdot \nr^\prime}}{(2\pi)^{\frac32}}\;
2\mu\, V(\nr^\prime) \;
\frac{\psi_{\nK_{\rm CM},\nk}(\nr^\prime)}{(2\pi)^{\frac32}}.
\label{perturbed_state_5}
\end{eqnarray}

In eq.~(\ref{perturbed_state_5}) the plane wave for the CM motion
cancels on both sides, and what remains is an integral B-G equation
for the "single" particle relative wave function $\psi_{\nK_{\rm
    CM},\nk}(\nr)$, which can also be written as
\begin{eqnarray}
&&\left\langle \nr \right| \left. \psi_{\nK_{\rm CM},\nk} \right\rangle =
\left\langle \nr \right| \left. \nk \right\rangle + 
\int d^3k^\prime \; \frac{Q(\nK_{\rm CM},\nk^\prime)}{k^2-k^{\prime\,2}}\;
\left\langle \nr \right| \left. \nk^\prime \right\rangle \nonumber\\
&\times& \int d^3r^\prime \; 
\left\langle \nk^\prime \right| \left. \nr^\prime \right\rangle\;
2\mu\, V(\nr^\prime) \;
\left\langle \nr^\prime \right| \left. \psi_{\nK_{\rm CM},\nk} \right\rangle=
\nonumber\\
&=& \left\langle \nr \right| \left. \nk \right\rangle +
\left\langle \nr \right| \int d^3k^\prime \; 
\frac{Q(\nK_{\rm CM},\nk^\prime)}{k^2-k^{\prime\,2}}\; 
\left| \nk^\prime \right\rangle\, \left\langle \nk^\prime \right|
\nonumber\\
&\times&\underbrace{\left( 
\int d^3r^\prime \; 2\mu\, V(\nr^\prime) \; 
\left| \nr^\prime \right\rangle \, \left\langle \nr^\prime \right|
\right)}  \left| \psi_{\nK_{\rm CM},\nk} \right\rangle 
\label{single_relative_perturbed_wf_appendix} \\
&& {\rm spectral}\; {\rm resolution} \;  {\rm of}\; {\rm the}\nonumber\\
&&\qquad 2\mu\,V \; {\rm operator} \nonumber
\end{eqnarray}

Finally, in eq.~(\ref{single_relative_perturbed_wf_appendix}), the bra
$\left\langle \nr \right|$ is arbitrary and appears on both sides of
the equation. That bra can be removed from both sides and what remains
is an integral equation for the ket $\left| \psi_{\nK_{\rm CM},\nk}
\right\rangle$, which is precisely the
eq.~(\ref{corr_relative_ket_equation}) of Sect.~\ref{sect:general}.

\section{Formal derivation of the integral B-G equation
for the radial part of the relative wave function}\label{derivation_radial}

Our aim in this appendix is to obtain eq.~(\ref{radial_wf}) of
Sect.~\ref{subsec:radial} by performing a partial wave expansion of
eq.~(\ref{corr_relative_ket_equation}) in Sect.~\ref{sect:general},
but having substituted the general Pauli-blocking function $Q(\nK_{\rm
  CM},\nk^\prime)$ by its angular average $\overline{Q}(K_{\rm
  CM},k^\prime)$ given in eq.~(\ref{Qbar_function}).

Until now, all the discussion given in
appendix~\ref{formal_derivation} has omitted the spin of the single
particle states or the total spin of the two-nucleon system. The
latter can be totally ascribed to the relative kets $\left|
\psi_{\nK_{\rm CM},\nk} \right\rangle$ and $\left| \nk \right\rangle$
in eq.~(\ref{corr_relative_ket_equation}).  This last equation is an
integral equation for the perturbed ket $\left| \psi_{\nK_{\rm
    CM},\nk} \right\rangle$, i.e, the same state appears on the
left-hand side of the equation and on the right-hand one.

Additionally, it is a well-known fact that the N-N potential conserves
the total spin $S$ of the nucleon pair, its total angular momentum $J$
and the third component of the latter $M$, but neither the third
component of the total spin $M_S$ nor the orbital angular momenta,
which can get mixed by the tensor force of the N-N potential. We start
from eq.~(\ref{corr_relative_ket_equation}) by substituting the
Pauli-blocking function by its angular average, putting the spin and
its third component on the unperturbed and perturbed states and now
the resolution of the identity in terms of the momentum eigenkets is
$\mathbf{I}=\sum_{S^\prime, M^\prime_S}\int d^3k^\prime
\left|\nk^\prime;S^\prime M^\prime_S\right\rangle \left\langle
\nk^\prime;S^\prime M^\prime_S\right|$:

\begin{eqnarray}
&&\left| \psi_{\nk}; S M_S \right\rangle_{K_{\rm CM}} =
\left| \nk; S M_S \right\rangle + \sum_{S^\prime,M^\prime_S}
 \int d^3k^\prime \;
\frac{\overline{Q}(K_{\rm CM},k^\prime)}{k^2-k^{\prime\,2}}\nonumber\\
&&\left| \nk^\prime;S^\prime M^\prime_S \right\rangle\, 
\left\langle \nk^\prime; S^\prime M^\prime_S \right| 2\mu\, V  
\left| \psi_{\nk}, S M_S \right\rangle_{K_{\rm CM}}.
\label{B-G_equation_relative_ket}
\end{eqnarray}

We know that for a large family of N-N potentials they have the
properties of being local, preserving the total spin $S$, the total
angular momentum $J$ and its third component $M$, but that due to the
tensor force, they mix orbital angular momenta. This means that we can
write the spectral resolution of the potential as
\begin{equation}\label{spectral_resolution_V}
V=\int^\infty_0 dr\; r^2 \sum_{l,l^\prime}\sum_{S,J,M} V^{SJ}_{ll^\prime}(r)
\left|r; lS; JM\right\rangle \left\langle r; l^\prime S; JM\right|,
\end{equation}
where the basis $\left| r; lS; JM\right\rangle$ is given in terms of
the eigenbasis of position and spin, $\left|\nr; S M_S\right\rangle$,
as:
\begin{equation}\label{change_eigenbasis}
\left| r; lS; JM\right\rangle=\sum_{m,M_S}\int d\Omega_{\hat{r}}\;
Y_{lm}(\hat{r}) \left\langle lm;SM_S\right|\left.JM\right\rangle\,
\left|\nr;SM_S\right\rangle.
\end{equation}
The spectral resolution of the potential given in
eq.~(\ref{spectral_resolution_V}) ensures that its matrix elements
between eigenkets of the form given in eq.~(\ref{change_eigenbasis})
is
\begin{eqnarray}
&&\left\langle r^\prime;l_1 S_1; J_1 M_1\right|V\left| r^{\prime\prime};
l_2 S_2; J_2 M_2\right\rangle=\delta_{S_1,S_2}\,\delta_{J_1,J_2}\,
\delta_{M_1,M_2}\nonumber\\
&&\frac{1}{r^\prime\,r^{\prime\prime}}\;\delta(r^\prime-r^{\prime\prime})\;
V^{S_1J_1}_{l_1l_2}(r^\prime),\label{pot_mat_element}
\end{eqnarray}
which is the obvious result for a local radial potential which
preserves spin, total angular momentum and its third component, but it
is not necessarily diagonal in the orbital angular momentum.  To
derive eq.~(\ref{pot_mat_element}) we have used the orthogonality
condition of the eigenbasis $\left| r; lS; JM\right\rangle$:
\begin{equation}\label{orthonormality_condition_rlsjm}
\left\langle r^\prime;l_1 S_1; J_1 M_1\right|\left.r;lS;JM\right\rangle=
\frac{1}{r\,r^\prime}\;\delta(r-r^\prime)\,\delta_{l,l_1}\delta_{S,S_1}
\delta_{J,J_1}\delta_{M,M_1},
\end{equation}
which in turn can be obtained by evaluating the bra-ket product with
the expansion of the eigenbasis given in
eq.~(\ref{change_eigenbasis}), and using the more obvious
orthogonality condition of the eigenbasis of position and spin
$\left|\nr; S M_S\right\rangle$.

Introducing the spectral resolution of the potential,
eq.~(\ref{spectral_resolution_V}), into the bra-ket product of
eq.~(\ref{B-G_equation_relative_ket}), we obtain
\begin{eqnarray}
 &&\left\langle \nk^\prime; S^\prime M^\prime_S \right| 2\mu\, V  
\left| \psi_{\nk}, S M_S \right\rangle_{K_{\rm CM}}=
\int^\infty_0 dr\;r^2 
\sum_{l,l^\prime}
\sum_{S^{\prime\prime}JM}
2\mu V^{S^{\prime\prime}J}_{ll^\prime}(r)\nonumber\\
&&\frac{4\pi}{(2\pi)^\frac32}\,
\delta_{S^{\prime\prime},S^\prime}
\;i^{-l}j_l(k^\prime r)\sum_m
Y_{lm}(\hat{k}^\prime)
\left\langle lm;S^{\prime\prime}
M^\prime_S\right|\left.JM
\right\rangle\nonumber\\
&&\left(\int d^3r^\prime
\left\langle r;l^\prime 
S^{\prime\prime};JM\right|\left.
\nr^\prime\right\rangle
\left\langle \nr^\prime\right|
\left. \psi_{\nk},S M_S
\right\rangle_{K_{\rm CM}}\right),
\label{pot_mat_elem_free_perturb}
\end{eqnarray}
where in the last piece between parenthesis we have introduced a
resolution of the identity in the form $\int d^3r^\prime
\left|\nr^\prime\right\rangle \left\langle \nr^\prime\right|$, and we
have also used that
\begin{eqnarray}
 &&\left\langle \nk^\prime;
 S^\prime M^\prime_S\right|
 \left. r;lS^{\prime\prime};JM
 \right\rangle=\frac{4\pi}{(2\pi)^\frac32}\,
\delta_{S^{\prime\prime},S^\prime}
\;i^{-l}j_l(k^\prime r)\nonumber\\
&&\sum_m Y_{lm}(\hat{k}^\prime)
\left\langle lm;S^{\prime\prime}
M^\prime_S\right|\left.JM
\right\rangle.\label{braket_eigen_momspin_rad_coupled}
\end{eqnarray}
This last expression can be easily obtained by multiplying
eq.~(\ref{change_eigenbasis}) from the left by the bra $\left\langle
\nk^\prime; S^\prime M^\prime_S\right|$, using the Rayleigh expansion
for the plane wave and carrying out the calculations.

In eq.~(\ref{pot_mat_elem_free_perturb}) we can substitute
eq.~(\ref{perturbed_wf_S_Ms}) for the correlated wave function in
coordinate representation, and the braket product
\begin{equation}\label{braket_prod_r_coupled_lsjm_pos_eigenket}
 \left\langle r;l^\prime S^{\prime\prime};JM\right|
 \left. \nr^\prime \right\rangle=
 \frac{1}{r\,r^\prime}\;
 \delta(r-r^\prime)\;
 \mathcal{Y}^\dagger_{l^\prime 
 S^{\prime\prime}JM}(\hat{r}^\prime),
\end{equation}
where this last equation can be easily obtained from
eq.~(\ref{change_eigenbasis}) by multiplying from the left by a
position eigenstate, carrying out the calculations and taking its
complex conjugate.

Carrying out the calculations of the piece between parenthesis of
eq.~(\ref{pot_mat_elem_free_perturb}) we obtain finally:
\begin{eqnarray}
 &&\int d^3r^\prime
\left\langle r;l^\prime 
S^{\prime\prime};JM\right|\left.
\nr^\prime\right\rangle
\left\langle \nr^\prime\right|
\left. \psi_{\nk},S M_S
\right\rangle_{K_{\rm CM}}=\frac{4\pi}{(2\pi)^\frac32}
\sum_{l^{\prime\prime}
m^{\prime\prime}}
\nonumber\\
&& i^{l^{\prime\prime}}
u^{SJ}_{k,l^{\prime\prime}\,
l^\prime}(r)\;
Y^*_{l^{\prime\prime}m^{\prime\prime}}(\hat{k})\;
\delta_{S,S^{\prime\prime}} \left\langle 
l^{\prime\prime} m^{\prime\prime};S M_S 
\right|\left. JM
\right\rangle,
\label{intd3rprima}
\end{eqnarray}
where to obtain the above result we have integrated over $r^\prime$
with the aid of the Dirac delta function of
eq.~(\ref{braket_prod_r_coupled_lsjm_pos_eigenket}); and we have also
used the orthogonality properties of the spin-angular wave functions,
namely:
\begin{equation}\label{orthogonality_spin_angular_wf}
 \int d\Omega_{\hat{r}^\prime}\;
 \mathcal{Y}^{*}_{l^\prime S^{\prime\prime}JM}(\hat{r}^\prime)\;
 \mathcal{Y}_{l S J^\prime M^\prime}(\hat{r}^\prime)=
 \delta_{l^\prime,l}\,
 \delta_{S^{\prime\prime},S}\,
\delta_{J, J^\prime}\,
\delta_{M, M^\prime},
\end{equation}
in order to carry out some discrete sums over $J^\prime,M^\prime,l$ in
the expansion of the perturbed wave function in partial waves, given
by eq.~(\ref{perturbed_wf_S_Ms}).

Introducing the result of eq.~(\ref{intd3rprima}) into
eq.~(\ref{pot_mat_elem_free_perturb}), we obtain finally
\begin{eqnarray}
 &&\left\langle \nk^\prime; S^\prime M^\prime_S \right| 2\mu\, V  
\left| \psi_{\nk}, S M_S \right\rangle_{K_{\rm CM}}=\frac{(4\pi)^2}{(2\pi)^3}\,
\delta_{S,S^\prime}
\sum_{l,l^\prime m}
\sum_{JM}\sum_{l^{\prime\prime}m^{\prime\prime}}\nonumber\\
&&i^{l^{\prime\prime}-l}
\;Y_{lm}(\hat{k}^\prime)
\;Y^*_{l^{\prime\prime}m^{\prime\prime}}(\hat{k})\left\langle
lm;SM^\prime_S\right|
\left. JM\right\rangle
\left\langle l^{\prime\prime}m^{\prime\prime};SM_S
\right|\left. JM
\right\rangle\nonumber\\
&&\int^\infty_0 dr^\prime
\; r^{\prime\,2}\;
U^{SJ}_{l l^\prime}(r^\prime)\;j_l(k^\prime
r^\prime)\;u^{SJ}_{k,l^{\prime\prime}\,l^\prime}(r^\prime),
\label{2muV_matrix_elem_final}
\end{eqnarray}
where $U^{SJ}_{l l^\prime}=2\mu V^{SJ}_{l l^\prime}$ is the reduced
potential.

If we now introduce the matrix element so far calculated in
eq.~(\ref{2muV_matrix_elem_final}) in the B-G equation for the
relative ket, eq.~(\ref{B-G_equation_relative_ket}), multiply from the
left by the position eigenbra $\left\langle \nr\right|$, and
substitute eqs.~(\ref{unperturbed_wf_S_Ms}) and
(\ref{perturbed_wf_S_Ms}) for the expansions of the free and perturbed
wave functions in coordinates representation, we have
eq.~(\ref{perturbed_wf_S_Ms}) on the left-hand side; while on the
right-hand side we have:
\begin{eqnarray}
 &&{\rm LHS}=\frac{4\pi}{(2\pi)^\frac32}
 \sum_{JM}\sum_{l\,l^\prime m}
 i^{l^\prime}\,
 j_{l^\prime}(kr)\,
 \delta_{l^\prime l}\,
 Y^*_{l^\prime m}(\hat{k}) \left\langle
 l^\prime m;SM_S\right|
 \left. JM\right\rangle
 \nonumber\\
&\times& \mathcal{Y}_{lSJM}(\hat{r})+
\sum_{S^\prime,M^\prime_S}\int d^3k^\prime\,
\frac{\overline{Q}(K_{\rm CM},k^\prime)}{k^2-k^{\prime\,2}}
\left(\frac{4\pi}{(2\pi)^\frac32}
\sum_{J^\prime M^\prime}
\sum_{l_1 l^\prime_1 m_1} i^{l^\prime_1}
\right.\nonumber\\
&&\left. j_{l^\prime_1}(k^\prime r)\,
\delta_{l^\prime_1 l_1}
Y^*_{l^\prime_1 m_1}(\hat{k}^\prime)
\left\langle l^\prime_1
m_1;S^\prime M^\prime_S
\right|\left. J^\prime
M^\prime\right\rangle
\mathcal{Y}_{l_1 S^\prime J^\prime M^\prime}(\hat{r})
\right)\nonumber\\
&&\left(
\frac{(4\pi)^2}{(2\pi)^3}\,
\delta_{S,S^\prime}
\sum_{l,l^\prime m}
\sum_{JM}\sum_{l^{\prime\prime}m^{\prime\prime}}
i^{l^{\prime\prime}-l}
\;Y_{lm}(\hat{k}^\prime)
\;Y^*_{l^{\prime\prime}m^{\prime\prime}}(\hat{k})\right.
\nonumber\\
&&\left.\left\langle
lm;SM^\prime_S\right|
\left. JM\right\rangle
\left\langle l^{\prime\prime}m^{\prime\prime};SM_S
\right|\left. JM
\right\rangle
\int^\infty_0 dr^\prime
\; r^{\prime\,2}\;
U^{SJ}_{l l^\prime}(r^\prime)\right.\nonumber\\
&&\left.j_l(k^\prime
r^\prime)\;u^{SJ}_{k,l^{\prime\prime}\,l^\prime}(r^\prime)
\right).\label{rhs_bg_eq_relative}
\end{eqnarray}
In the second term of the above equation we can carry out the sum over
$S^\prime$ because of the presence of the Kronecker delta
$\delta_{S,S^\prime}$.  In addition, we can carry out the integration
over the angles of $\hat{k}^\prime$ exploiting the orthogonality of
the spherical harmonics of $\hat{k}^\prime$, thus obtaining a
$\delta_{l^\prime_1 l}\, \delta_{m_1 m}$ and carry out the additional
sums over $l^\prime_1$ and $m_1$, obtaining at the end:
\begin{eqnarray}
  &&{\rm LHS}=\frac{4\pi}{(2\pi)^\frac32}
 \sum_{JM}\sum_{l\,l^\prime m}
 i^{l^\prime}\,
 j_{l^\prime}(kr)\,
 \delta_{l^\prime l}\,
 Y^*_{l^\prime m}(\hat{k}) \left\langle
 l^\prime m;SM_S\right|
 \left. JM\right\rangle
 \nonumber\\
&\times& \mathcal{Y}_{lSJM}(\hat{r})+
\frac{(4\pi)^3}{(2\pi)^\frac92}
\sum^{S}_{M^\prime_S=-S}
\sum_{J^\prime M^\prime}
\sum_{JM}
\sum_{l,l^\prime m}
\sum_{l^{\prime\prime}m^{\prime\prime}}
\sum_{l_1}
i^{l^{\prime\prime}}\,
\delta_{l, l_1}
\nonumber\\
&&\left\langle 
lm;S M^\prime_S
\right|\left. J^\prime
M^\prime\right\rangle
\left\langle
lm;SM^\prime_S\right|
\left. JM\right\rangle
\left\langle l^{\prime\prime}m^{\prime\prime};SM_S
\right|\left. JM
\right\rangle
\nonumber\\
&&Y^*_{l^{\prime\prime}m^{\prime\prime}}(\hat{k})\;
\mathcal{Y}_{l_1 S J^\prime M^\prime}(\hat{r})
\int^\infty_0
dk^\prime\;
k^{\prime\,2}\,
\frac{\overline{Q}(K_{\rm CM},k^\prime)}{k^2-k^{\prime\,2}}
j_{l}(k^\prime r)
\nonumber\\
&&\int^\infty_0 dr^\prime\; 
r^{\prime\,2}\;
U^{SJ}_{l l^\prime}(r^\prime)\;
j_l(k^\prime
r^\prime)\;u^{SJ}_{k,l^{\prime\prime}\,l^\prime}(r^\prime).\label{rhs_bg_eq_relative2}
\end{eqnarray}
Again, in the second term of eq.~(\ref{rhs_bg_eq_relative2}) we can
perform easily the sum over $l_1$. Furthermore, the sum over $m$ and
$M^\prime_S$ for fixed $(l,S,J,M,J^\prime,M^\prime)$ only involves two
Clebsch-Gordan coefficients and its result is
$\delta_{JJ^\prime}\,\delta_{MM^\prime}$, and we can then carry out
the sum over $J^\prime$ and $M^\prime$, obtaining:
\begin{eqnarray}
  &&{\rm LHS}=\frac{4\pi}{(2\pi)^\frac32}
 \sum_{JM}\sum_{l\,l^\prime m}
 i^{l^\prime}\,
 j_{l^\prime}(kr)\,
 \delta_{l^\prime l}\,
 Y^*_{l^\prime m}(\hat{k}) \left\langle
 l^\prime m;SM_S\right|
 \left. JM\right\rangle
 \nonumber\\
&\times& \mathcal{Y}_{lSJM}(\hat{r})+
\frac{(4\pi)^3}{(2\pi)^\frac92}
\sum_{JM}
\sum_{l,l^\prime}
\sum_{l^{\prime\prime}m^{\prime\prime}}
i^{l^{\prime\prime}}
\left\langle l^{\prime\prime}m^{\prime\prime};SM_S
\right|\left. JM
\right\rangle\;
Y^*_{l^{\prime\prime}m^{\prime\prime}}(\hat{k})\nonumber\\
&&\mathcal{Y}_{l S J M}(\hat{r})
\int^\infty_0 dr^\prime\; 
\frac{kr^{\prime}}{kr}\;
U^{SJ}_{l l^\prime}(r^\prime)\;
u^{SJ}_{k,l^{\prime\prime}\,l^\prime}(r^\prime)
\nonumber\\
&&\int^\infty_0 dk^\prime\;
(k^{\prime}r)\,j_{l}(k^\prime r)\;
\frac{\overline{Q}(K_{\rm CM},k^\prime)}{k^2-k^{\prime\,2}}\;
(k^\prime r^\prime)\,j_l(k^\prime
r^\prime).\label{rhs_bg_eq_relative3} 
\end{eqnarray}
Finally, the second term of eq.~(\ref{rhs_bg_eq_relative3}) can be
arranged in the final form:
\begin{eqnarray}
  &&{\rm LHS}=\frac{4\pi}{(2\pi)^\frac32}
 \sum_{JM}\sum_{l\,l^\prime m}
 i^{l^\prime}\,
 j_{l^\prime}(kr)\,
 \delta_{l^\prime l}\,
 Y^*_{l^\prime m}(\hat{k}) \left\langle
 l^\prime m;SM_S\right|
 \left. JM\right\rangle
 \nonumber\\
&\times& \mathcal{Y}_{lSJM}(\hat{r})+
\frac{(4\pi)}{(2\pi)^\frac32}
\sum_{JM}
\sum_{l,l^{\prime\prime}}
\sum_{l^{\prime}m}
i^{l^{\prime}}
\left\langle l^{\prime}m;SM_S
\right|\left. JM
\right\rangle\;
Y^*_{l^{\prime}m}(\hat{k})\nonumber\\
&&\mathcal{Y}_{l S J M}(\hat{r})
\int^\infty_0 dr^\prime\;
U^{SJ}_{l l^{\prime\prime}}(r^\prime)\;
\frac{\widetilde{u}^{SJ}_{k,l^{\prime}\,l^{\prime\prime}}(r^\prime)}{kr}
\nonumber\\
&&\underbrace{\frac{2}{\pi}\int^\infty_0 dk^\prime\;
\hat{j}_{l}(k^\prime r)\;
\frac{\overline{Q}(K_{\rm CM},k^\prime)}{k^2-k^{\prime\,2}}\;
\hat{j}_l(k^\prime
r^\prime)},
\label{rhs_bg_eq_relative4}\\
&&\qquad\qquad\qquad\quad \widetilde{G}^{K_{\rm CM}}_{k,l}(r,r^\prime)\nonumber
\end{eqnarray}
where in eq.~(\ref{rhs_bg_eq_relative4}) we have used the definition
of the Green's function for the radial B-G equation, given in
eq.~(\ref{Green_function_Kcm}); and the normalization of the perturbed
radial wave function given in eq.~(\ref{normalization_radial_wf}) has
also been used. Finally, also in the second term of
eq.~(\ref{rhs_bg_eq_relative4}) the labels $l^\prime \leftrightarrow
l^{\prime\prime}$ and $m^{\prime\prime}\rightarrow m$ have been
renamed in the sums.

Therefore, at the end, we have on both sides of the equation:
\begin{eqnarray}
 &&\frac{4\pi}{(2\pi)^{\frac32}}\sum_{J,M}\; 
\sum_{l,l^\prime,m} 
i^{l^\prime}\, 
u^{SJ}_{k,l^\prime\,l}(r)\, 
Y^*_{l^\prime m}(\hat{k})
\left\langle l^\prime m; S M_S\right| \left. J M \right\rangle\,
\mathcal{Y}_{l S J M} (\hat{r})\nonumber\\
&&=\frac{4\pi}{(2\pi)^\frac32}
 \sum_{JM}\sum_{l\,l^\prime m}
 i^{l^\prime}\,
 j_{l^\prime}(kr)\,
 \delta_{l^\prime l}\,
 Y^*_{l^\prime m}(\hat{k}) \left\langle
 l^\prime m;SM_S\right|
 \left. JM\right\rangle
\mathcal{Y}_{lSJM}(\hat{r})\nonumber\\
&&+\frac{4\pi}{(2\pi)^\frac32}
\sum_{JM}
\sum_{l l^{\prime}m}
\sum_{l^{\prime\prime}}
i^{l^{\prime}}
\left\langle l^{\prime}m;SM_S
\right|\left. JM
\right\rangle
Y^*_{l^{\prime}m}(\hat{k})\;\mathcal{Y}_{l S J M}(\hat{r})
\nonumber\\
&&\int^\infty_0 dr^\prime\; \widetilde{G}^{K_{\rm CM}}_{k,l}(r,r^\prime)\;
U^{SJ}_{l l^{\prime\prime}}(r^\prime)\;
\frac{\widetilde{u}^{SJ}_{k,l^{\prime}\,l^{\prime\prime}}(r^\prime)}{kr}.\label{two_sides_bg_eq}
\end{eqnarray}

Obviously, the factors cancel on both sides, and to obtain the
equation for the radial part we have to get rid of all the angular
dependencies on $\hat{k}$ and $\hat{r}$.  To this end, we can multiply
from the left by the spin-angular wave function
$\mathcal{Y}^\dagger_{l_1 S_1 J_1 M_1}(\hat{r})$ and to integrate over
the solid angle of $\hat{r}$. Using the orthogonality properties of
these functions we can perform trivially the sums over $l$,$J$ and
$M$:
\begin{eqnarray}
 &&\delta_{S,S_1}
 \sum_{l^\prime m} 
i^{l^\prime}\, 
u^{SJ_1}_{k,l^\prime\,l_1}(r)\, 
Y^*_{l^\prime m}(\hat{k})
\left\langle l^\prime m; S M_S\right| \left. J_1 M_1 \right\rangle
\nonumber\\
&&=\delta_{S,S_1}
\sum_{l^\prime m}
 i^{l^\prime}\,
 j_{l^\prime}(kr)\,
 \delta_{l^\prime l_1}\,
 Y^*_{l^\prime m}(\hat{k}) \left\langle
 l^\prime m;SM_S\right|
 \left. J_1M_1\right\rangle
 \nonumber\\
&&+\delta_{S,S_1}
\sum_{l^{\prime}m}
\sum_{l^{\prime\prime}}
i^{l^{\prime}}
\left\langle l^{\prime}m;SM_S
\right|\left. J_1M_1
\right\rangle
Y^*_{l^{\prime}m}(\hat{k})
\nonumber\\
&&\int^\infty_0 dr^\prime\; \widetilde{G}^{K_{\rm CM}}_{k,l_1}(r,r^\prime)\;
U^{SJ_1}_{l_1 l^{\prime\prime}}(r^\prime)\;
\frac{\widetilde{u}^{SJ_1}_{k,l^{\prime}\,l^{\prime\prime}}(r^\prime)}{kr}.\label{two_sides_bg_eq2} 
\end{eqnarray}

We can take without loss of generality that $S_1=S$ to get rid of the
Kronecker deltas.  Then, we can multiply on both sides of
eq.~(\ref{two_sides_bg_eq2}) by $Y_{lm^\prime}(\hat{k})$ and integrate
over the solid angle of $\hat{k}$. Using the orthogonality of the
spherical harmonics, we can carry out the sum over $l^\prime$ and $m$,
thus obtaining finally:
\begin{equation}\label{two_sides_bg_eq3}
u^{SJ}_{k,l\,l^\prime}(r)= j_{l}(kr)\,
 \delta_{l, l^\prime}
+\int^\infty_0 dr^\prime\; \widetilde{G}^{K_{\rm CM}}_{k,l^\prime}(r,r^\prime)
\sum_{l^{\prime\prime}}
U^{SJ}_{l^\prime l^{\prime\prime}}(r^\prime)\;
\frac{\widetilde{u}^{SJ}_{k,l\,l^{\prime\prime}}(r^\prime)}{kr}, 
\end{equation}
where in eq.~(\ref{two_sides_bg_eq3}) we have canceled the
Clebsch-Gordan coefficients after having carried out the sum over
$l^\prime$ and $m$ because they were the same on both sides of the
equation. And finally we have renamed the free indices $J_1\rightarrow
J$, and $l_1\rightarrow l^\prime$ on both sides of the equation after
having performed the sums. Finally, if we multiply both sides by $kr$,
we obtain the B-G equation for the perturbed (correlated) relative
radial wave function (see eq.~(\ref{radial_wf})):
\begin{equation}\label{two_sides_bg_eq_final}
\widetilde{u}^{SJ}_{k,l\,l^\prime}(r)= \hat{j}_{l}(kr)\;
 \delta_{l, l^\prime}
+\int^\infty_0 dr^\prime\; \widetilde{G}^{K_{\rm CM}}_{k,l^\prime}(r,r^\prime)
\sum_{l^{\prime\prime}}
U^{SJ}_{l^\prime l^{\prime\prime}}(r^\prime)\;
\widetilde{u}^{SJ}_{k,l\,l^{\prime\prime}}(r^\prime).  
\end{equation}

\bibliography{betheg}

\end{document}